%% file: HIG-22-006_temp.tex
\pdfoutput=1
\documentclass[11pt,twoside,a4paper,cmspaper,final,collab]{cms-tdr}
\def\svnVersion{1d01f66}\def\svnDate{2024/10/09}\def\cmsCernNoTag{CERN-EP-2024-064}\def\cmsCernDate{\today}\def\cmsMessage{Published in the Journal of High Energy Physics as \href{http://dx.doi.org/10.1007/JHEP10(2024)061}{\doi{10.1007/JHEP10(2024)061}.}}
\begin{document}\cmsNoteHeader{HIG-22-006}

\newlength\cmsTabSkip\setlength{\cmsTabSkip}{1ex}
\providecommand{\cmsTable}[1]{\resizebox{\textwidth}{!}{#1}}

\newcommand{\PQj}{{\HepParticle{j}{}{}}\xspace}
\newcommand{\PHo}{{\HepParticle{H}{1}{}}\xspace}
\newcommand{\PHt}{{\HepParticle{H}{2}{}}\xspace}
\newcommand{\Pello}{{\HepParticle{\ell}{1}{}}\xspace}
\newcommand{\Pellt}{{\HepParticle{\ell}{2}{}}\xspace}
\newcommand{\btags}{\PQb tags\xspace}
\newcommand{\muR}{\ensuremath{\mu_{\mathrm{R}}}\xspace}
\newcommand{\muF}{\ensuremath{\mu_{\mathrm{F}}}\xspace}
\newcommand{\ttbb}{\ensuremath{\ttbar\bbbar}\xspace}
\newcommand{\ttH}{\ensuremath{\ttbar\PH}\xspace}
\newcommand{\ttV}{\ensuremath{\ttbar\PV}\xspace}
\newcommand{\ttW}{\ensuremath{\ttbar\PW}\xspace}
\newcommand{\tW}{\ensuremath{\PQt\PW}\xspace}
\newcommand{\ttbarj}{\ensuremath{\ttbar\text{+jets}}\xspace}
\newcommand{\PZj}{\ensuremath{\PZ\text{+jets}}\xspace}
\newcommand{\pp}{\ensuremath{\Pp\Pp}\xspace}
\newcommand{\ptvecX}[1]{\ensuremath{\ptvec(#1)}\xspace}
\newcommand{\pvecX}[1]{\ensuremath{\vec{p}(#1)}\xspace}
\newcommand{\kl}{\ensuremath{\kappa_{\lambda}}\xspace}
\newcommand{\kv}{\ensuremath{\kappa_{\PV}}\xspace}
\newcommand{\kz}{\ensuremath{\kappa_{\PZ}}\xspace}
\newcommand{\kw}{\ensuremath{\kappa_{\PW}}\xspace}
\newcommand{\kvv}{\ensuremath{\kappa_{2\PV}}\xspace}
\newcommand{\kww}{\ensuremath{\kappa_{2\PW}}\xspace}
\newcommand{\kzz}{\ensuremath{\kappa_{2\PZ}}\xspace}
\newcommand{\HH}{\ensuremath{\PH\PH}\xspace}
\newcommand{\bbbb}{\ensuremath{\bbbar\bbbar}\xspace}
\newcommand{\bbww}{\ensuremath{\bbbar\PW\PW}\xspace}
\newcommand{\bbtt}{\ensuremath{\bbbar\PGt\PGt}\xspace}
\newcommand{\bbZZ}{\ensuremath{\bbbar\PZ\PZ\xspace}}
\newcommand{\VVHH}{\ensuremath{\PV\PV\HH}\xspace}
\newcommand{\VVH}{\ensuremath{\PV\PV\PH}\xspace}
\newcommand{\VHH}{\ensuremath{\PV\HH}\xspace}
\newcommand{\WHH}{\ensuremath{\PW\HH}\xspace}
\newcommand{\ggZHH}{\ensuremath{\Pg\Pg\to\PZ\HH}\xspace}
\newcommand{\ZHH}{\ensuremath{\PZ\HH}\xspace}
\newcommand{\ttHH}{\ensuremath{\ttbar\HH}\xspace}
\newcommand{\bbgg}{\ensuremath{\bbbar\gamma\gamma}\xspace}
\newcommand{\Zll}{\ensuremath{\PZ\to\Pell\Pell}\xspace}
\newcommand{\Znn}{\ensuremath{\PZ\to\PGn\PAGn}\xspace}
\newcommand{\Wln}{\ensuremath{\PW\to\Pell\PAGn}\xspace}
\newcommand{\Vqq}{\ensuremath{\PZ/\PW\to\PQq\PAQq/\PQq\PAQq^\prime}\xspace}
\newcommand{\FH}{\ensuremath{\text{FH}}\xspace}
\newcommand{\twoL}{\ensuremath{\text{2L}}\xspace}
\newcommand{\oneL}{\ensuremath{\text{1L}}\xspace}
\newcommand{\met}{\ensuremath{\text{MET}}\xspace}
\newcommand{\lone}{\ensuremath{\text{L1}}\xspace}
\newcommand{\ptmisslone}{\ensuremath{p_{\mathrm{T,\lone}}^{\text{miss}}}\xspace}
\newcommand{\rhh}{\ensuremath{r_{\HH}}\xspace}
\newcommand{\Nb}{\ensuremath{N_{\PQb}}\xspace}
\newcommand{\ggF}{\ensuremath{\Pg\Pg\text{F}}\xspace}
\newcommand{\Dbb}{\ensuremath{D_{\bbbar}}\xspace}

\newcommand{\ptX}[1]{\ensuremath{\pt(#1)}\xspace}
\newcommand{\ptz}{\ptX{\PZ}}
\newcommand{\ptw}{\ptX{\PW}}
\newcommand{\ptv}{\ptX{\PV}}
\newcommand{\ptho}{\ptX{\PHo}}
\newcommand{\ptht}{\ptX{\PHt}}
\newcommand{\pthh}{\ptX{\HH}}
\newcommand{\ptlo}{\ptX{\Pello}}

\newcommand{\BDTSVB}{\ensuremath{\text{BDT}_{\text{SvB}}}\xspace}
\newcommand{\NNSVB}{\ensuremath{\text{NN}_{\text{SvB}}}\xspace}
\newcommand{\mreg}{\ensuremath{m_\text{reg}}\xspace}

\newcommand{\mV}{\ensuremath{m_{\PV}}\xspace}
\newcommand{\mH}{\ensuremath{m_{\PH}}\xspace}
\newcommand{\mHone}{\ensuremath{m_{\PHo}}\xspace}
\newcommand{\mHtwo}{\ensuremath{m_{\PHt}}\xspace}
\newcommand{\mHH}{\ensuremath{m_{\PH\PH}}\xspace}
\newcommand{\mHHgen}{\ensuremath{m_{\PH\PH}^{\text{gen}}}\xspace}
\newcommand{\mHminH}{\ensuremath{\mHone{-}\mHtwo}\xspace}
\newcommand{\mtop}{\ensuremath{m_{\PQt}}\xspace}
\newcommand{\ptcomma}[1]{\ensuremath{\pt{}_{,#1}}\xspace}
\newcommand{\ptlone}{\ptcomma{\lone}}
\newcommand{\etcomma}[1]{\ensuremath{\et{}_{,#1}}\xspace}
\newcommand{\etlone}{\etcomma{\lone}}
\newcommand{\htlone}{\ensuremath{\HT{}_{,\lone}}\xspace}
\newcommand{\deta}{\ensuremath{\Delta\eta}\xspace}
\newcommand{\dphi}{\ensuremath{\Delta\phi}\xspace}
\newcommand{\abseta}{\ensuremath{\abs{\eta}}\xspace}
\newcommand{\DeepJet}{\ensuremath{\textsc{DeepJet}}\xspace}
\newcommand{\ParticleNet}{\ensuremath{\textsc{ParticleNet}}\xspace}
\newcommand{\iso}{\ensuremath{I_{\text{PF}}}\xspace}
\newcommand{\ptcPF}{\ensuremath{\pt^{\text{charged PF}}}\xspace}
\newcommand{\ptnPF}{\ensuremath{\pt^{\text{neutral PF}}}\xspace}
\newcommand{\ptgPF}{\ensuremath{\pt^{\PGg\text{ PF}}}\xspace}
\newcommand{\ptPU}{\ensuremath{\pt^{\text{PU}}(\Pell)}\xspace}
\newcommand{\PQjsub}[1]{\ensuremath{\PQj_{\,#1}}\xspace}
\newcommand{\mjonejtwo}{\ensuremath{m_{\PQjsub{1}\PQjsub{2}}}\xspace}
\newcommand{\mjj}[1]{\ensuremath{m_{\PQj\PQj,#1}}\xspace}
\newcommand{\mll}{\ensuremath{m_{\Pell\Pell}}\xspace}
\newcommand{\DHH}{\ensuremath{D_{\HH}}\xspace}
\newcommand{\delHH}{\ensuremath{\delta_{\HH}}\xspace}
\newcommand{\BDTcat}{\ensuremath{\text{BDT}_{\text{Cat.}}}\xspace}
\newcommand{\DRHot}{\ensuremath{\DR(\PHo,\PHt)}\xspace}
\newcommand{\DEtaHot}{\ensuremath{\deta(\PHo,\PHt)}\xspace}
\newcommand{\DPhiHot}{\ensuremath{\dphi(\PHo,\PHt)}\xspace}
\newcommand{\DPhiVHt}{\ensuremath{\dphi(\PV,\PHt)}\xspace}
\newcommand{\HTex}{\ensuremath{\HT^{\text{ex}}}\xspace}
\newcommand{\Njets}{\ensuremath{N_{\text{jets}}}\xspace}
\newcommand{\Dbbbb}{\ensuremath{D_{\bbbar,1},D_{\bbbar,2}}\xspace}
\newcommand{\minDbbbb}{\ensuremath{\min(\Dbbbb)}\xspace}
\newcommand{\Ssm}{\ensuremath{\mathrm{S}_{\mathrm{SM}}}\xspace}
\newcommand{\Bkg}{\ensuremath{\mathrm{B}}\xspace}
\newcommand{\SoverB}{\ensuremath{\log_{10}\big(100(\Ssm/\Bkg)\big)}\xspace}

\cmsNoteHeader{HIG-22-006}
\title{Search for Higgs boson pair production with one associated vector boson in proton-proton collisions at \texorpdfstring{$\sqrt{s}=13\TeV$}{sqrt(s)=13 TeV}}

\date{\today}

\abstract{
A search for Higgs boson pair (\HH) production in association with a vector boson \PV (\PW or \PZ boson) is presented. The search is based on proton-proton collision data at a center-of-mass energy of 13\TeV, collected with the CMS detector at the LHC, corresponding to an integrated luminosity of 138\fbinv. Both hadronic and leptonic decays of \PV bosons are used. The leptons considered are electrons, muons, and neutrinos. The \HH production is searched for in the \bbbb decay channel. An observed (expected) upper limit at 95\% confidence level of \VHH production cross section is set at 294 (124) times the standard model prediction. Constraints are also set on the modifiers of the Higgs boson trilinear self-coupling, \kl, assuming $\kvv = 1$, and vice versa on the coupling of two Higgs bosons with two vector bosons, \kvv. The observed (expected) 95\% confidence intervals of these coupling modifiers are $-37.7<\kl<37.2$ ($-30.1<\kl<28.9$) and $-12.2<\kvv<13.5$ ($-7.2<\kvv<8.9$), respectively.
}

\hypersetup{
    pdfauthor={CMS Collaboration},
    pdftitle={Search for Higgs boson pair production with one associated vector boson in proton-proton collisions at sqrt(s) = 13 TeV},
    pdfsubject={CMS},
    pdfkeywords={CMS, HH, VHH}}

\maketitle

\section{Introduction}

Since the discovery of the Higgs boson (\PH) with a mass (\mH) of 125\GeV by the ATLAS and CMS Collaborations at the CERN LHC in
2012~\cite{Aad:2012tfa,Chatrchyan:2012xdj,CMS:2013btf}, the focus of the Higgs boson experimental and phenomenological communities
has shifted towards precise measurements of the properties of this particle and its interactions.
The measured properties of this Higgs boson so far are consistent with the standard model (SM)
predictions~\cite{CMS:2022dwd,ATLAS:2022vkf}. The production of a pair of Higgs bosons (\HH) is a rare process
that provides unique access to so far unmeasured Higgs boson couplings.
Of particular interest is the trilinear self-interaction coupling $\lambda$,
that is embedded in the \HH production in the SM, where $\lambda_{\mathrm{SM}}=\mH^2/(2v^2)$ is 0.13 for $\mH=125\GeV$ and a vacuum expectation value, $v$, of 246\GeV.

The study of coupling modifications provides a probe
of the shape of the scalar Higgs potential.
Variations from their SM values are parametrized as $\kl=\lambda/\lambda_{\mathrm{SM}}$.
The implications of a beyond-the-SM (BSM) Higgs potential are profound. With a BSM potential, the electroweak
phase transition in the early universe may not be a smooth transition from unbroken to broken electroweak symmetry as
predicted by the SM, but rather, the transition would be abrupt~\cite{COHEN1991727}. In the presence of charge-conjugation and parity violation,
the origins of baryon asymmetry could arise from electroweak
baryogenesis~\cite{Morrissey:2012db,Noble:2007kk}. These models often require deviations from the SM expectation for $\lambda$,
and measuring \kl can provide valuable insights into the underlying
physics and the existence of new phenomena beyond the SM.

In the SM, there are two other couplings whose modification could enhance some \HH production channels involving vector bosons \PV (denoting either a \PW or \PZ boson)~\cite{PhysRevD.95.073006,epjp_i2019}:
the coupling of two vector bosons with a single Higgs boson ($c_{\VVH}$), and two vector bosons with two Higgs bosons ($c_{\VVHH}$).
Their coupling modifiers relative to the SM are \kv and \kvv, respectively.
In this paper, we investigate the \VVHH couplings in two scenarios: the case where the couplings to \PW and \PZ are scaled together and the case where the couplings to \PW and \PZ are scaled independently.
In the former case, the coupling modifier \kvv represents both \kww and \kzz.
In the latter case, they are uncorrelated.
Measurements of single Higgs boson production by the ATLAS and CMS Collaborations have constrained both \kw and \kz independently
with a precision of 6--8\%, and measured values are compatible with the SM predictions within that
precision~\cite{CMS:2022dwd,ATLAS:2022vkf}. No process arising from either $c_{\VVHH}$ coupling
has been observed yet.

The \HH production modes at the LHC are, in order of decreasing SM cross section, gluon-gluon fusion
(\ggF), vector boson fusion (VBF), vector boson associated production (\VHH), and top quark associated
production (\ttHH).
In the SM with $\mH=125\GeV$,
the cross section of different \HH production modes are shown in the Table~\ref{tab:HH_XS}.
The \HH cross section is small due to the destructive interference of the contributing Feynman diagrams at leading order (LO).
The total SM cross sections for \VHH production is $\sigma_{\VHH}=0.865\,^{+5.4\%}_{-5.0\%}\unit{fb}$, computed at next-to-next-to-LO (NNLO) in quantum chromodynamics (QCD)~\cite{LHCHiggsCrossSectionWorkingGroup:2016ypw, Baglio_2013}, and is approximately half the cross section of VBF \HH production.

\begin{table}[htp!]
\centering
\topcaption{The cross sections and uncertainties of different \HH production modes~\cite{Baglio_2021,Dreyer_2018,LHCHiggsCrossSectionWorkingGroup:2016ypw, Baglio_2013}, where PDF is the parton distribution function, \alpS is the strong coupling constant, and \mtop is the top quark mass.}
\label{tab:HH_XS}
\renewcommand\arraystretch{1.2}
\cmsTable{\begin{tabular}{ccccc}
    \hline
    Production mode & Cross section (fb) & Scale uncertainty & $\text{PDF}{+}\alpS$ uncertainty  & \mtop uncertainty \\
    \hline
    \ggF & 31.05    & ${+2.2\%}/{-5.0\%}$ & $\pm3\%$  & ${+4\%}/{-18\%}$ \\
    VBF & 1.726       & ${+0.03\%}/{-0.04\%}$ & $\pm2.1\%$  & \NA\\
    \ZHH & 0.363    & ${+3.4\%}/{-2.7\%}$ & $\pm 1.9\%$ & \NA \\
    $\PWp\HH$ & 0.329 & ${+0.32\%}/{-0.41\%}$ & $\pm 2.2\%$ & \NA\\
    $\PWm\HH$ & 0.173 & ${+1.2\%}/{-1.3\%}$ & $\pm 2.8\%$ & \NA\\
    \ttHH & 0.775   & ${+1.5\%}/{-4.3\%}$ & $\pm 3.2\%$ & \NA\\
    \hline
\end{tabular}}
\end{table}

{\tolerance=800
The CMS Collaboration has searched for the \HH production process in the \bbgg~\cite{HHbbggRun2}, \bbbb~\cite{CMS:2022cpr,HH4bboostedRun2},
\bbtt~\cite{HHbbtautaRun2}, \bbZZ~\cite{HHbbZZ4lRun2}, and multilepton~\cite{HHmultileptonRun2} final states. A combination
of these searches is presented in Ref.~\cite{CMS:2022dwd}, where the \HH production cross section is constrained
at 95\% confidence level (\CL) to 3.4 times the cross section predicted by the SM. In this combination, \kl and \kvv
are constrained from searches that target both \ggF and VBF \HH production channels.
The allowed values at 95\% \CL of \kl and \kvv are
$-1.24<\kl<6.49$ (assuming $\kvv=1$) and $0.67<\kvv<1.38$ (assuming $\kl=1$).  Moreover, the $\kvv=0$ coupling strength is excluded
with a significance of 6.6 standard deviations while assuming SM values for all other couplings~\cite{CMS:2022dwd}.
\par}

{\tolerance=800
The ATLAS Collaboration has searched for the \HH production process in the \bbgg~\cite{ATLAS:2021ifb}, \bbtt~\cite{ATLAS:2022xzm},
\bbbb~\cite{ATLAS:2023qzf}, and \bbww~\cite{Aad:2019yxi} final states.
These \HH production searches have been combined with the primary single
Higgs boson decay channels~\cite{ATLAS:2022vkf} in a global search for \kl and \kvv
couplings~\cite{ATLAS:2022jtk}.
Various \HH analyses have been combined in a global search for \kl and \kvv couplings~\cite{ATLAS:2022jtk},
the former of which gains sizable contribution from the combination from single Higgs boson measurements~\cite{ATLAS:2022vkf}.
The 95\% \CL limit on the inclusive \HH production cross section is 2.4 times
the SM expectation, while the 95\% \CL allowed range for \kl is $-0.4<\kl<6.3$.
\par}

\begin{figure}[!ht]
\centering
\includegraphics[width=0.32\textwidth]{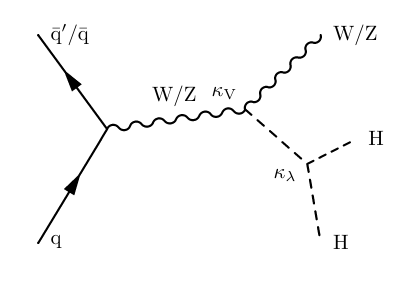}%
\hfill%
\includegraphics[width=0.32\textwidth]{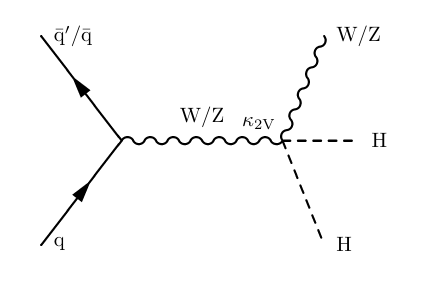}%
\hfill%
\includegraphics[width=0.32\textwidth]{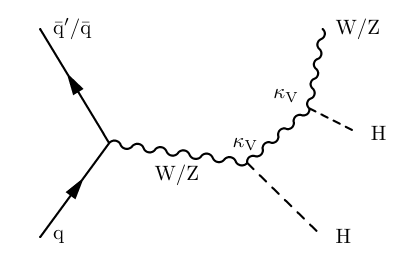}%
\caption{The three leading-order quark-initiated Feynman diagrams above result in a final state with two Higgs bosons and a \PW or \PZ boson. The left diagram requires one \kv coupling vertex and one \kl coupling vertex. The middle diagram requires only one \kvv coupling vertex, and the right diagram requires two \kv coupling vertices.}
\label{fig:feynman}
\end{figure}

Figure~\ref{fig:feynman} shows representative Feynman diagrams for \VHH production illustrating the dependence on the Higgs boson coupling modifiers.
These three types of Feynman diagrams exhibit constructive interference when all the coupling modifiers are positive,
leading to a significant enhancement in the cross section for \VHH production~\cite{PhysRevD.95.073006,epjp_i2019}.
The \VHH production cross section continues to steadily increase for positive coupling modifiers ($\kl>0$ and $\kvv>0$),
whereas other \HH production channels have sizably reduced cross sections at approximately $\kl=2$ because of the destructive interference.
As a result, \VHH stands out as an interesting channel for studying deviations from the predictions of the SM,
given its sustained enhancement and absence of a minimum in the region $\kl > 0$.
Notably, in the range of $4<\kl<7$ where the matrix element
level interference is destructive for leading production mechanisms, \VHH has constructive interference, and
the sensitivity of the \VHH search is near other subleading \HH searches.

The ATLAS Collaboration recently published a search for \VHH production~\cite{ATLAS:2022fpx} providing a 95\% \CL observed (expected)
limit on \VHH production at 183 (87) times the SM cross section. Only the leptonic channels were considered and 95\% \CL limits were set on the coupling modifiers \kl and \kvv.
The observed (expected) allowed ranges from the ATLAS \VHH search are $-34.4<\kl<33.3$
($-24.1<\kl<22.9$) (assuming $\kvv=1$) and $-8.6<\kvv<10.0$ ($-5.7<\kvv<7.1$) (assuming $\kl=1$).

This paper reports on a search for \VHH production in the final states with both Higgs bosons decaying into a bottom quark-antiquark pair, with a total branching
fraction of $\mathcal{B}(\HH \to \bbbb) = 33.9\pm 0.9\%$~\cite{LHCHiggsCrossSectionWorkingGroup:2016ypw}.
The analysis is based
on data from proton-proton (\pp) collisions produced by the CERN LHC at $\sqrt{s} = 13\TeV$.
To compensate for the low cross section of \VHH production with SM coupling strengths,
this analysis includes all the leptonic decay modes of the \PZ and \PW bosons
except for those into tau leptons and is the first analysis to include a fully hadronic \PV decay channel in the search for nonresonant \VHH production.
These decay modes encompass \Znn, \Wln, \Zll, and \Vqq, respectively, where \Pell corresponds to electrons or muons.

Experimentally, the \PV boson decay modes listed above are identified by their specific characteristics,
including the presence of a large missing transverse momentum (\ptvecmiss),
the presence of one charged lepton, the presence of two charged leptons,
and the presence of two hadronic showers (jets).
The vector \ptvecmiss is defined as the projection onto the plane perpendicular to the beam axis of the
negative vector momenta sum of all reconstructed particles in an event. Its magnitude is referred to as \ptmiss.
These characteristics help in distinguishing and categorizing the different decay modes.

The Higgs boson decaying to \bbbar can be reconstructed as two small-radius jets. In the case where the Higgs boson has
a large Lorentz boost, a single, large-radius jet provides better reconstruction efficiency than the small-radius jet
reconstruction. All channels in this paper utilize the four small-radius jets topology to select events with two Higgs boson candidates,
while two of the channels (large \ptmiss and one charged lepton channels) utilize the two large-radius jets
topology as well.

This paper is organized as follows: Section~\ref{sec:detector} provides a description of the CMS detector.
In Section~\ref{sec:samples}, the data sets and simulated event samples utilized in the study are presented.
The event reconstruction techniques are discussed in Section~\ref{sec:objects}. Section~\ref{sec:selection} outlines the event selection criteria used to identify the events of interest.
The categorization scheme and overall analysis strategy are presented in Section~\ref{sec:cate}, providing insights into the methodology employed.
The estimation and modeling of background contributions are addressed in Section~\ref{sec:background}.
Systematic uncertainties and their impact on the analysis are discussed in Section~\ref{sec:syst}.
The final analysis results are presented in Section~\ref{sec:results}.
Finally, Section~\ref{sec:summary} provides a comprehensive summary and conclusion of the analysis. The tabulated results are provided in a HEPData record~\cite{HEPData}.

\section{The CMS detector}
\label{sec:detector}

The central feature of the CMS apparatus is a superconducting solenoid of 6\unit{m} internal diameter,
providing a magnetic field of 3.8\unit{T}. Within the solenoid volume are a silicon pixel and strip
tracker, a lead tungstate crystal electromagnetic calorimeter (ECAL), and a brass and scintillator
hadron calorimeter (HCAL), each composed of a barrel and two endcap sections. Forward calorimeters
extend the pseudorapidity ($\eta$) coverage provided by the barrel and endcap detectors. Muons are detected in
gas-ionization chambers embedded in the steel flux-return yoke outside the solenoid. A more detailed
description of the CMS detector, together with a definition of the coordinate system used and the
relevant kinematic variables, can be found in Ref.~\cite{CMS:2008xjf, CMS:2023gfb}.

Events of interest are selected using a two-tiered trigger system. The first level (L1), composed of
custom hardware processors, uses information from the calorimeters and muon detectors to select events
at a rate of around 100\unit{kHz} within a fixed latency of about 4\mus~\cite{CMS:2020cmk}. The second
level, known as the high-level trigger (HLT), consists of a farm of processors running a version of
the full event reconstruction software optimized for fast processing, and reduces the event rate to
around 1\unit{kHz} before data storage~\cite{CMS:2016ngn}.

\section{Data and simulated samples}
\label{sec:samples}

The search for the \VHH signature entails exploration through four distinct channels: the missing transverse energy (MET), 1-lepton (\oneL), 2-lepton (\twoL), and fully hadronic (\FH) channels.
Each is designed to probe specific decay modes of \PW and/or \PZ bosons: \Znn, \Wln, \Zll, and \Vqq.
Therefore, dedicated trigger strategies are implemented.

The selection requirements at the trigger level are listed in Table~\ref{tab:trigger}.
QCD multijet events, characterized by the exclusive production of jets through strong interactions,
are large cross section backgrounds for the \FH and \met channels at lower transverse momentum (\pt) scales. On the other hand,
backgrounds for the \oneL and \twoL channels are mostly from electroweak processes with much lower cross sections.
Thus, \FH and \met channels' triggers require more stringent \pt requirements than those in the \oneL and \twoL channels.

\begin{table}[ht!]
\centering
\topcaption{Kinematic thresholds for L1 triggers and for the HLT are listed for each analysis channel with variations per year as needed.
HLT reconstruction is very similar to that for the offline reconstruction.
The L1 reconstruction does not include any information from tracking in the inner detector.
Transverse energy from ECAL plus HCAL systems is referred to as \etlone.
The scalar sum of \etlone from all energy deposits over a threshold of 30\GeV is denoted by \HT.
The scalar sum of $\et{}_{,\lone}$ from all energy deposits over a threshold of 30\GeV is \HT.
The \pt, $\et{}_{,\lone}$, and \HT thresholds are reported in \GeVns.
The multiplicities of \PQb-tagged jets used in the FH triggers are reported as $n$ \btags.}
\label{tab:trigger}
\renewcommand\arraystretch{1.2}
\cmsTable{\begin{tabular}{llcc}
    \hline
    Channel     & Year       & L1 trigger                                             & HLT \\
    \hline
    \met        & 2016       & $\ptmisslone>110$ or 120                           & $\ptmiss>170$ \\
    & 2017/2018  & $\ptmisslone>120$                                  & $\ptmiss>180$ \\[\cmsTabSkip]
    1 electron  & 2016       & $\etlone> 27$                                     & $\pt(\Pe)>32$ \\
    & 2017/2018  & $\etlone> 30$                                     & $\pt(\Pe)>35$ \\[\cmsTabSkip]
    1 muon      & 2017       & $\ptlone(\Pgm)>22$                              & $\pt(\Pgm)>24$ \\
    & 2016/2018  & $\ptlone(\Pgm)>25$                              & $\pt(\Pgm)>27$ \\[\cmsTabSkip]
    2 electrons & 2016--2018 & $\etlone(\Pe_1)>22$ and $\etlone(\Pe_2)>10$              & $\pt(\Pe_1)>22$ and $\pt(\Pe_2)>10$  \\[\cmsTabSkip]
    2 muons     & 2016--2018 & $\ptlone(\Pgm_1)>15$ and $\ptlone(\Pgm_2)>8$     & $\pt(\Pgm_1)>17$ and $\pt(\Pgm_2)>8$  \\[\cmsTabSkip]
    \FH         & 2016      &  $\HT>280$                          & 4 jets $\pt>45$, 3 \btags                                    \\
    &           &  $\HT>280$                                      & 4 jets $\pt> (90, 90, 30, 30)$, 3 \btags      \\[0.5\cmsTabSkip]
    & 2017      & $\HT>280$ and 4 $\times \et{}_{,\lone}> (70,55,40,35)$  & $\HT>300$ and 4 jets $\pt > (75,60,45,40)$, 3 \btags                 \\[0.5\cmsTabSkip]
    & 2018      & $\HT>320$ and 4 $\times \et{}_{,\lone}> (70,55,40,40) $                            & $\HT>330$ and 4 jets $\pt > (75,60,45,40)$, 3 \btags                \\
    \hline
\end{tabular}}
\end{table}

Selection criteria for the triggers used in this analysis evolved with data-taking conditions in the 2016 (36.3\fbinv),
2017 (41.5\fbinv), and 2018 (59.8\fbinv) data sets, where the number between parentheses refer to integrated luminosities~\cite{CMS-LUM-17-003,CMS-PAS-LUM-17-004,CMS-PAS-LUM-18-002}.
The \met channel requires a large ($>$120\GeV; $>$110\GeV for early 2016) L1 \ptmiss signature from coarsely reconstructed calorimeter energy deposits.
The minimum \ptmiss threshold at the HLT is 170 (180)\GeV for 2016 (2017--2018) data.
Anomalous high-\ptmiss events can be due to a variety of reconstruction failures, detector malfunctions, or noncollision backgrounds.
Such events are rejected by event filters that are designed to identify more than 85--90\% of the spurious high-\ptmiss events with a mistagging rate less than 0.1\%~\cite{CMS:2019ctu}.
In the \oneL channel, a single electron or muon is required, while the \twoL channel requires two electrons or muons. The \pt criteria for these leptons vary from year to year, with muons having less stringent requirements compared to electrons due to the higher likelihood of jets being misidentified as electrons rather than muons.

The trigger strategy in the \FH channel targets directly the decay products of the Higgs boson.
Events are selected at L1 using triggers requiring the presence of at least four jets in the tracker acceptance ($\abseta<2.5$) and large \HT, defined as the scalar sum of the  \pt of the reconstructed jets in the event.
During the 2016 data taking, events are required to have $\HT > 280\GeV$.
In the 2017 data set, events are required to have $\HT > 280\GeV$ and the four leading jets are required to pass staggered \pt thresholds of 70, 55, 40, and 35\GeV.
In the 2018 data set, the \HT requirement was raised to 320\GeV and the lowest jet \pt threshold was raised to 40\GeV.

Events in the \FH channel are selected in the HLT using a combination of triggers requiring the presence of jets coming from the hadronization of \PQb quarks (\PQb jets).
Events are required to have at least four jets, at least three of which are identified as arising from a bottom quark (\PQb tagged).
In the 2016 data set, events are required to have either at least four jets with transverse momentum $\pt > 45\GeV$, or two or more jets with $\pt > 90\GeV$ and two or more jets with $\pt > 30\GeV$.
In the 2017 data set, an \HT requirement of 300\GeV was added to match the threshold at L1, and the four highest-\pt jets were required to pass staggered \pt thresholds of 75, 60, 45, and 40\GeV.
The \HT threshold was raised to 330\GeV in 2018.
The \PQb tagging was performed in the HLT using the \textsc{csv} algorithm~\cite{BTV-16-002} in 2016--2017, and with the \textsc{DeepCSV} algorithm~\cite{Bols:2020bkb} in 2018.

In the leptonic channels (MET, \oneL, and \twoL), the background model is based on shapes derived from Monte Carlo (MC) simulation.
However, in the \FH channel, the background model comprises two components: MC simulation for \ttbar events and a data-driven multijet background.
The MC simulated samples of \VHH production are generated at LO in a fixed-order perturbative QCD calculation of up to four
noncollinear high-\pt partons with \MGvATNLO (v2.6.5)~\cite{MGatNLO}. The \pp interaction simulation is supplemented with
parton showering and multiparton interactions with \PYTHIA (v8.240)~\cite{Sjostrand:2014zea}.
The NNPDF3.1~\cite{EXT:NNPDF-2017} PDF is used in the simulation. The underlying event description tuned with CMS data is CP5~\cite{CMS:2019csb}.

\begin{figure}[!ht]
\centering
\includegraphics[width=0.45\textwidth]{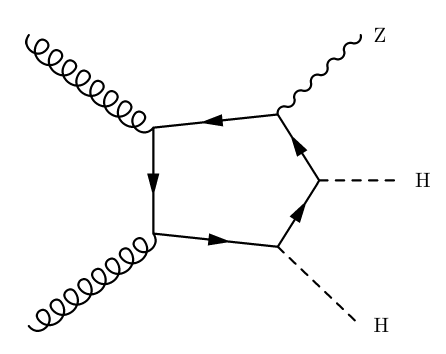}%
\hspace*{0.05\textwidth}%
\includegraphics[width=0.45\textwidth]{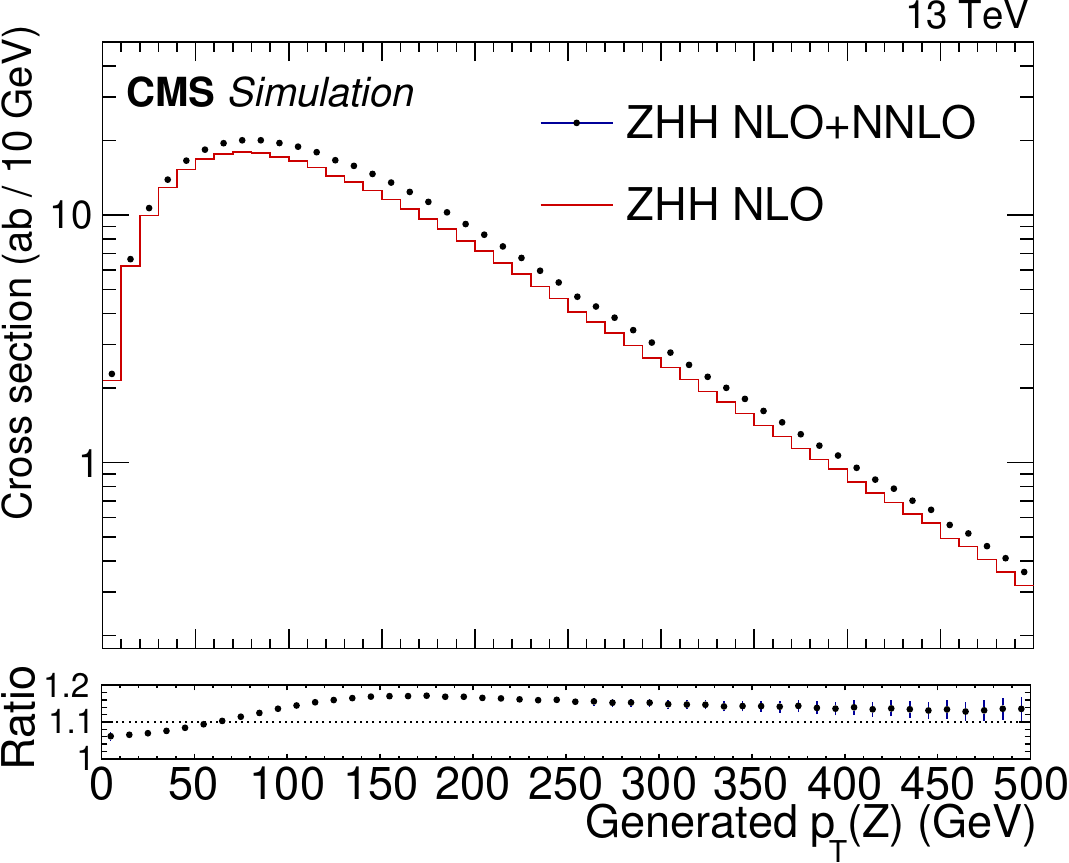}%
\caption{Left: representative Feynman diagram for \ggF \ZHH production, which represents approximately 14\% of the total cross section for \ZHH production.  Right: distribution of \ptz with and without \ggZHH process. The ratio is applied to NLO to incorporate the \ggZHH cross section enhancement. }
\label{fig:ggZHH}
\end{figure}

Additionally, samples at next-to-LO (NLO) and NNLO are generated to study the impact of higher-order corrections on kinematics.
While the kinematic distributions at NLO are found to be similar to LO, the NNLO \ggF \ZHH process (\ggZHH) exhibits, on average, higher \ptz compared to \ZHH production at NLO. A representative Feynman diagram for \ggZHH is illustrated in Fig.~\ref{fig:ggZHH} (left panel).
The LO simulated signals are first scaled to NLO in perturbative QCD with a constant.
To incorporate the \ggZHH contributions, the LO \ZHH samples are further corrected. They are reweighted as functions of \ptz to differentially account for the \ggZHH cross section enhancement. The comparison between NLO and NLO+\ggZHH is presented in Fig.~\ref{fig:ggZHH} (right panel), and the ratios shown in the lower panel represent the functions used to reweight the NLO signals.

In order to correctly infer the kinematic properties and the inclusive cross section of a \VHH signal with
arbitrary coupling modifiers (\ie, \kl, \kvv, and \kv), at least six samples, which are linearly independent in
the three-dimensional coupling space, must be generated and combined (\ie, weighted and summed).
Assuming the matrix elements of the three LO Feynman diagrams illustrated in Fig.~\ref{fig:feynman} are denoted as $A$, $B$, and $C$ under the SM, the cross section follows a relationship defined by:
\begin{equation}
\sigma( \kl, \kvv, \kv) \propto \abs{\kv\kl A + \kvv B + {\kv}^2 C}^2.
\end{equation}
Six is the minimum number of samples required, arising from the three LO Feynman diagrams yielding three squared and three interference terms. In regions of coupling space beyond the generated values, improved statistical precision is facilitated by generating additional samples beyond the minimum. This is achieved using the Moore--Penrose inverse method~\cite{Moore, Penrose}, where this analysis uses eight independent samples that are generated and combined with appropriate weights.

The \ttbarj process simulated in \POWHEG (v2.0)~\cite{Nason:2004rx,Frixione:2007vw,Alioli:2010xd,Jezo:2015aia}
is the dominant background in the \met and \oneL channels, and is a significant background in the \twoL channel.
A dedicated \ttbb MC sample is generated in a four-flavor scheme (4FS) with dedicated \textsc{powheg-box-res} and \textsc{OpenLoops} programs~\cite{Buccioni_2019,Jezo:2018yaf}, where the \PQb quark is not considered a sea quark but
kinematic effects due to the \PQb quark mass are included.
A complementary sample of simulated \ttbarj events is generated using a five-flavor scheme (5FS) with \POWHEG (v2.0)~\cite{Nason:2004rx,Frixione:2007vw,Alioli:2010xd,Jezo:2015aia}, where the \PQb quark is included in the sea but considered massless.
For both samples, the \pp interaction simulation is supplemented with \PYTHIA (v8.230) CP5 tune~\cite{Sjostrand:2014zea,CMS:2019csb} and NNPDF3.1 PDF~\cite{EXT:NNPDF-2017}.
The \ttbb process is defined as events having additional \Pb jets at the particle level that do not originate from the top quark decays and that fulfill the acceptance requirements of $\pt>20\GeV$ and $\abseta<2.4$. The other events that do not fulfill this requirement are included in the \ttbar process, and thus \ttbb and \ttbar are mutually exclusive.
The \ttbb process in the 4FS sample is combined with the \ttbar process in the 5FS sample to yield the complete \ttbarj process.

While \PZj events are a negligible background to the \FH and \oneL channels, the Drell--Yan and \PZj processes are
substantial background in the \twoL and \met channels. Drell--Yan production is generated with
\MGvATNLO (v2.6.5)~\cite{MadGraph} using the MLM~\cite{Alwall:2007fs} matching prescription at LO with up to 4 jets in the matrix element
and is corrected to NLO kinematics with \ptz reweighting.
The samples are
normalized to the cross section at NNLO~\cite{nnlovjet,Ferrera:2013yga,Brein:2012ne,Ferrera:2014lca}.

Other background processes generated by the \MGvATNLO (v2.6.5)~\cite{MadGraph} are \ttV (where \PV is \PW or \PZ and FxFx matching~\cite{Frederix:2012ps} is used for \ttW), and single top $s$-channel.
The single top $t$-channel is generated with \textsc{comphep} (v4.4)~\cite{BOOS2004250}.
The \POWHEG (v2.0)~\cite{Nason:2004rx,Frixione:2007vw,Alioli:2010xd,Jezo:2015aia} generator is used to simulate events for single top quark \tW and \ttH production.
All the samples are interfaced with \PYTHIA (v8.240)~\cite{Sjostrand:2014zea} CP5~\cite{CMS:2019csb} tune and NNPDF3.1~\cite{EXT:NNPDF-2017} PDF for parton showering, fragmentation, and hadronization.
In the MC that has Higgs boson production, the \mH is set to be 125\GeV. A full detector simulation is performed for all MC samples with \GEANTfour~\cite{EXT:GEANT4-2002}.

Processes involving a \PW boson with additional jets, two top quarks with two vector bosons, two vector bosons, three vector bosons, one top quark with a \PZ boson, and one Higgs boson with a vector boson are negligible in all analysis channels based on simulations at NLO.

\section{Event reconstruction}
\label{sec:objects}

Global event reconstruction, utilizing the particle-flow (PF) algorithm~\cite{CMS:2017yfk}, is designed to reconstruct and identify individual particles within an event (PF candidates) by optimizing information from subdetectors.
Particle identification, in this process, plays an important role in the determination of the particle direction and energy.
Photons are identified as ECAL energy clusters not associated with charged particle trajectories.
Electrons are identified as primary tracks with corresponding ECAL energy clusters, accounting for bremsstrahlung photons within the tracker material.
Muons are recognized as tracks consistent with either muon system hits or tracks with calorimeter deposits in line with the muon hypothesis.
Charged hadrons are identified as tracks distinct from electrons or muons,
while neutral hadrons are recognized as HCAL energy clusters without associated charged hadron trajectories or as an aggregate of ECAL and HCAL energy exceeding the expected charged hadron energy deposit.

The primary vertex (PV) is taken to be the vertex corresponding to the hardest scattering in the event, evaluated
using tracking information alone~\cite{CMS-TDR-15-02}.
For each event, hadronic jets are clustered from the reconstructed PF particles using the infrared and collinear
safe anti-\kt algorithm~\cite{Cacciari:2008gp, Cacciari:2011ma} with a distance parameter of 0.4 (denoted small-radius
jets).
As the Lorentz boost increases, hadron pairs from single-particle decays become more collimated,
rendering the reconstruction of separate small-radius jets inefficient as the outer perimeters start to overlap.
To enhance efficiency in this high signal-to-background scenario,
an alternative reconstruction of hadronic jets with the anti-\kt algorithm with a distance parameter of 0.8
(denoted large-radius jets) is used.

For small-radius jets, jet momentum is determined as the vectorial sum of all particle momenta in the jet, and is found from simulation to be,
on average, within 5--10\% of the true momentum over the whole \pt spectrum and detector acceptance. Additional
\pp interactions within the same or nearby bunch crossings (pileup) can contribute additional tracks and
calorimetric energy depositions to the jet momentum. To mitigate this effect, charged particles identified to be
originating from pileup vertices are discarded and an offset correction is applied to correct for remaining contributions.
Jet energy corrections are derived from simulation to bring the measured response of jets to that of particle level jets on
average. In situ measurements of the momentum balance in dijet, photon${}+{}$jets, $\PZ+{}$jets, and
multijet events are used to account for any residual differences in the jet energy scale between data and
simulation~\cite{CMS:2016lmd}. The jet energy resolution amounts typically to 15--20\% at 30\GeV, 10\% at 100\GeV, and 5\%
at 1\TeV~\cite{CMS:2016lmd}. Additional selection criteria are applied to each jet to remove jets potentially dominated by
anomalous contributions from various subdetector components or reconstruction failures. All selected jets in this analysis
have an additional \PQb jet energy regression applied~\cite{breg}.
The regression improves the resolution of the dijet mass of reconstructed Higgs boson candidates in signal simulation by 10--15\%, as a function of the dijet \pt.

Reconstructed jets must have sufficient fractions of energy from various PF candidates
(\ie, neutral hadron, charged hadron, and neutral electromagnetic) required via \textsc{JetID} thresholds~\cite{Sirunyan:2020foa} and the medium working point
for the \textsc{PileupJetID}~\cite{Sirunyan:2020foa}.
Since \ptvecmiss represents the negative vector sum of momenta from all reconstructed particles in an event, the jet energy corrections are propagated to \ptvecmiss for the energy balance.

For large-radius jets, the pileup-per-particle identification algorithm~\cite{Sirunyan:2020foa,Bertolini:2014bba} is used to
mitigate the effect of pileup at the reconstructed-particle level, making use of local shape information, event
pileup properties, and tracking information. A local shape variable is defined that distinguishes between
collinear and soft diffuse distributions of other particles surrounding the particle under
consideration; the
former is attributed to particles originating from the hard scatter and the latter to particles originating from
pileup interactions. Charged particles originating from pileup vertices are discarded. For each
neutral particle, a local shape variable is evaluated using the surrounding charged particles compatible with the
PV within the tracker acceptance ($\abseta < 2.5$), and using both charged and neutral particles in
the region outside of the tracker coverage. The momenta of the neutral particles are then rescaled according to
their probability to originate from the PV deduced from the local shape variable, obviating
the need for jet-based pileup corrections~\cite{Sirunyan:2020foa}.

Small-radius \PQb jets are identified with a deep neural network (\DeepJet) trained on
all PF inputs~\cite{Bols:2020bkb,CMS-DP-2018-058},
achieving an expected selection efficiency of 80\% with misidentification rate of 1 (15)\% of
light-flavor (charm-flavor) jets per selected jet for the medium working point~\cite{BTV-16-002}.
The number of \PQb-tagged small-radius jets, denoted by \Nb, is defined as the number of jets passing a certain working point of this tagger.
For the channels with large-radius jets, a graph neural network called \ParticleNet~\cite{Qu:2019gqs} is used to identify large-radius jets originating from the hadronization of two \PQb quarks.
The \ParticleNet algorithm outputs \Dbb, which represents the probability that a given jet originates from the hadronization of a \PQb quark pair.
For the tight, medium, and loose working points,
\ParticleNet achieves expected selection efficiencies of 60, 80, and 90\%,
with misidentification rate of 0.3, 1, and 2\%, respectively~\cite{HH4bboostedRun2}.

\section{Event selection}
\label{sec:selection}

In the scenario of SM couplings, only about 40 events
would be produced in the \bbbb decay channel of \VHH production for the integrated luminosity of 138\fbinv without any selection applied.
In the leptonic channels, to maximize signal efficiency, offline thresholds on the objects used
in the trigger selection are close to the trigger thresholds.
The trigger efficiencies are measured in data, and simulation is corrected to match data as a
function of relevant object \pt, $\eta$, and azimuthal angle~$\phi$.
The FH channel uses offline jet thresholds below the trigger thresholds; including events that pass the trigger but are below the region that is fully efficient leads to a significant increase in signal acceptance.
The efficiency of the \HT (jet-level \pt and \PQb-tagging) requirements used in the trigger are measured as a function of the offline \HT (jet \pt).
These efficiency measurements, referred to as trigger turn-ons, are measured in data and simulation using leptonic (electron${}+{}$muon) \ttbar events, triggered by the leptons.
The \HT and jet-level trigger turn-ons are combined to correct the per-event trigger efficiency in simulation to match that of data.

In order to reduce contribution from backgrounds with misidentified leptons,
electrons and muons in the \oneL and \twoL channels are required to be isolated.
The specific isolation variable computed for this analysis is:
\begin{equation}
\iso \equiv \frac{1}{\ptX{\Pell}} \left( \sum_{\text{in cone}} \ptcPF + \max\left[ 0, \sum_{\text{in cone}} \ptnPF +  \sum_{\text{in cone}} \ptgPF - \ptPU \right] \right),
\end{equation}
where \ptcPF, \ptnPF, and \ptgPF are the reconstructed \pt of the
charged, neutral, and photon PF candidates inside the cone centered on (but not including) the
lepton track.  The \ptPU term is a pileup contribution, which is subtracted to remove momenta from particles not originating from  the lepton's \pp collision. The separation from the lepton track is measured as $\DR = \sqrt{\smash[b]{(\deta)^2 + (\dphi)^2}}$, where \deta and \dphi are the $\eta$ and $\phi$ differences, respectively.
The cone size is $\DR = 0.4$ (0.3) for electrons (muons), and the isolation variable is normalized by the lepton \pt.
In the \oneL channel, leptons must have $\iso<0.06$, which is a stringent selection criterion
to remove any QCD multijet background in this channel. Furthermore, to ensure compatibility with a \PW boson,
the lepton and \ptvecmiss must satisfy the requirement $\dphi(\ptvecX{\Pell},\ptvecmiss)<2.0$.
In the \twoL channel, the isolation requirement is $\iso<0.15$ (0.25) for electrons (muons), which is much less
stringent due to the intrinsic absence of substantial reducible backgrounds in the \twoL channel.

Electrons are required to pass selection criteria based on the ECAL energy shape, compatibility of the associated track momentum and ECAL energy, and the
absence of energy deposition in the HCAL.  These criteria are optimized with a boosted decision tree
(BDT)~\cite{CMS:2020uim,CMS-DP-2020-021}.  In the \twoL (\oneL) channel, the medium (tight) BDT working point is
required. Muons in both \oneL and \twoL channels are required to pass the ``tight'' identification requirements, as detailed in Ref.~\cite{CMS:2018rym}.

To further eliminate reducible backgrounds, such as QCD multijet events in \met and \oneL channels and \ttbar events in the \twoL channel,
the vector boson is reconstructed using \ptvecmiss, $\ptvecX{\Pell}+\ptvecmiss$, $\pvecX{\Pello}+\pvecX{\Pellt}$, and
$\pvecX{\PQj_1}+\pvecX{\PQj_2}$ in the \MET, \oneL, \twoL, and \FH channels where $\PQj_1$ and $\PQj_2$ stand for the leading and subleading jets ordered by their \pt,
respectively. Kinematic thresholds
on leptons, jets, \ptmiss, and reconstructed vector bosons are described in Table~\ref{tab:selection}.

\begin{table}[ht!]
\centering
\topcaption{Thresholds on kinematic variables for all selected objects are listed for each channel.
Objects are always required to be within the acceptance of the CMS subdetectors, which is
$\abseta<2.5$ for electrons and 2.4 for all other objects, as well as outside of barrel-endcap
transition regions near $\abseta\sim1.5$. The dijet mass of the two jets with lower \PQb
tagging scores than the Higgs candidate jets in the \FH channel is denoted \mjj{\PV}.}
\label{tab:selection}
\renewcommand\arraystretch{1.4}
\cmsTable{\begin{tabular}{cccc}
    \hline
    \multirow{2}{2.4cm}{\centering Channel} & Vector boson decay & Vector boson & \multirow{2}{2.4cm}{\centering Jet selection} \\[-5pt]
    & products selection & selection &  \\
    \hline
    \multirow{2}{2.4cm}{\centering \met small-radius} & & \multirow{2}{4.0cm}{\centering $\ptmiss>150\GeV$}& \multirow{2}{4.0cm}{\centering $\geq$4 small-radius jets\\ with $\pt>35\GeV$} \\
    & & & \\[\cmsTabSkip]
    \multirow{2}{2.4cm}{\centering \met large-radius} & & \multirow{2}{4.0cm}{\centering $\ptmiss>250\GeV$} & \multirow{2}{4.0cm}{\centering $\geq$2 large-radius jets\\ with $\pt>200\GeV$} \\
    & & & \\[\cmsTabSkip]
    \multirow{6}{*}{\oneL} & \multirow{6}{4.0cm}{\centering $\ptX{\Pe}>32$ (28)\GeV\\
    2018/2017 (2016)\\
    OR\\
    $\ptX{\PGm}>25\GeV$
    } & \multirow{6}{*}{$\ptw>125\GeV$}
    & \multirow{6}{4.0cm}{\centering $\geq$3 small-radius jets\\
    with $\pt>25\GeV$ AND\\
    $\geq$4 small-radius jets\\
    with $\pt>15\GeV$\\
    OR\\
    $\geq$2 large-radius jets\\
    with $\pt>200\GeV$
    } \\
    & & & \\
    & & & \\
    & & & \\
    & & & \\
    & & & \\[\cmsTabSkip]
    \multirow{3}{*}{\twoL} & $\ptX{\PGm_{1[2]}}>20$ $[20]$\GeV & & \multirow{3}{4.0cm}{\centering
    $\geq$4 small-radius jets \\
    with $\pt>20\GeV$ } \\
    & OR & $\ptX{\Pell\Pell}>50\GeV$  & \\
    & $\ptX{\Pe_{1[2]}}>25$ $[20]$\GeV & & \\[\cmsTabSkip]
    \multirow{4}{*}{\FH} & \multirow{4}{*}{$\ptX{\PQj_i}>20\GeV$} & \multirow{4}{*}{$65<\mjj{\PV}<105\GeV$} & \multirow{4}{4.0cm}{\centering $\geq$4 small-radius jets\\
    with $\pt>40\GeV$ and\\
    $\geq$6 small-radius jets\\
    with $\pt>20\GeV$
    } \\
    & & & \\
    & & & \\
    & & & \\ \hline
\end{tabular}}
\end{table}

When jet \pt is inaccurately estimated, it can result in large \ptmiss.
These events are vetoed from the \met channel by requiring that the \dphi between
\ptvecmiss and all Higgs boson decay candidate jets must be above a \ptvecmiss-dependent threshold:
\begin{equation}
\dphi>0.4\exp\big(4-\ptmiss/50\GeV\big)+0.07.
\end{equation}
Since the \ptvecmiss spatial resolution is worse at lower values of \ptmiss,
this function is designed to impose a larger \dphi criteria at lower values of
\ptvecmiss. This selection removes events with wrongly reconstructed \ptmiss
that are not modeled in the simulation.

Jets in all channels have standard quality selection criteria applied to avoid selecting reconstructed jets resulting
from noise in the HCAL or ECAL~\cite{Sirunyan:2020foa}. All leptonic channels require at least four small-radius jets
or two large-radius jets. These jets are ordered by \DeepJet \PQb tagger score or \Dbb. The leading four
small-radius jets or two large-radius jets are selected as the Higgs boson candidate jets. In the \FH channel, at least six
small-radius jets are needed. The four jets with the highest \DeepJet score are considered to be the Higgs boson candidate jets and at least three of them must pass a threshold of 0.6 in the \DeepJet \PQb tagging
score optimized to maximize the expected significance; this threshold is more stringent than the medium working point used in the leptonic channels.
In other jets (excluding the Higgs boson candidates), at least one pair must be able to form a dijet with invariant mass
$65<\mjj{\PV}<105\GeV$. Among the candidates, the pair with leading \pt is selected to reconstruct the vector boson.

\begin{figure}[!ht]
\centering
\includegraphics[width=0.8\textwidth]{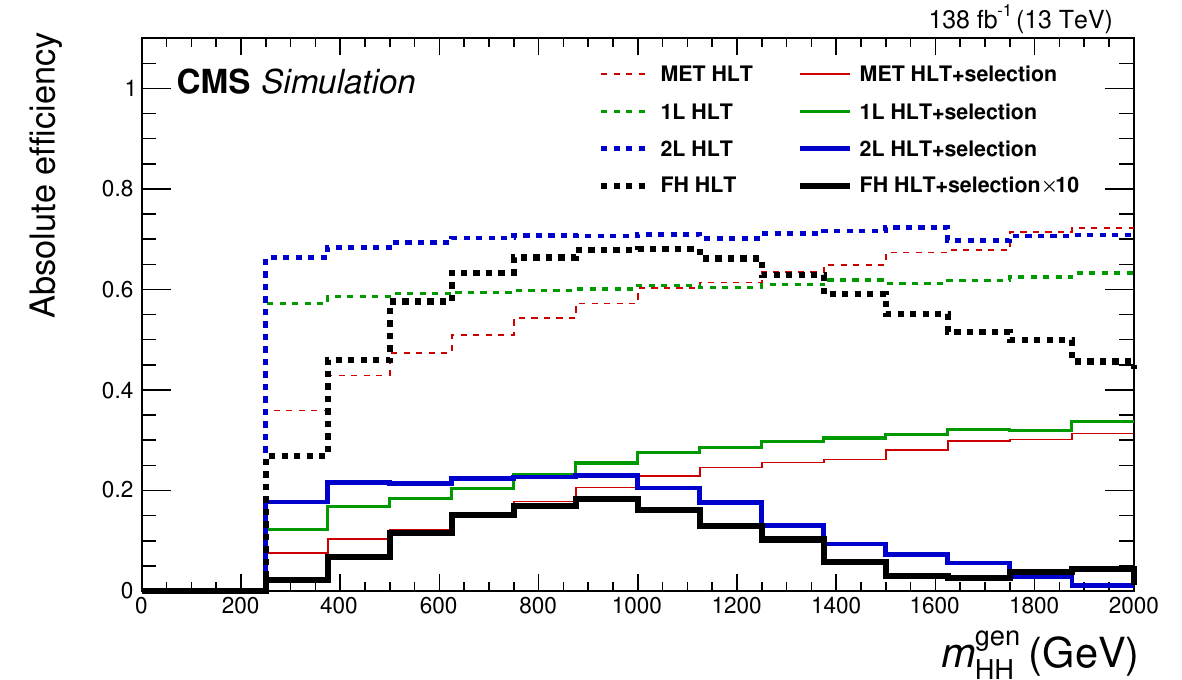}
\caption{The SM \VHH efficiencies of trigger selections (dashed lines) and full selections (solid lines) are shown for all four analysis channels.
Decays involving tau lepton decays are not considered for the \oneL and \twoL channel efficiencies.
The full selection efficiency in the \FH channel is scaled up by 10 for visibility.
Both sets of efficiencies are absolute efficiencies (acceptance times selections efficiencies).}
\label{fig:eff}
\end{figure}

Figure~\ref{fig:eff} shows the dependence of the absolute efficiencies (acceptance times trigger and analysis selections efficiencies) on the
mass of \HH at generator level, \mHHgen.  The leptonic channels generally have high efficiency after HLT selection because no jets are required.
The \FH selection requires four jets at the HLT and thus its efficiency is generally lower.
After applying all the analysis selections, the \oneL and \met channel still maintain relatively high efficiencies at high \mHHgen because of the inclusion of the large-radius jet regions.

{\tolerance=800
In channels featuring four small-radius jets, there exist three combinations of dijet pairings for reconstructing the two Higgs boson candidates.
To optimize the selection of signal events and prevent biasing kinematic features of background processes,
a metric is employed that aims to balance the masses of the two Higgs boson candidates.
For each pairing, the variable \DHH is calculated as
$\DHH=\abs{\mjj{1}-c \mjj{2}}$,
where \mjj{1} and \mjj{2} represent the masses of the Higgs boson candidates~\cite{CMS:2022cpr},
and $c$ is a correction factor between \mjj{1} and \mjj{2}.
The pairing with the smallest value of \DHH is chosen.
\par}

\section{Categorization and analysis strategy}
\label{sec:cate}

Categorization is a critical aspect of the \VHH analysis. The kinematic distributions of the \VHH process vary differently for \kl or \kvv enhancements. The general strategy is to separate the regions that are most important for the two types of signal enhancements and to optimize in each region separately. Instead of creating categories based on simple kinematic properties, a categorization BDT (\BDTcat) is used in most channels to define two signal regions: one with enhanced signal for large \kvv couplings and another for large \kl couplings. In each of these categories, a multivariate algorithm, either a BDT (\BDTSVB) or neural network (\NNSVB), is used as fit observable for signal vs background separation.

Events are subdivided into different
regions based on the invariant mass of the Higgs boson pair \mHH, the number of \PQb-tagged
jets \Nb, and further into regions enhanced in sensitivity to \kl and \kvv, and, for the
\twoL channel, a \ttbar-enhanced region. In the \twoL channel, we define an additional \ttbar-enriched
category by selecting all events that fall outside a \PZ boson mass window in the dilepton mass
\mll: $80<\mll<100\GeV$. This category constrains some systematic uncertainties
in the \ttbar process simulations. The regions where all four jets are \PQb-tagged are the most significant.

When correctly reconstructed, the \HH signal is concentrated in the region where both Higgs boson candidates
have masses near 125\GeV. There are reconstruction effects that cause the peak value of the Higgs boson
candidates to not be reconstructed at 125\GeV. The candidate mass is corrected by a factor $r$ so that the
peak value of the signal is at generator mass of 125\GeV. These simulation-based correction factors $r_1$ ($r_2$) for the first
(second) Higgs boson candidate are determined to be $r_1 = r_2 = 1$ for the leptonic channels, and $r_1 = 1.02$
and $r_2 = 0.98$ for the \FH channel. To quantify how close a given event is to the expected \HH signature, we
then define two variables \rhh and \delHH quantifying the distance of the reconstructed candidate masses
to the expected Higgs boson mass in simulation:
\begin{equation*}
  \rhh=\sqrt{(\mHone- 125\GeV \,r_1)^2 + (\mHtwo - 125\GeV \,r_2)^2},
\end{equation*}
and
\begin{equation*}
  \delHH=\sqrt{\Bigg(\frac{\mHone-125\GeV \,r_1}{\mHone}\Bigg)^2 + \Bigg(\frac{\mHtwo-125\GeV \,r_2}{\mHtwo}\Bigg)^2},
\end{equation*}
where \mHone and \mHtwo are the masses of the Higgs boson candidates.

Signal regions (SRs), control regions (CRs), and sideband regions (SBs) are defined by requirements on \rhh or \delHH:
For the leptonic channels, the SRs are defined by $\rhh<25\GeV$, the CRs by $25<\rhh<50\GeV$, and
the SBs by $50<\rhh<75\GeV$, similar to Ref.~\cite{CMS:2022cpr}.
For the \FH channel, the SR is defined by $\delHH<0.19$, while the region beyond that defines the SB.
In the leptonic channels, SRs with three or four \PQb jets have significant signal yields, and are thus used for signal extraction.
As expected, the category with $\Nb=4$ is the most
significant.  In the \twoL channel, $\Nb=4$ and 3 are optimized and statistically analyzed separately,
while in other leptonic channels they are validated separately but analyzed together. The electron and
muon channels are merged in \oneL and \twoL channels.  In the \twoL channel, data (and simulation) from all years
are merged to improve statistical precision in the background estimation.

\begin{figure}[!t]
\centering
\includegraphics[width=0.32\textwidth]{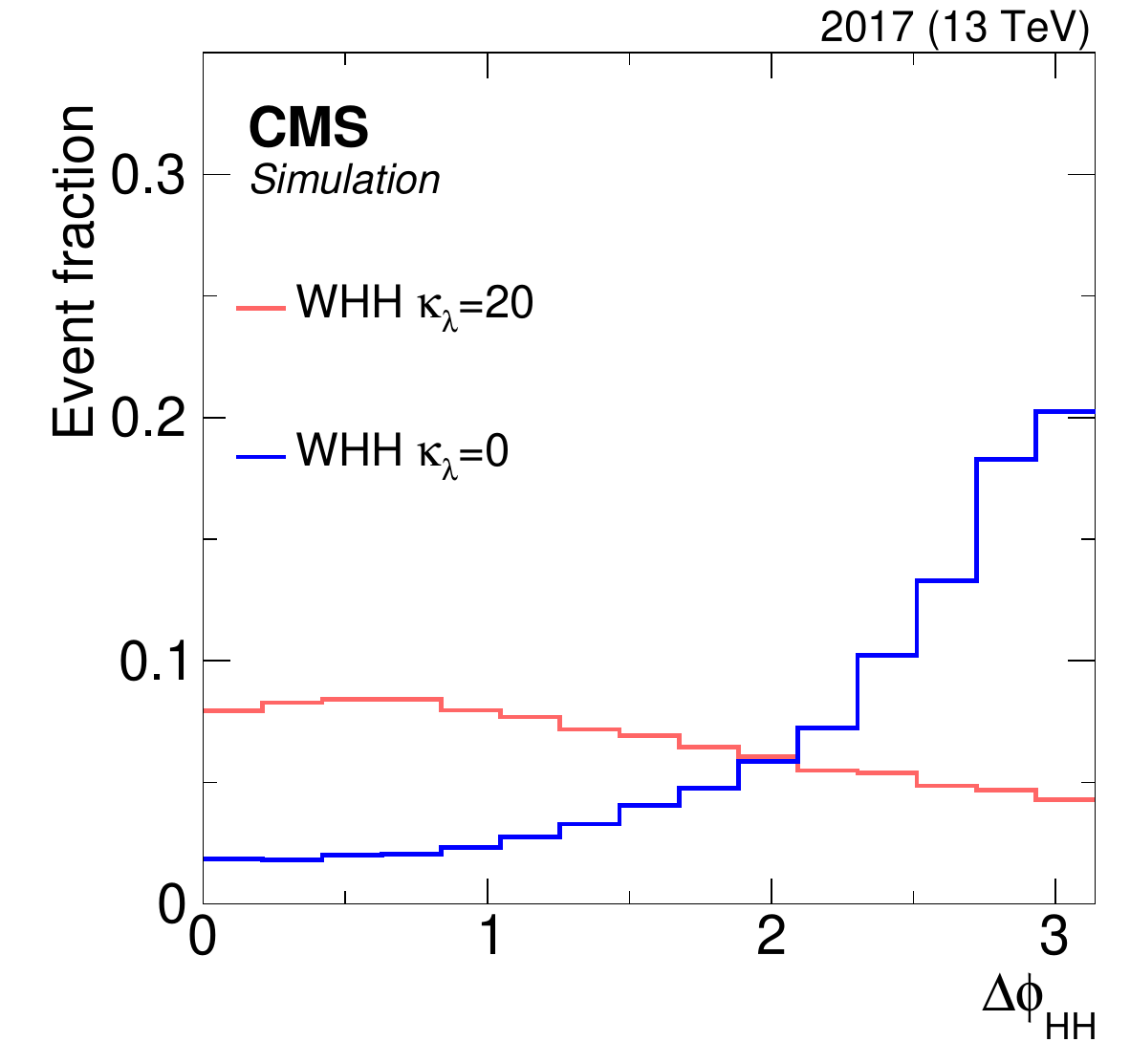}%
\hfill%
\includegraphics[width=0.32\textwidth]{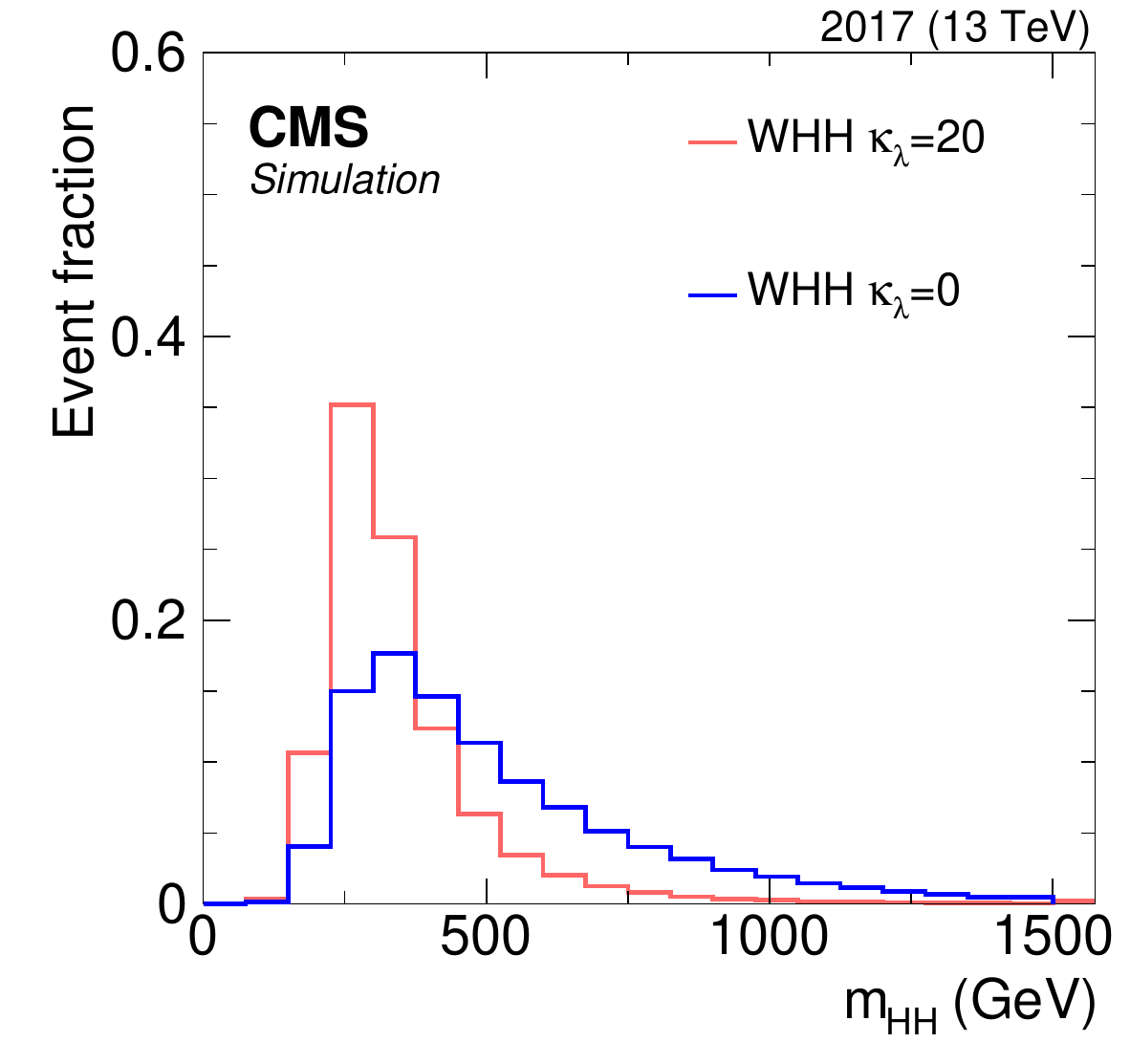}%
\hfill%
\includegraphics[width=0.32\textwidth]{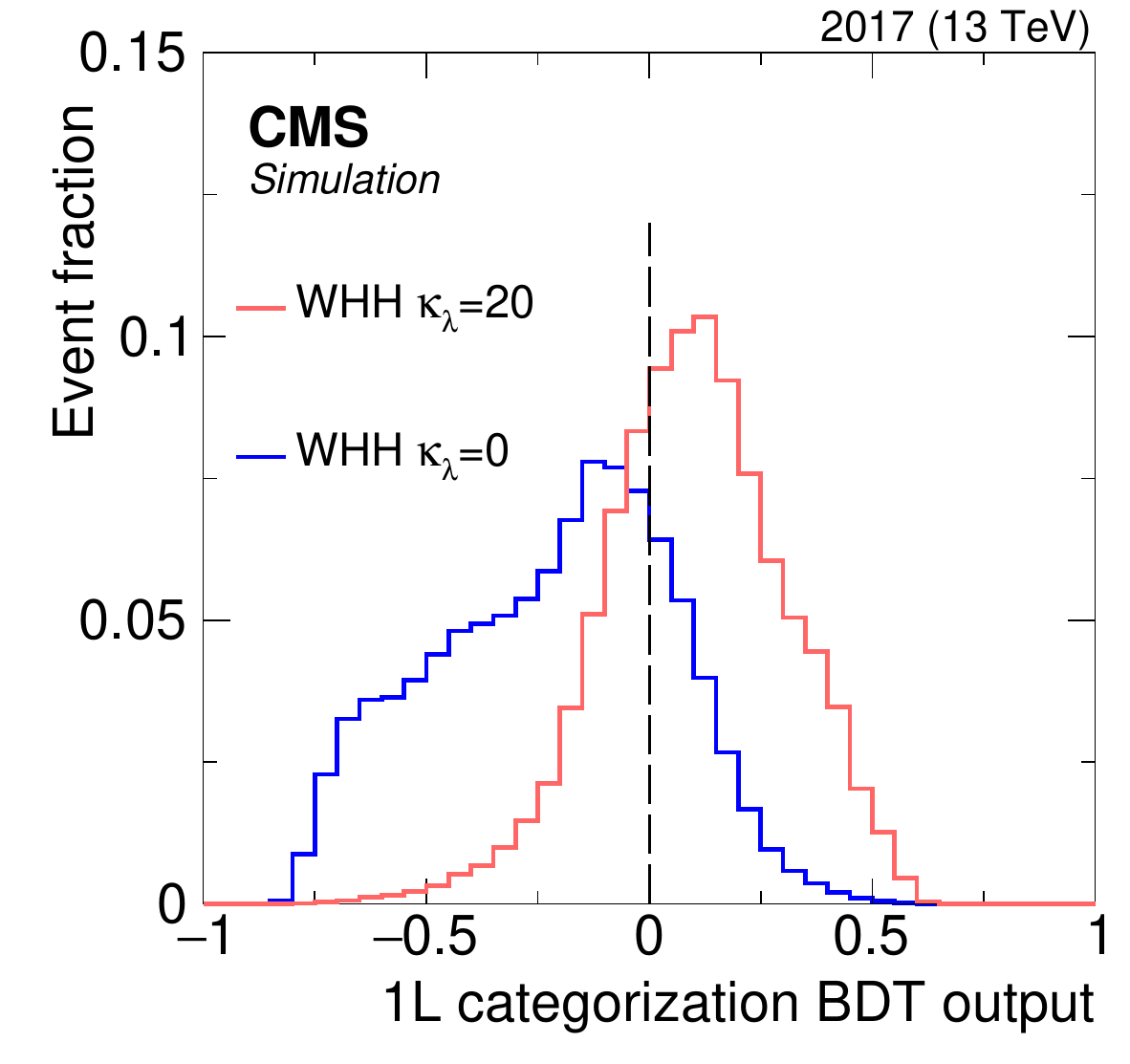}%
\caption{Kinematic distributions of the \HH signal for different coupling strengths.
Left and middle: azimuthal angle between the two reconstructed Higgs boson
candidates, $\dphi_{\HH}$, and the reconstructed \HH mass, \mHH, in the \oneL SR for two different coupling values, $\kl=20$ and 0.
Right: the categorization BDT output for the same two models.  The dashed vertical line shows where
the categorization boundary is set.}
\label{fig:catbdt}
\end{figure}

In all channels, we maximize the sensitivity for enhanced SM coupling strengths, particularly on \kl and \kvv, by
leveraging kinematic variables whose distributions are significantly different when one or the other coupling of
interest is larger than the SM prediction.  To optimize the separation of these kinematic regions, a BDT is trained to
separate events according to the signal kinematics into a category with enhanced \kl contribution ($\kl = 20$ in training) and one with
enhanced \kvv contribution ($\kl = 0$ in training) using the \textsc{Scikit-learn} software~\cite{pedregosa2011scikit}.
In the latter case, reducing the \kl coupling to zero
effectively enhances \kvv.  Other pairs of enhanced coupling samples were tested in the training of candidate
\BDTcat.  The selected pair, $\kl=20$ and 0,
was among several that were equally performant in the separation of signal kinematics.
The categorization BDT is used to define \kl-enriched and \kvv-enriched SRs, which are
distinct and optimized separately for signal versus background separation.
Figure~\ref{fig:catbdt} highlights one of many kinematic inputs to
a categorization BDT (from the \oneL channel as an example), as well as the BDT output, for two
simulated signal samples.
Separate BDTs are trained for each channel due to the varying mass resolution and kinematic effects of
the trigger and selection criteria. The categorization boundary value in the \BDTcat is optimized to
separate the signal models in the training. This procedure was validated by varying the boundary and
comparing final results.
The same procedure is implemented in each channel.
The variables used as input to the BDT trainings are
listed in Table~\ref{tab:catbdtinputs}.

\begin{table}[!p]
\centering
\topcaption{Variables used in the categorization BDTs for the separation of the \kl- and
\kvv-enriched regions, as well as in \BDTSVB and \NNSVB for extracting signal-like events. The $\checkmark$ symbol indicates that the BDTs include the variable.
These variables include the reconstructed Higgs boson with leading (\PHo) and subleading (\PHt) transverse momentum,
the Higgs boson candidate jets ordered by the \DeepJet \PQb tagging score (\PQjsub{1,2,3,4}),
the other jets (\PQjsub{\mathrm{other}}),
the dijets formed by any two jets ($\PQj_i\PQj_k$ with $i,k\in\{1,2,3,4\}$),
the scalar sum of the transverse energy of all the jets excluding \PQjsub{1,2,3,4} (\HTex),
the number of jets (\Njets),
the selected leptons in the \twoL channel (\Pello, \Pellt),
the $N$-subjettiness~\cite{EXT:Subjettiness-2011} ratio $\tau_2/\tau_1$ and $\tau_3/\tau_2$.
The small-radius (large-radius) regions are designated with an ``S'' (``L'') in parentheses.}
\label{tab:catbdtinputs}
\renewcommand\arraystretch{1.3}
\newcolumntype{C}[1]{>{\centering\let\newline\\\arraybackslash\hspace{0pt}}m{#1}}
\cmsTable{\begin{tabular}{l@{\hspace{6\tabcolsep}}C{1.5cm}C{1.5cm}C{1.5cm}@{\hspace{6\tabcolsep}}C{1.5cm}C{1.5cm}cC{1.5cm}@{\hspace{6\tabcolsep}}c}
    \hline
                    & \multicolumn{3}{c@{\hspace{6\tabcolsep}}}{\BDTcat}                      & \multicolumn{4}{c@{\hspace{6\tabcolsep}}}{\BDTSVB}                                 & \NNSVB \\
    Input variable   & \met/\oneL              & \twoL      &\FH        &\met(S)       & \oneL(S)      & \met/\oneL (L)& \twoL      & \FH \\
    \hline
    \ptv, \ptho     &   \checkmark            & \checkmark &\checkmark&  \checkmark  &   \checkmark  &   \checkmark  & \checkmark & \\
    \pthh           &   \checkmark            &            &          &  \checkmark  &   \checkmark  &   \checkmark  & \checkmark & \\
    \ptht           &   \checkmark            &            &          &  \checkmark  &   \checkmark  &   \checkmark  &            & \\
    \mHone          &   \checkmark            &            &          &  \checkmark  &   \checkmark  &   \checkmark  & \checkmark & \\
    \mHtwo          &   \checkmark            &            &          &  \checkmark  &   \checkmark  &   \checkmark  &            & \\
    \mHH            &   \checkmark            & \checkmark &\checkmark&  \checkmark  &   \checkmark  &   \checkmark  & \checkmark & \\
    \DRHot          &   \checkmark            &            &\checkmark&              &               &               &            & \\
    \DPhiVHt        &   \checkmark            &            &\checkmark&  \checkmark  &   \checkmark  &   \checkmark  &            & \\
    $\ptht/\ptho$   &   \checkmark            &\checkmark  &\checkmark&              &               &               &            & \\
    \DEtaHot        &   \checkmark            &            &\checkmark&  \checkmark  &               &               &            & \\
    \DPhiHot        &   \checkmark            &            &\checkmark&  \checkmark  &   \checkmark  &   \checkmark  &            & \\
    Energy of \PHo  &   \checkmark            &            &\checkmark&              &               &               & \checkmark & \\
    Energy of \PHt  &   \checkmark            &            &\checkmark&              &               &               &            & \\
    Energy of \HH   &   \checkmark            &            &\checkmark&              &               &               & \checkmark & \\
    $\eta_{\HH}$    &   \checkmark            &            &\checkmark&              &               &               &            & \\
    $\eta_{\PHo}$   &                         &            &\checkmark&  \checkmark  &               &               &            & \\
    $\phi(\PV)$     &                         &            &          &  \checkmark  &   \checkmark  &   \checkmark  &            & \\
    $s_{\text{\PQb-tag}}(\PQjsub{1,2,3,4})$ & &            &          &  \checkmark  &   \checkmark  &               &            & \\
    \HTex           &                         &            &          &  \checkmark  &               &               &            & \\
    \mV             &                         &            &          &  \checkmark  &               &               & \checkmark & \\
    \DRHot          &                         & \checkmark &          &  \checkmark  &               &               & \checkmark & \\
    \DPhiVHt        &                         & \checkmark &          &  \checkmark  &               &               & \checkmark & \\
    \DRHot          &                         &            &          &  \checkmark  &               &               & \checkmark & \\
    \Njets          &                         &            &          &  \checkmark  &               &               &            & \checkmark \\
    $\tau_2/\tau_1(\PHo, \PHt)$ &             &            &          &              &               &   \checkmark  &            & \\
    $\tau_3/\tau_2(\PHo, \PHt)$ &             &            &          &              &               &   \checkmark  &            & \\
$\ptX{\Pellt}/\ptlo$  &                       & \checkmark &          &              &               &               & \checkmark & \\
$\dphi(\Pello,\Pellt)$&                       & \checkmark &          &              &               &               & \checkmark & \\
$\deta(\Pello,\Pellt)$&                       & \checkmark &          &              &               &               & \checkmark & \\
$\DR(\PQjsub{1,\PHt},\PQjsub{2,\PHt})$&       & \checkmark &          &              &               &               &            & \\
$\DR(\PQjsub{1,\PHo},\PQjsub{2,\PHo})$&       & \checkmark &          &              &               &               &            & \\
$\ptlo/\mV$         &                         & \checkmark &          &              &               &               &            & \\
\ptlo               &                         & \checkmark &          &              &               &               &            & \\
$\ptX{\PQjsub{3,4}}$&                         &            &          &              &               &               & \checkmark & \\
$\HT^{\VHH}$        &                         &            &          &              &               &               & \checkmark & \\
$\ptv/\pthh$        &                         &            &          &              &               &               & \checkmark & \\
$\dphi(\PV,\HH)$    &                         &            &          &              &               &               & \checkmark & \\
$\ptlo/\mV$         &                         &            &          &              &               &               & \checkmark & \\
$\eta/\phi/m/\ptX{\PQjsub{1,2,3,4,\mathrm{other}}}$& &     &          &              &               &               &  & \checkmark\\
$m_{\PQj_i\PQj_k}$ & &              &          &              &               &               &  & \checkmark\\
$\dphi(\PQj_i,\PQj_k)$ & &              &          &              &               &               &  & \checkmark\\
Year                &                         &            &          &              &               &               &  & \checkmark\\
\hline
\end{tabular}}
\end{table}

In the large-radius jet analysis in the \met and \oneL channels,
events are divided into three regions using the \Dbb of each Higgs boson candidate jet:
$\minDbbbb>0.94$ defines high-purity regions (HP), $0.90<\minDbbbb<0.94$ defines
low-purity regions (LP), and $\minDbbbb<0.90$ and $\max(\Dbbbb)>0.80$ define
failing regions.
No \kl-enrichment is possible in this topology because the two large-radius jets tend to have
$\dphi\sim\pi$, which corresponds both to enhanced \kvv, shown in Fig.~\ref{fig:catbdt}, and larger values of \mHH.
These regions are therefore considered \kvv-enriched by construction, and thus no additional categorization based on
\BDTcat is considered. Separate \BDTSVB are optimized in the LP and HP regions.

For the signal extraction, the output of a dedicated machine learning classifier trained
to separate signal and background is used. In the leptonic channels, \BDTSVB~s are used to
separate signal-like events from background-like events. Simulated signal samples with corresponding enhancements
(\ie, \kvv-enriched or \kl-enriched) are used in the training for the different regions.
The variables used as input to the BDT are listed in Table~\ref{tab:catbdtinputs}.

In the \FH channel, the \NNSVB classifier is a neural network discriminant
with residual learning~\cite{ResNet}. The neural network contains multiple convolutional layers
where all combinations of the four Higgs boson candidate jets are considered.  A multi-head attention
block~\cite{MultiHeadAttention} in the neural network considers the kinematic variables associated with
additional jets.

In the SB regions, the observables are kinematic variables with significant systematic
uncertainties that can be constrained with a fit to data. The observable in all small-radius
leptonic SBs and in the \ttbar CR is the reconstructed \ptv. In the
large-radius SBs, the mass of the subleading large-radius jet is the observable. Overall, 59 distinct
regions are used. Table~\ref{tab:cats} summarizes the categories used in each channel.

\begin{table}[!tph]
\centering
\topcaption{A summary of the categorization used in all of the channels, where DY denotes Drell--Yan production. The first row outlines the variables used for the categorization.
HP and LP are regions defined based on \Dbb cuts: $\minDbbbb>0.94$ (HP), and $0.90<\minDbbbb<0.94$ (LP).
\Nb is the number of jets that pass \DeepJet \PQb tagging score medium working point.}
\label{tab:cats}
\renewcommand\arraystretch{1.3}
\cmsTable{\begin{tabular}{lllccccc}
    \hline
    \multicolumn{2}{l}{Variable for categorization}    & \BDTcat      &   \Nb, \Dbb   &  $\rhh$, \delHH, \mV  & Year split & $N$(regions) & Observable \\
    \hline
    \multicolumn{7}{l}{Signal regions}\\
    &MET small-radius     & \kl, \kvv      &   $\Nb\geq3$    &  SR+CR   & Per year & 6 & \BDTSVB \\
    &MET large-radius                   & \kvv           &   HP, LP    &  SR+CR   & Per year & 6 & \BDTSVB \\
    &1L small-radius    & \kl, \kvv      &   $\Nb\geq3$    &  SR+CR   & Per year & 6 & \BDTSVB \\
    &1L large-radius                   & \kvv           &   HP, LP    &  SR+CR   & Per year & 6 & \BDTSVB \\
    &2L               & \kl, \kvv       &   $\Nb=3$ or 4  & SR,CR  & Combined & 8 & \BDTSVB \\
    &FH                                   & \kl, \kvv       &   $\Nb=4$  & SR  & Per year & 6 & \NNSVB \\
    \multicolumn{7}{l}{Control regions}\\
    &MET small-radius   & \NA           &   $\Nb\geq3$    &  SB   & Per year & 3 & \ptv \\
    &MET large-radius   & \NA           &   HP, LP    &  SB   & Per year & 6 &  \mHtwo \\
    &1L small-radius   & \NA           &   $\Nb\geq3$    &  SB   & Per year & 3 & \ptv \\
    &1L large-radius   & \NA           &   HP, LP    &  SB   & Per year & 6 & \mHtwo \\
    &2L (DY)           & \NA       &   $\Nb=3$ or 4  & DY CR  & Combined & 2 & \ptv \\
    &2L (TT)           & \NA       &   $\Nb\geq3$  & \ttbar CR  & Combined & 1 & \ptv \\
    \hline
\end{tabular}}
\end{table}

\section{Background estimation methods}
\label{sec:background}

\begin{figure}[!thbp]
\centering
\includegraphics[width=0.31\textwidth]{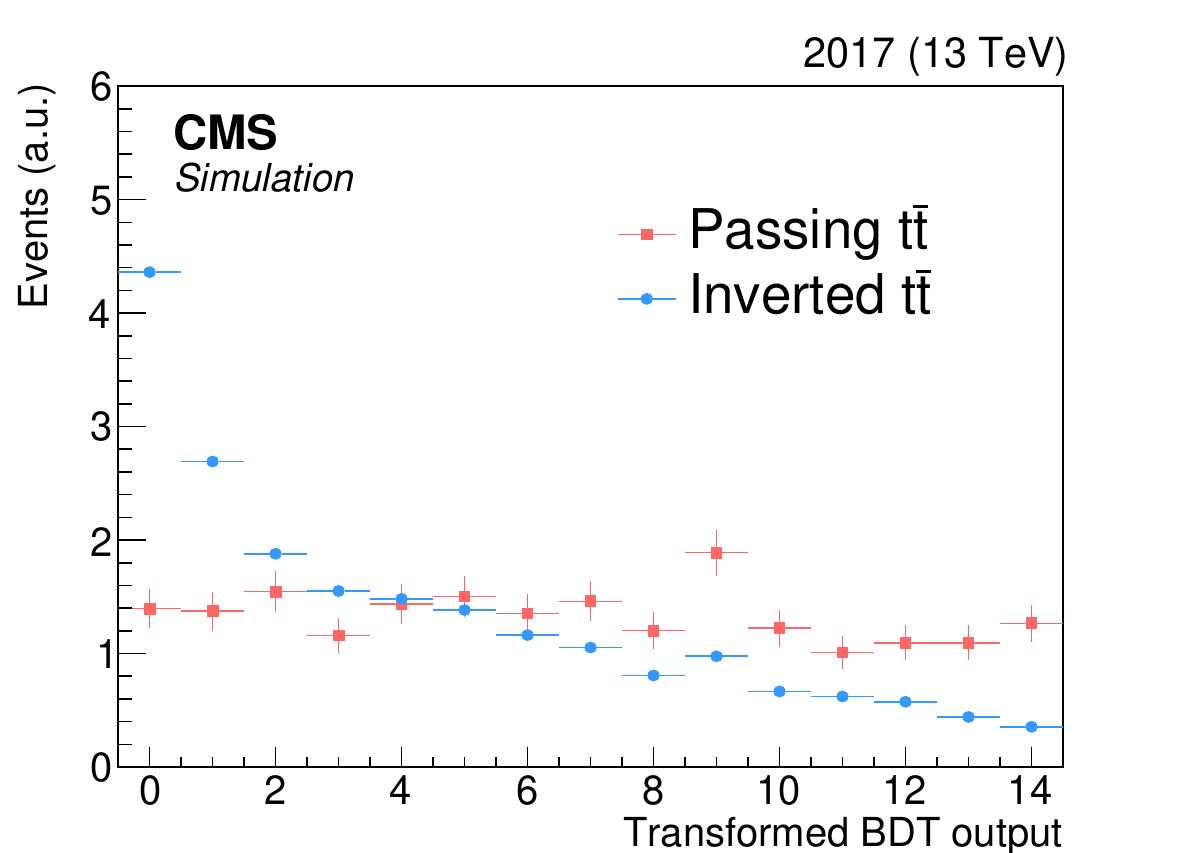}%
\hfill%
\includegraphics[width=0.31\textwidth]{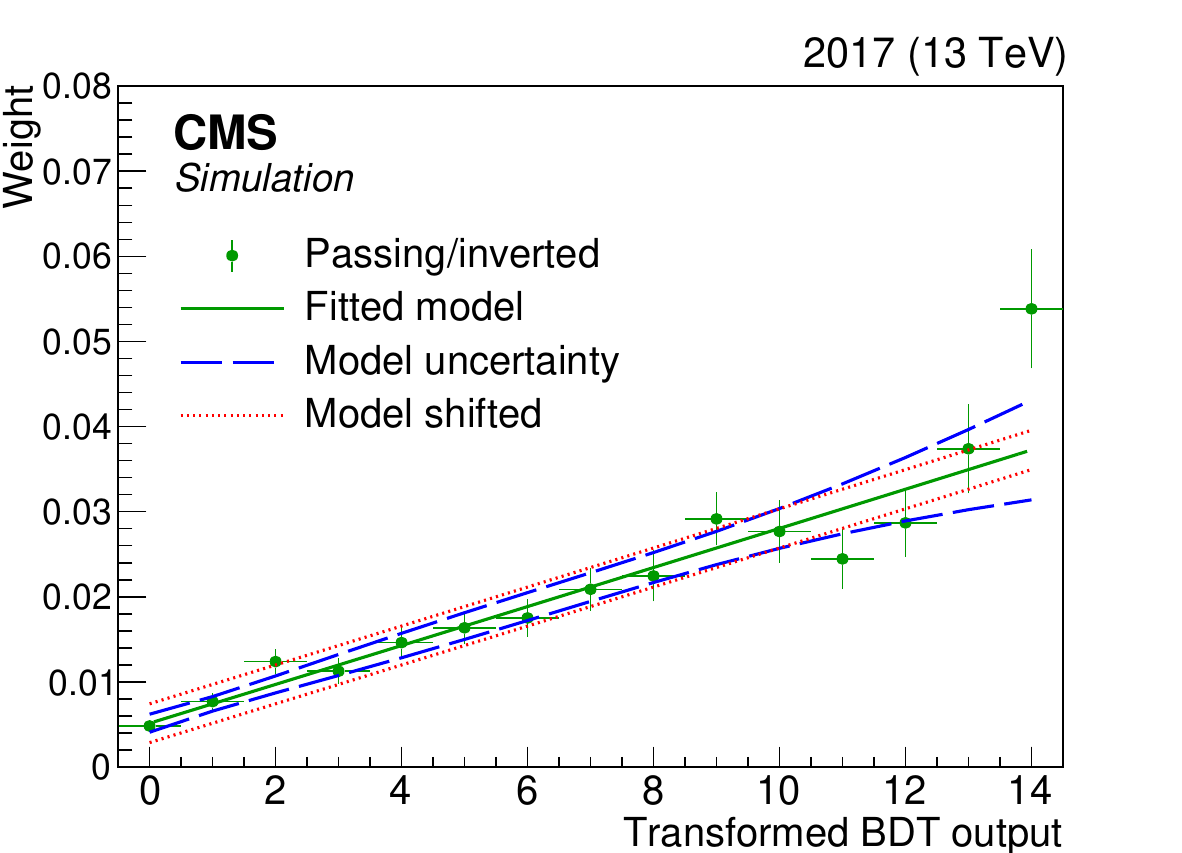}%
\hfill%
\includegraphics[width=0.31\textwidth]{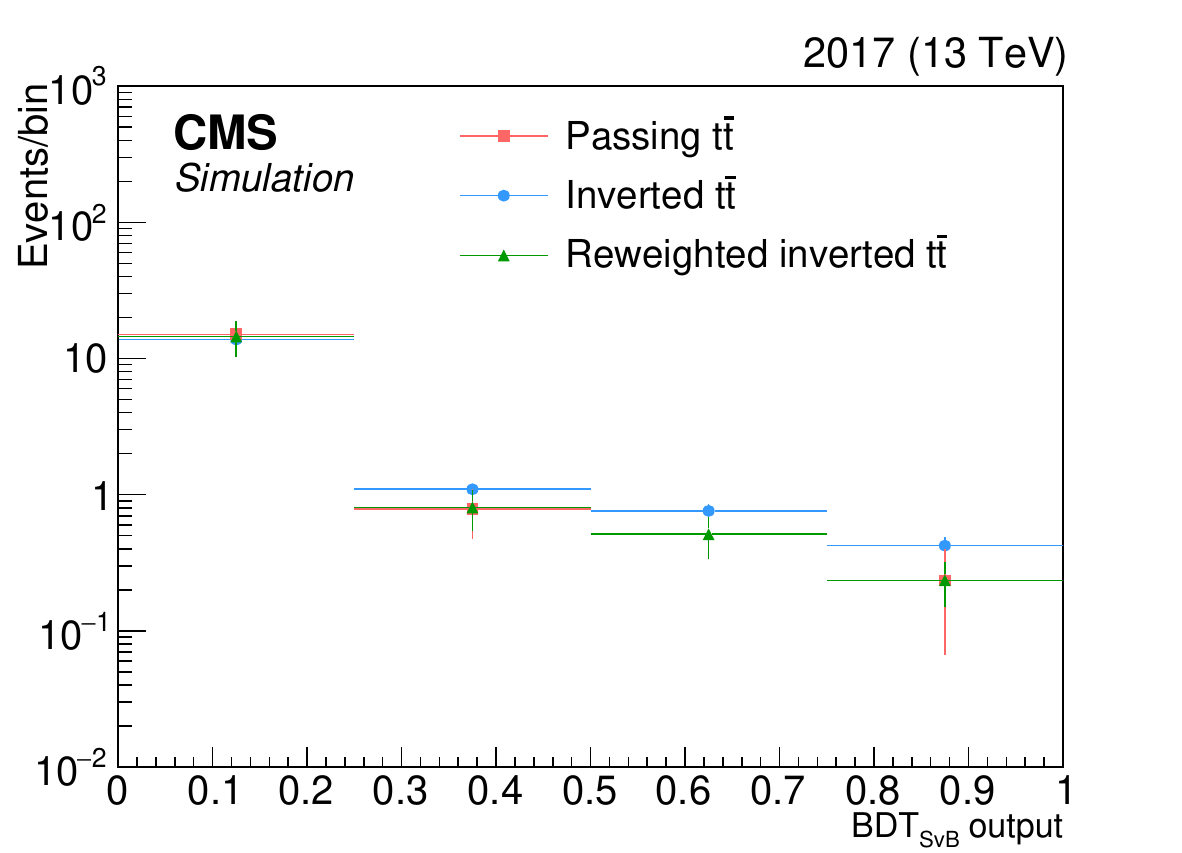}%
\caption{Left:
a reweighting BDT in the \oneL LP region for the \ttbar process that is transformed such that the
limited-precision passing \ttbar sample, shown as red squares, is approximately evenly distributed across all bins. In blue circles
is the same process where the \PQb tagging selections are inverted.  Middle:  the ratio is shown of
passing \ttbar to inverted \ttbar (green points) as a function of the transformed reweighting BDT score.
The solid line is the second-order polynomial fit of the green points, which is used for the reweighting.
In dotted red and dashed blue are the associated systematic uncertainties, which are obtained from
shifting the BDT score bin in evaluation of the model and the evaluation of the fit uncertainties on the weight, respectively.
These systematic variations account for finite binning and limited statistical precision of the passing events, and they enhance the flexibility of the model. Right: the distribution of \BDTSVB from passing (red squares), inverted (blue points), and reweighted inverted \ttbar (green triangles) sample in the \oneL LP region. The inverted \ttbar is normalized to make it the same yield as the reweighted inverted \ttbar. The uncertainty of reweighted inverted \ttbar distribution includes both statistical and systematic uncertainties on the reweighting.
}
\label{fig:reweight}
\end{figure}

A variety of background estimate strategies are employed in this analysis. In the \FH channel, a data-driven technique is used to create a background estimate from a region with an inverted \PQb tagging selection directly from reweighted data. In the leptonic channels, simulation is used to model the various background processes.

Because of the limited statistical precision of background simulations, a reweighting method exploiting a BDT is
used in some regions~\cite{Rogozhnikov_2016}. The technique is to invert part of the event selection to define an ``inverted'' region
where MC events are numerous and reweight events from the inverted region such that analysis variables, as well as the
correlations among them, match those of the signal selection. Similar approaches were also adapted in other analyses, as demonstrated in Refs.~\cite{CMS:2013ncv,ATLAS:2020fcp}. In practice, the BDTs are trained to distinguish between
the inverted and signal regions. The BDT output scores are used to produce the reweighting function between the inverted and signal
regions. The BDT output distribution is binned such that each bin has approximately the same yield in the sample passing
the full selection. The ratio of passing events to inverted events as a function of the BDT output with this binning is used to
derive a parametrized weighting to be applied to the inverted region. The parametrization is performed using a first- or
second-order polynomial to provide a smooth reweighting function. Two systematic uncertainties are
assessed for these reweightings: shifting the parametrization left and right along BDT scores and evaluating the fit
uncertainty directly. In all cases, all of the input variables of the reweighted inverted samples are compared with the distributions of the passing samples as a closure test. The distributions match within statistical uncertainty and the systematic uncertainties previously described generously cover any discrepancy. Figure~\ref{fig:reweight} shows this procedure for the \ttbar process simulation in the \oneL large-radius SR.

The primary backgrounds in the \met and \oneL channels are \ttbb, inclusive \ttbar, and single \PQt production. In the
small-radius jet analyses, the statistical precision is sufficient to use the simulation with nominal selection to directly
produce the templates for these processes in the analysis regions. The number of selected simulated events in the large-radius
channels is so small that the reweighting procedure is used in these regions instead.
The inverted region, which is the previously mentioned failing region, is trained separately against both HP and LP regions.
The small-radius \twoL channel uses this technique for main backgrounds
(Drell--Yan $\Pell\Pell$ and inclusive \ttbar) in all regions, as well as for $\Nb=3$ and 4 regions separately. The inverted region for the \twoL channel is a region with exactly two \PQb tagged jets.

\begin{figure}[!bh]
\centering
\includegraphics[width=0.31\textwidth]{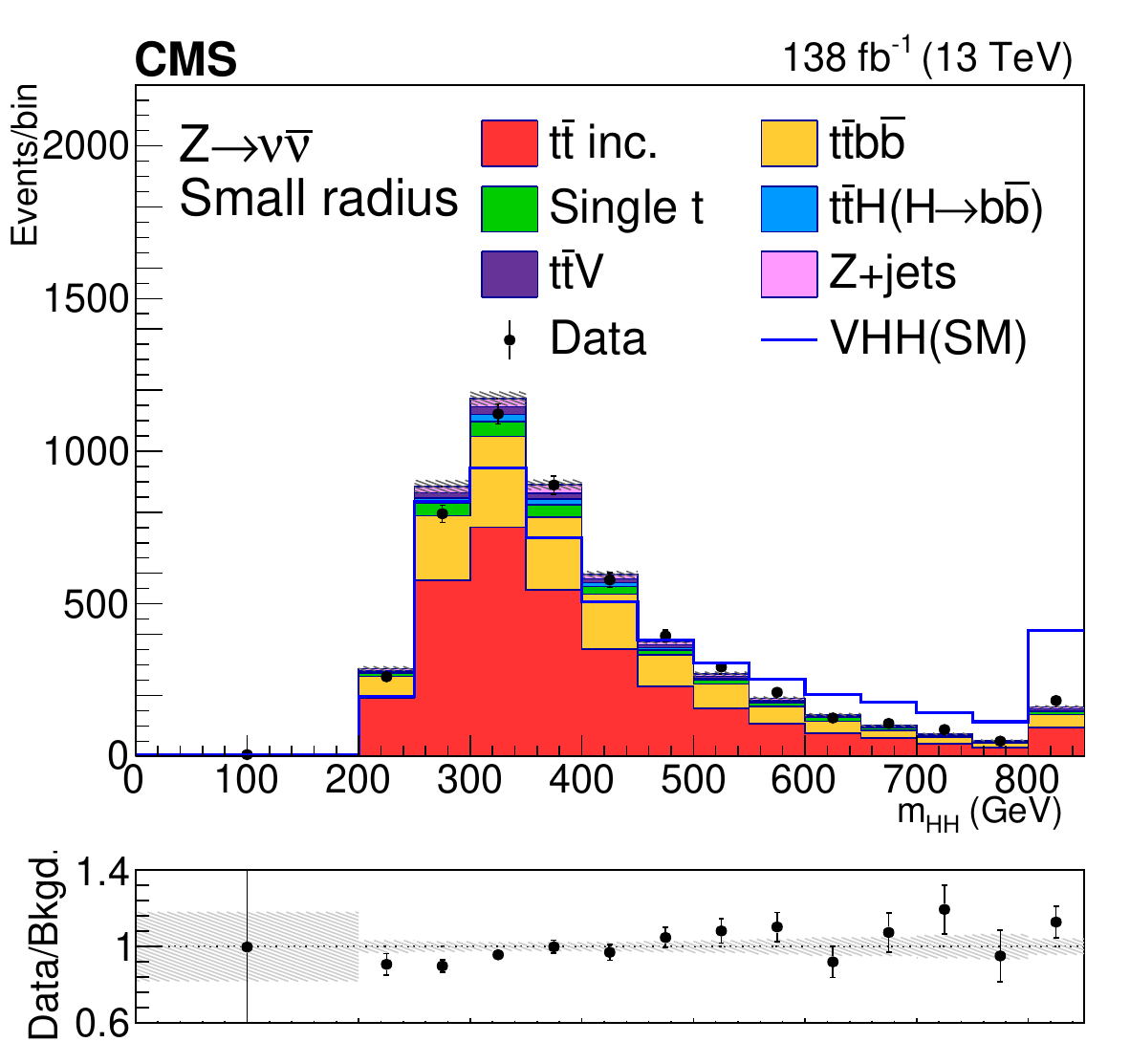}%
\hfill%
\includegraphics[width=0.31\textwidth]{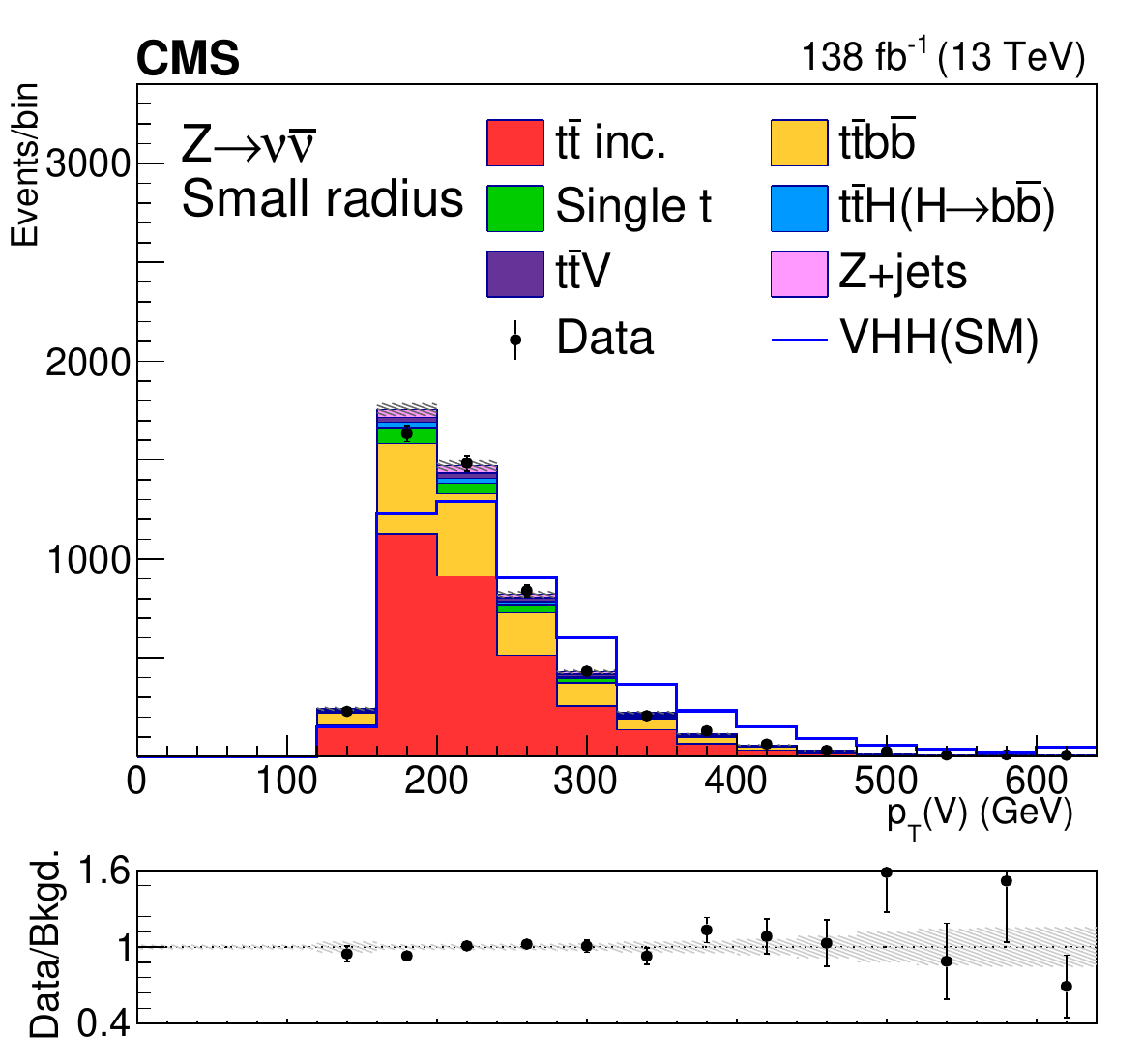}%
\hfill%
\includegraphics[width=0.31\textwidth]{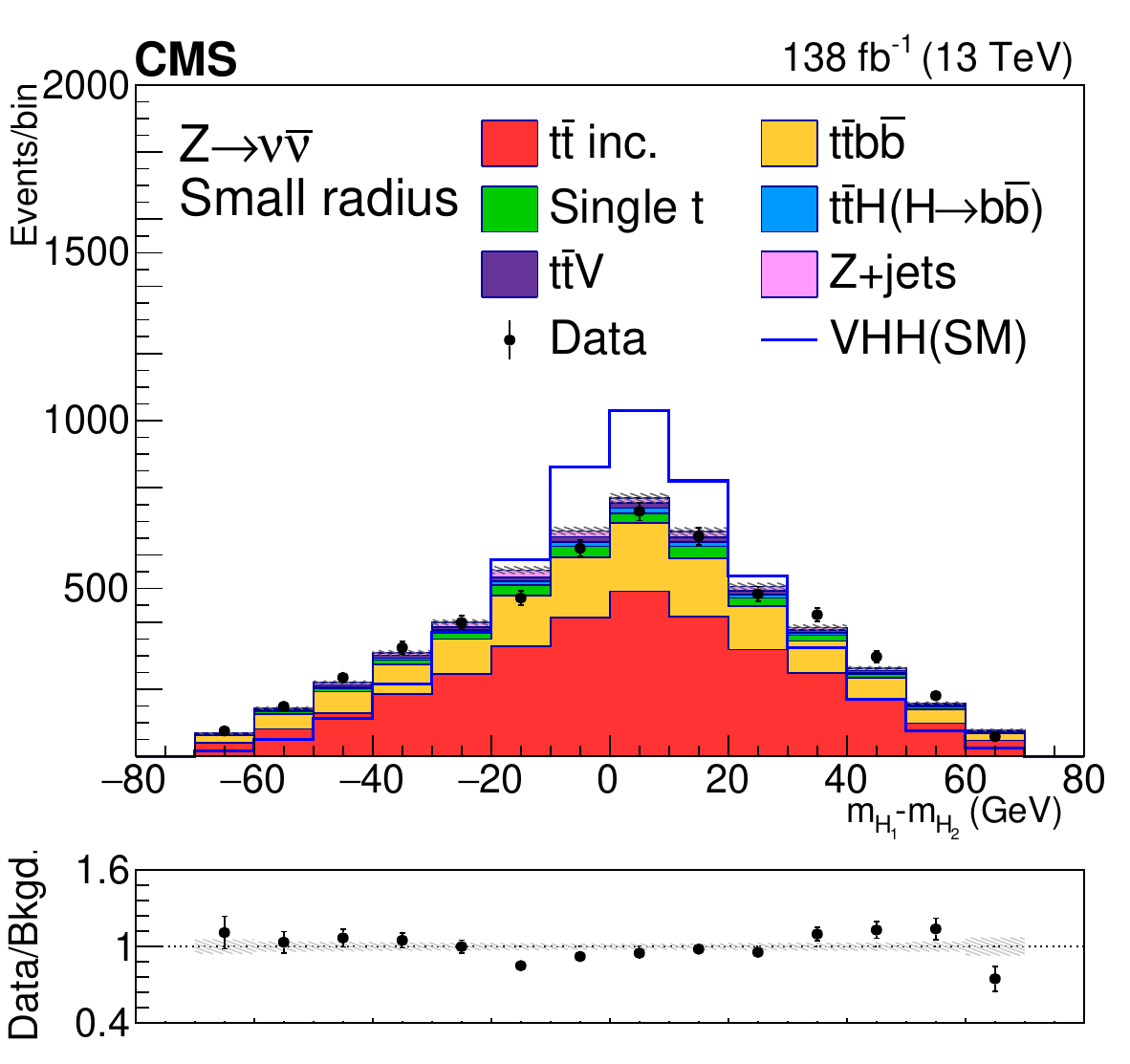} \\
\includegraphics[width=0.31\textwidth]{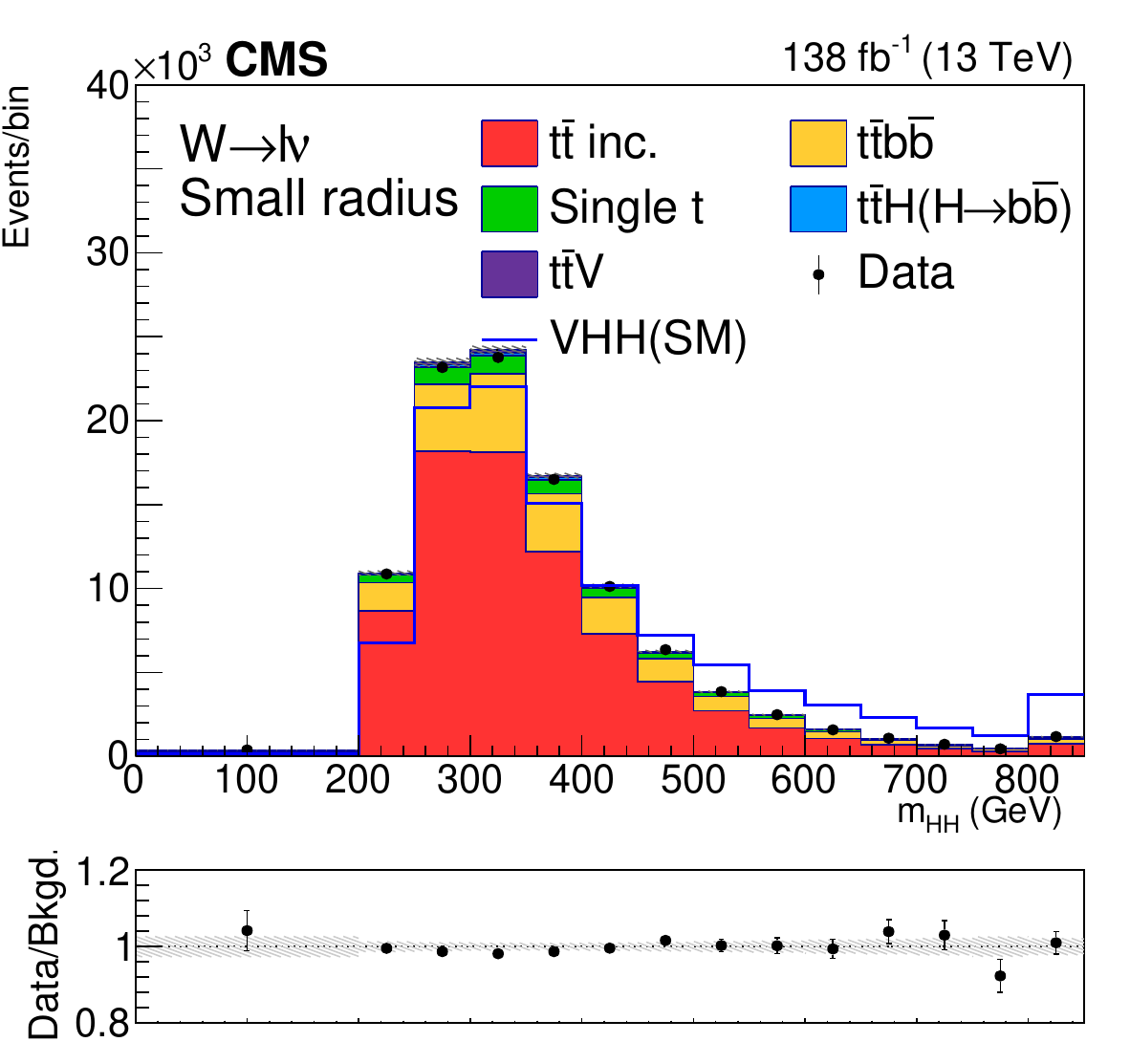}%
\hfill%
\includegraphics[width=0.31\textwidth]{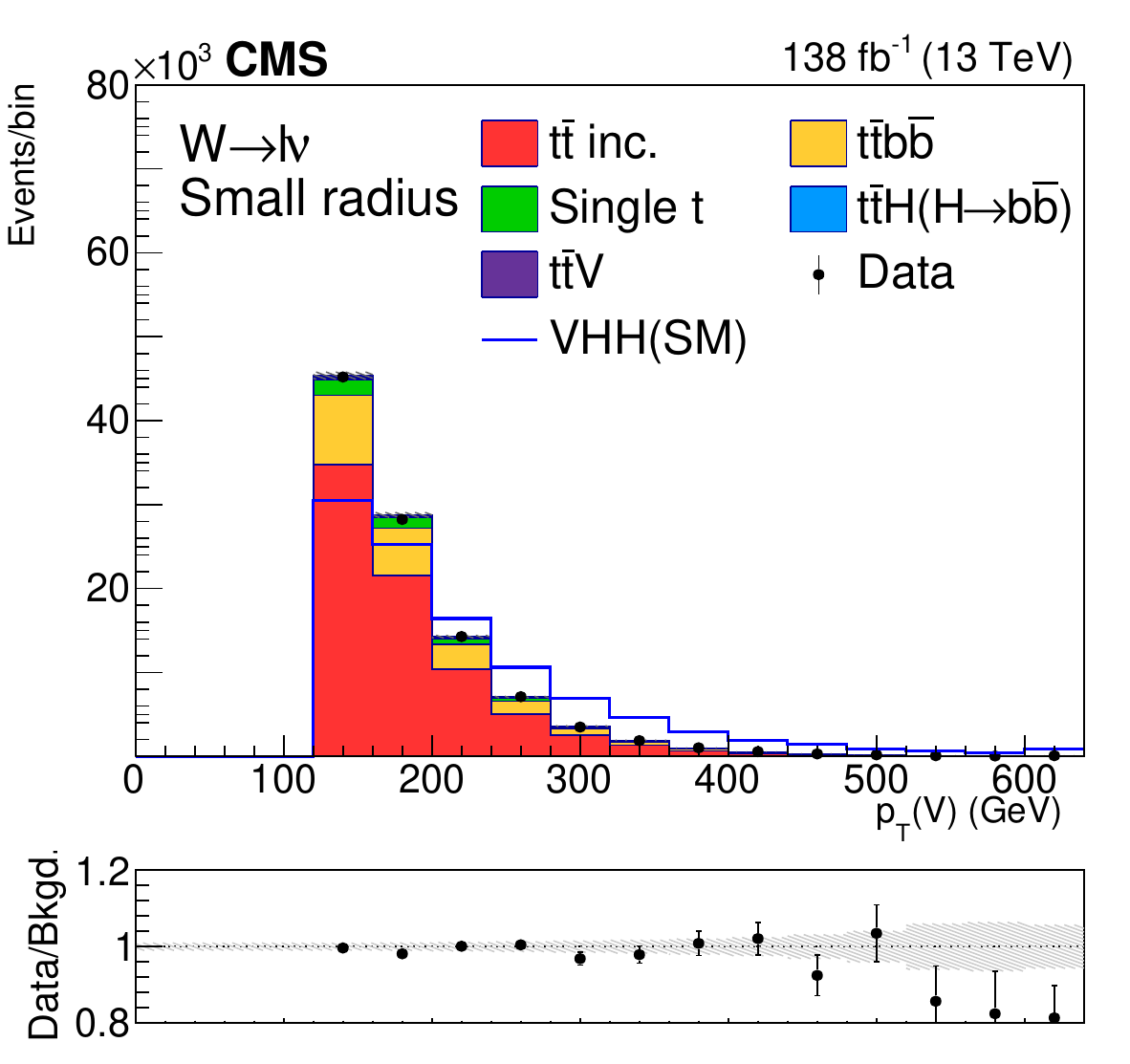}%
\hfill%
\includegraphics[width=0.31\textwidth]{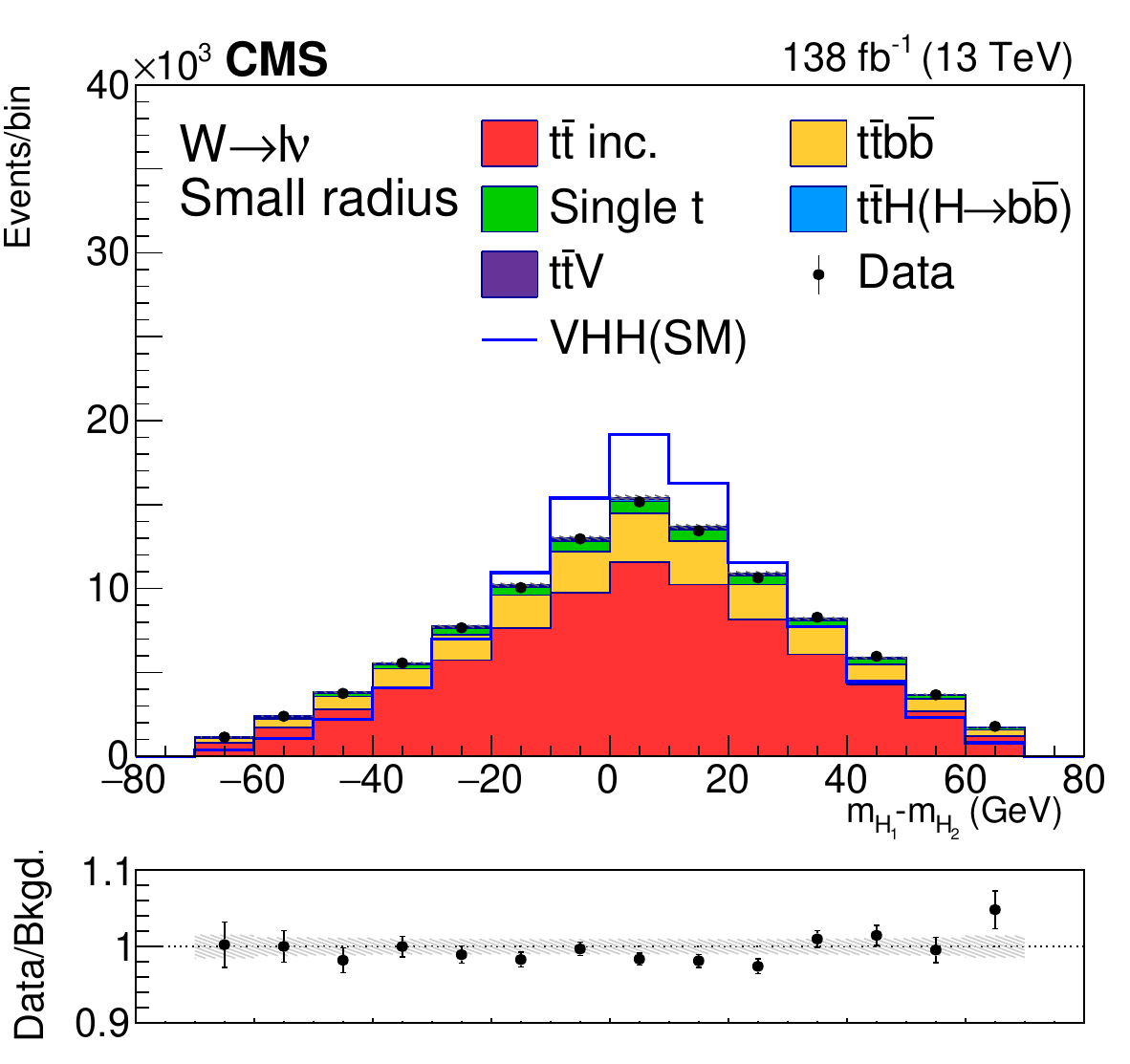}%
\caption{Postfit distributions of kinematic variables in the small-radius jet regions. The upper (lower) row shows the \met (\oneL) channel. The variables in each channel are \mHH, \ptv, and \mHminH. The fit is done with the background-only hypotheses and the final bin in each plot includes overflows. The ratios of data to the total expected background are shown in the lower panel of each plot and the hatched bands are the combined statistical and systematic uncertainties of total background.
The blue lines are SM signal distributions, which are scaled to have the same number of events as the background.}
\label{fig:postfit_bkg_R_0l1l}
\end{figure}

\begin{figure}[!ht]
\centering
\includegraphics[width=0.31\textwidth]{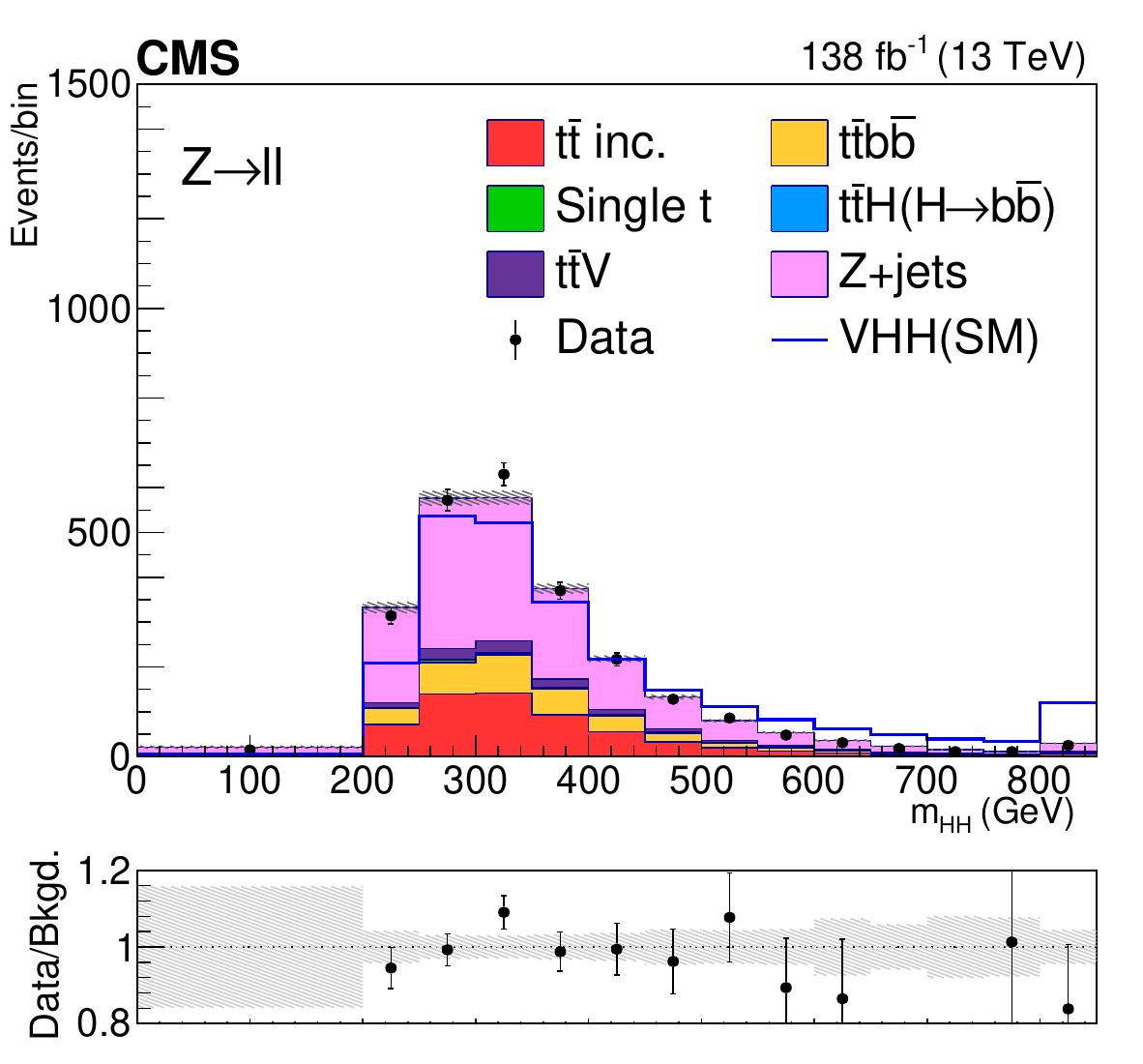}%
\hfill%
\includegraphics[width=0.31\textwidth]{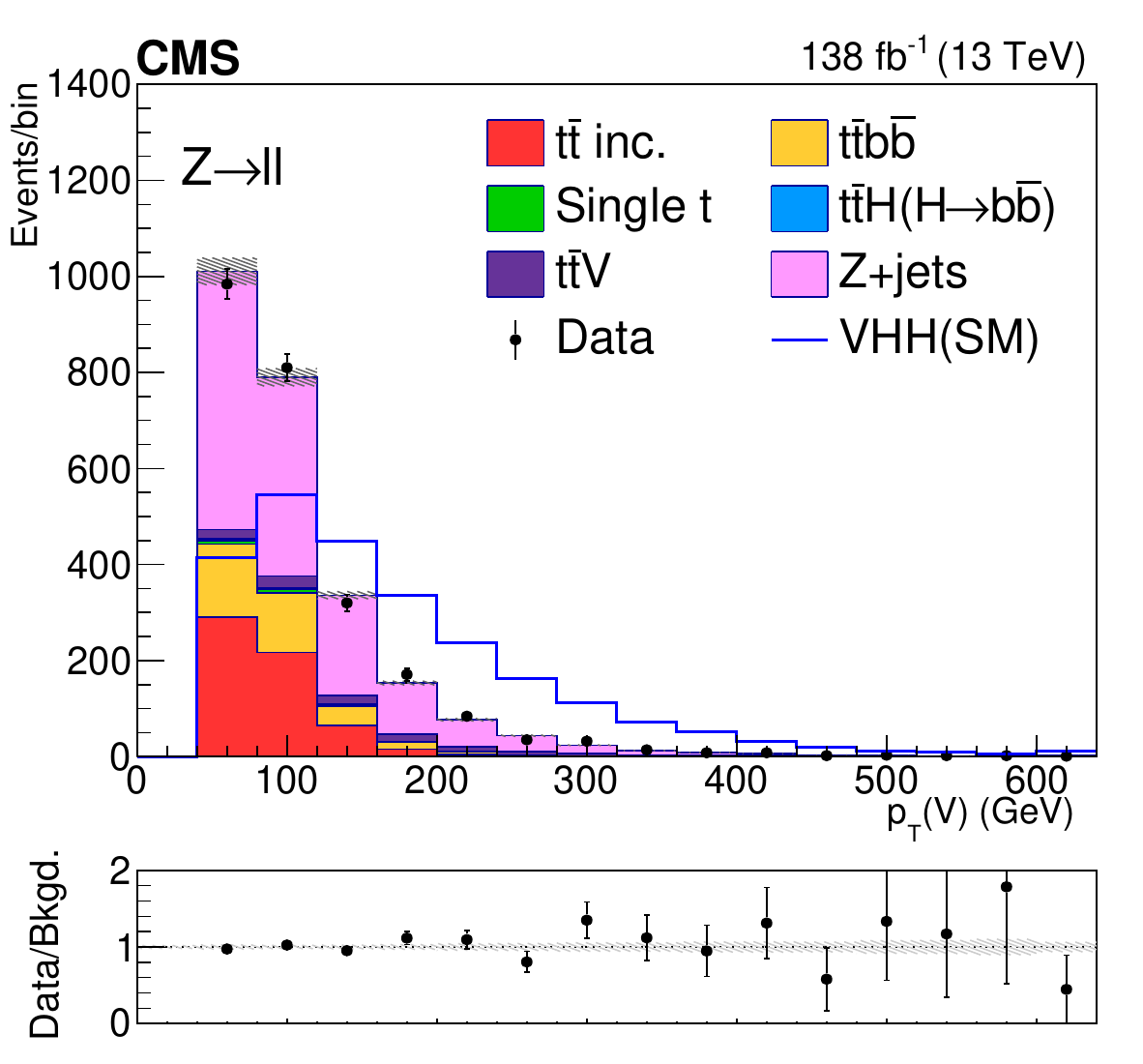}%
\hfill%
\includegraphics[width=0.31\textwidth]{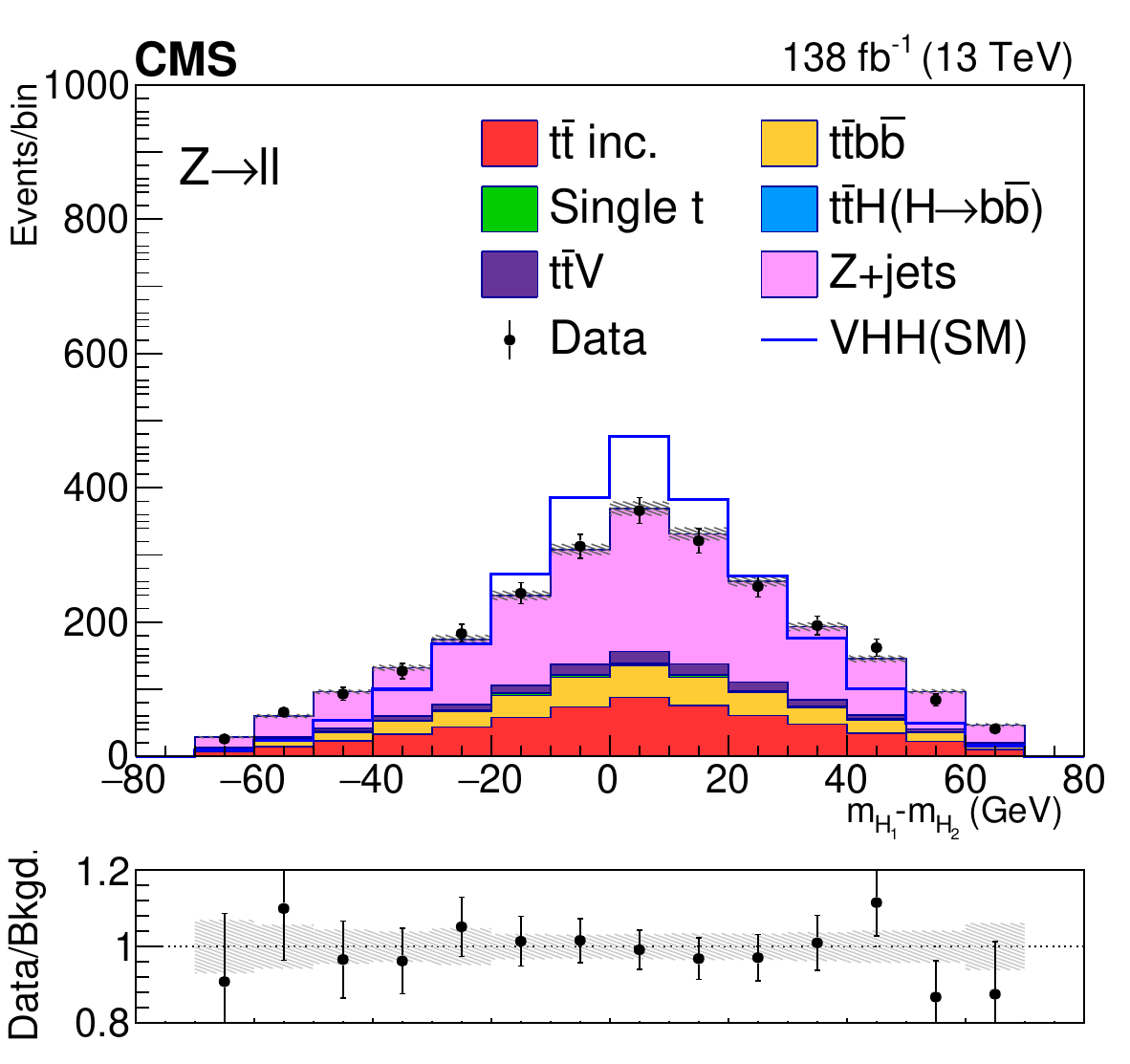} \\
\includegraphics[width=0.31\textwidth]{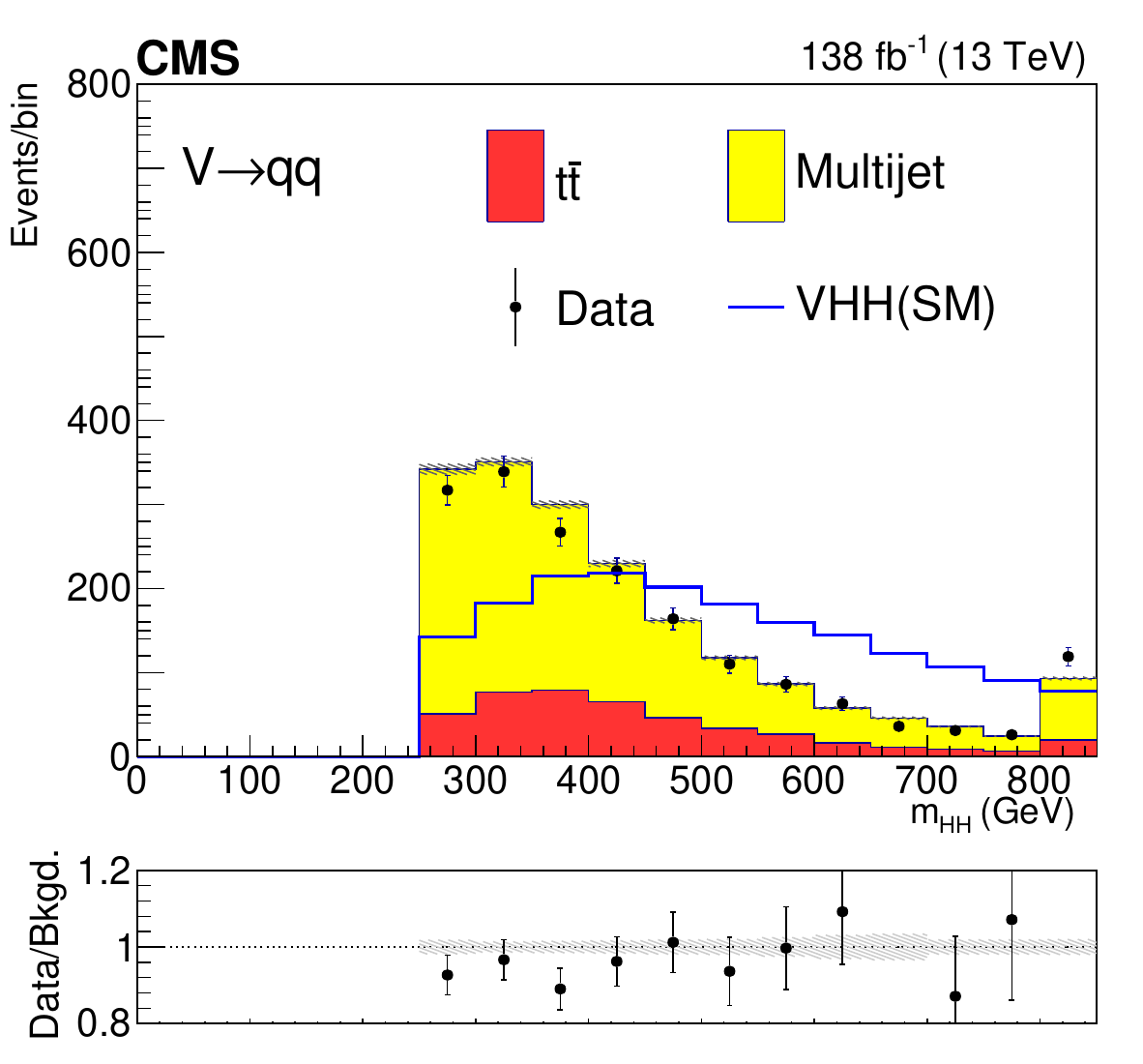}%
\hfill%
\includegraphics[width=0.31\textwidth]{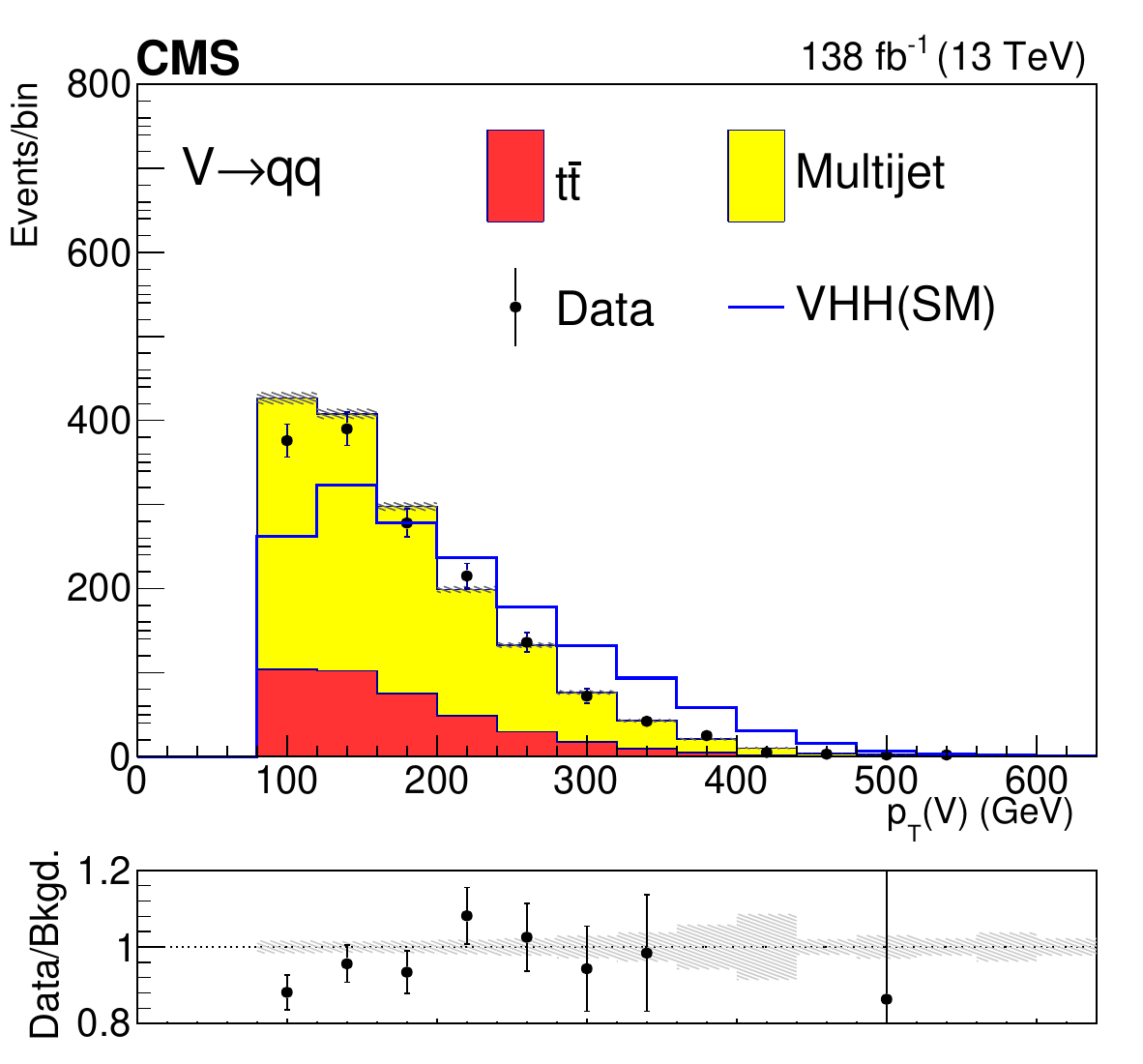}%
\hfill%
\includegraphics[width=0.31\textwidth]{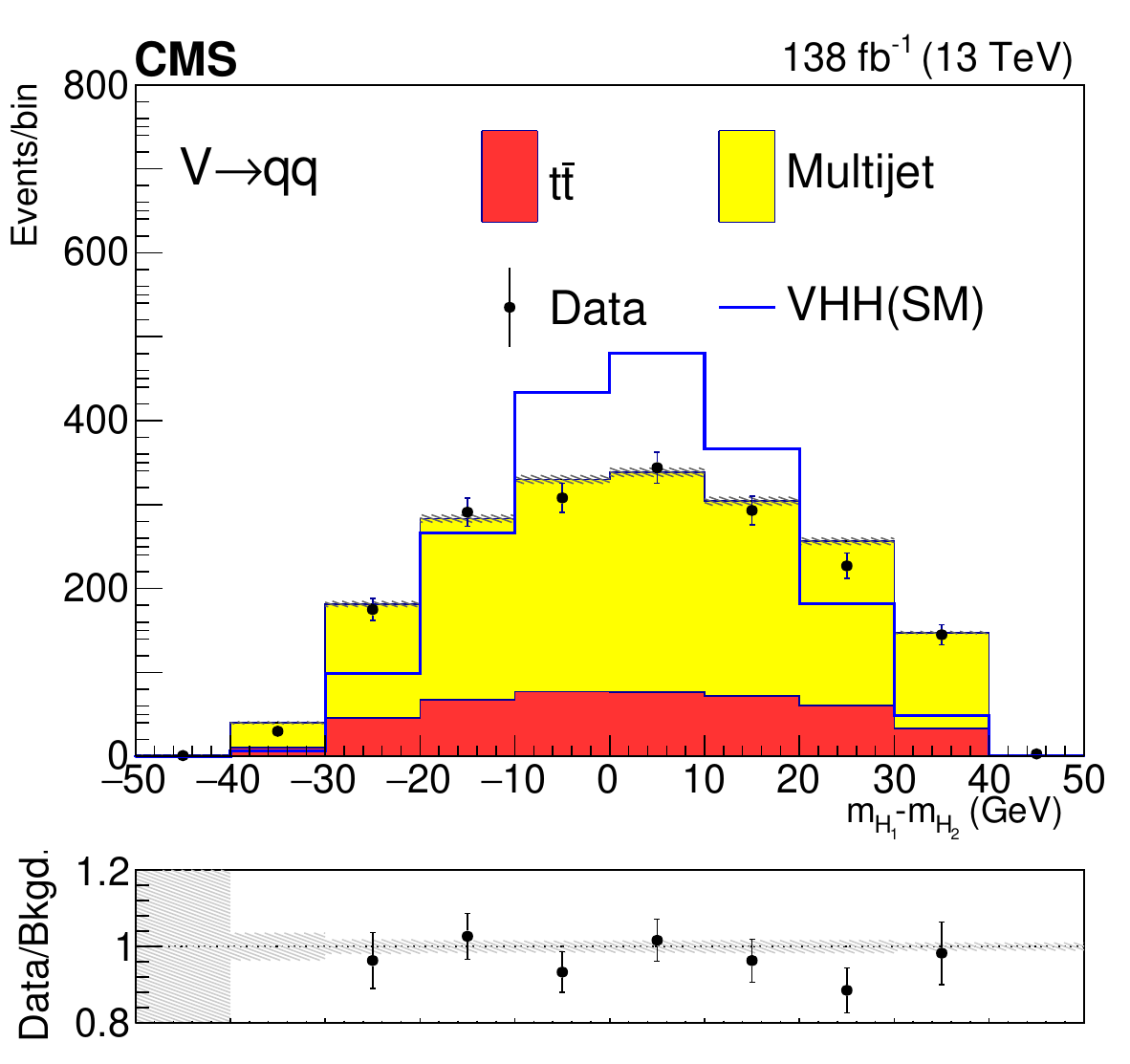}%
\caption{Postfit distributions of kinematic variables in the small-radius jet regions. The  upper (lower) row shows the \twoL (\FH) channel. The variables in each channel are \mHH, \ptv, and \mHminH. The fit is done with the background-only hypotheses and the final bin in each plot includes overflows. The ratios of data to the total expected background are shown in the lower panel of each plot and the hatched bands are the combined statistical and systematic uncertainties of total background.
The blue lines are SM signal distributions, which are scaled to have the same number of events as the background.}
\label{fig:postfit_bkg_R_2lfh}
\end{figure}

\begin{figure}[!ht]
\centering
\includegraphics[width=0.31\textwidth]{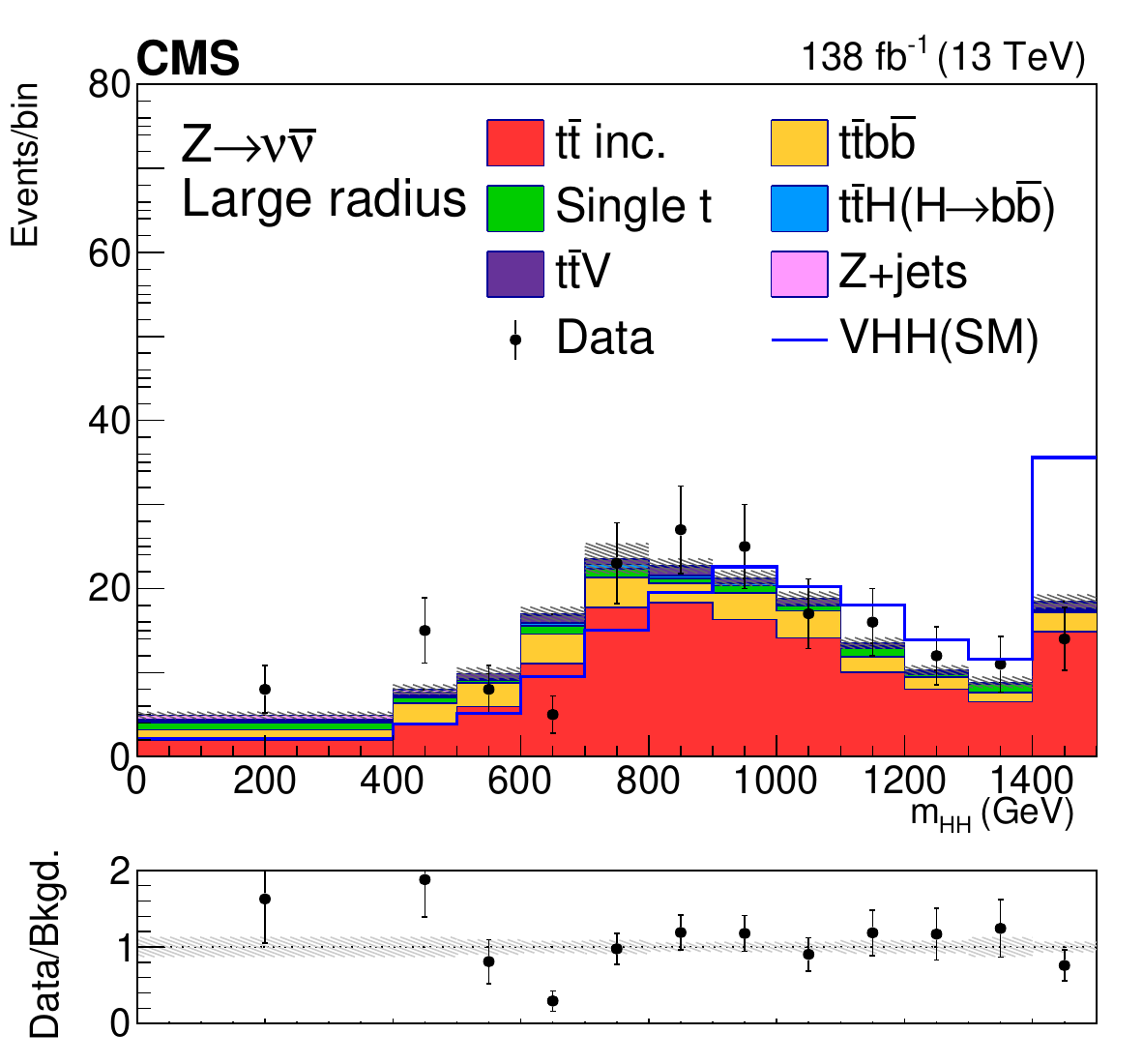}%
\hfill%
\includegraphics[width=0.31\textwidth]{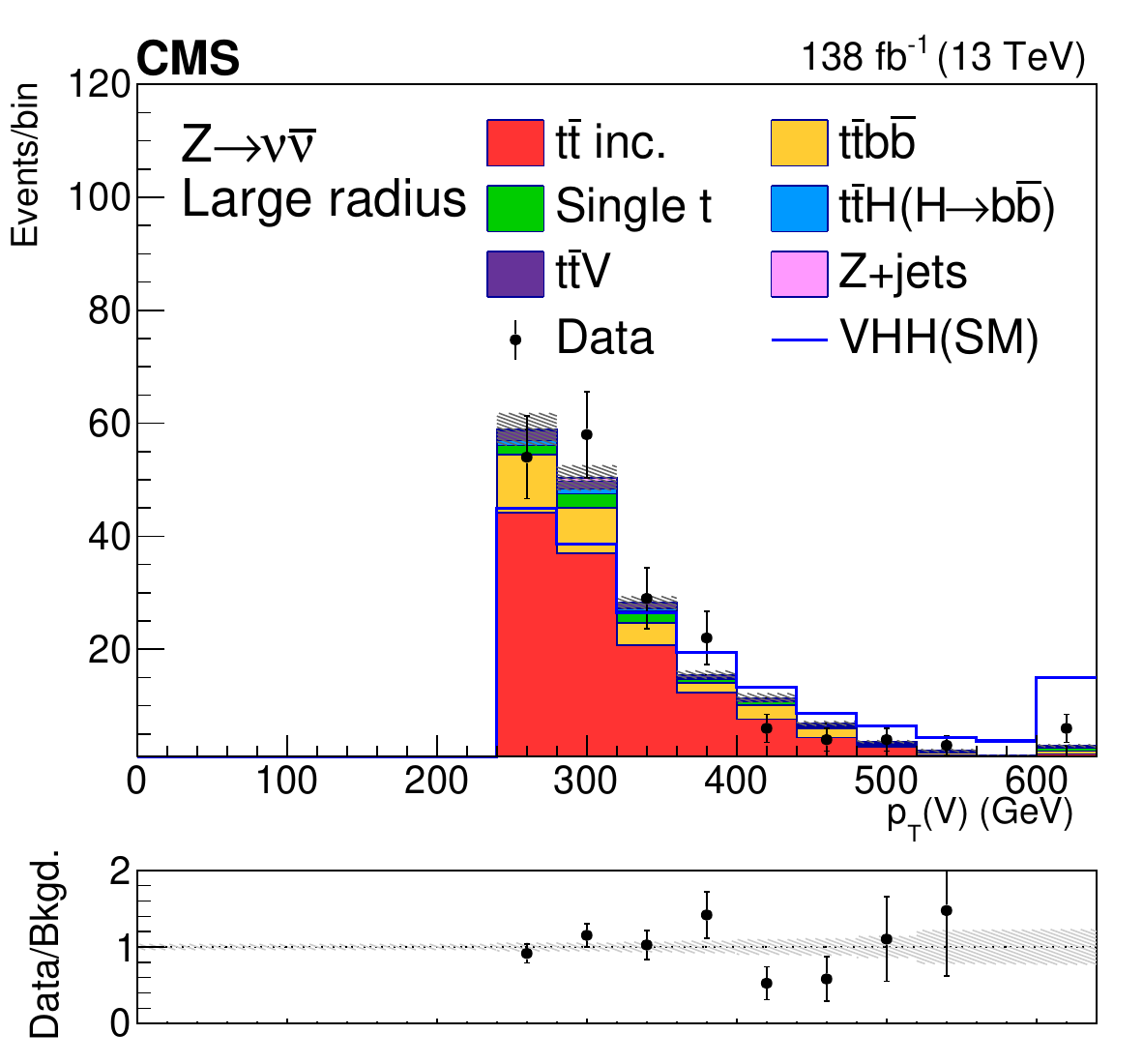}%
\hfill%
\includegraphics[width=0.31\textwidth]{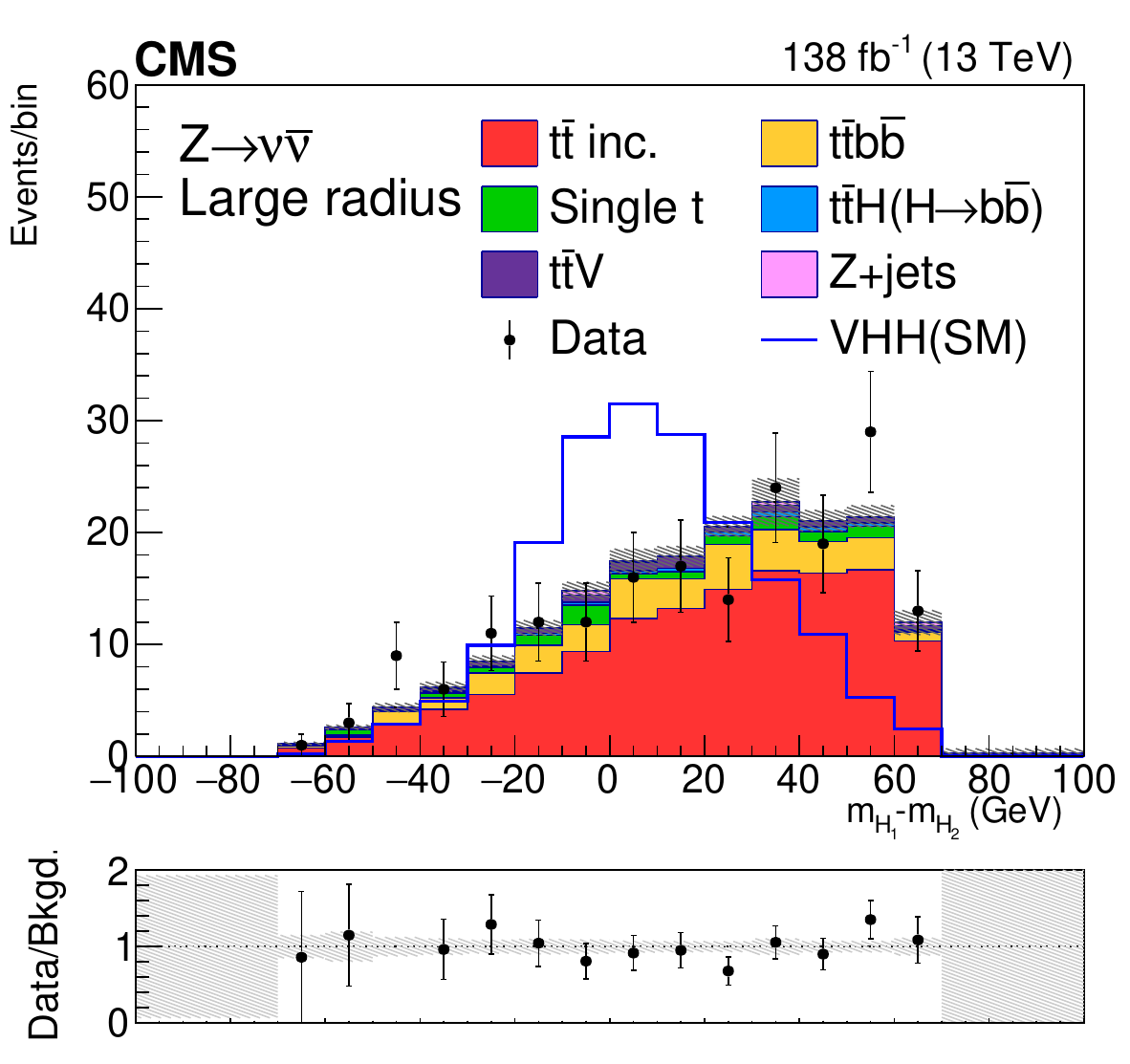} \\
\includegraphics[width=0.31\textwidth]{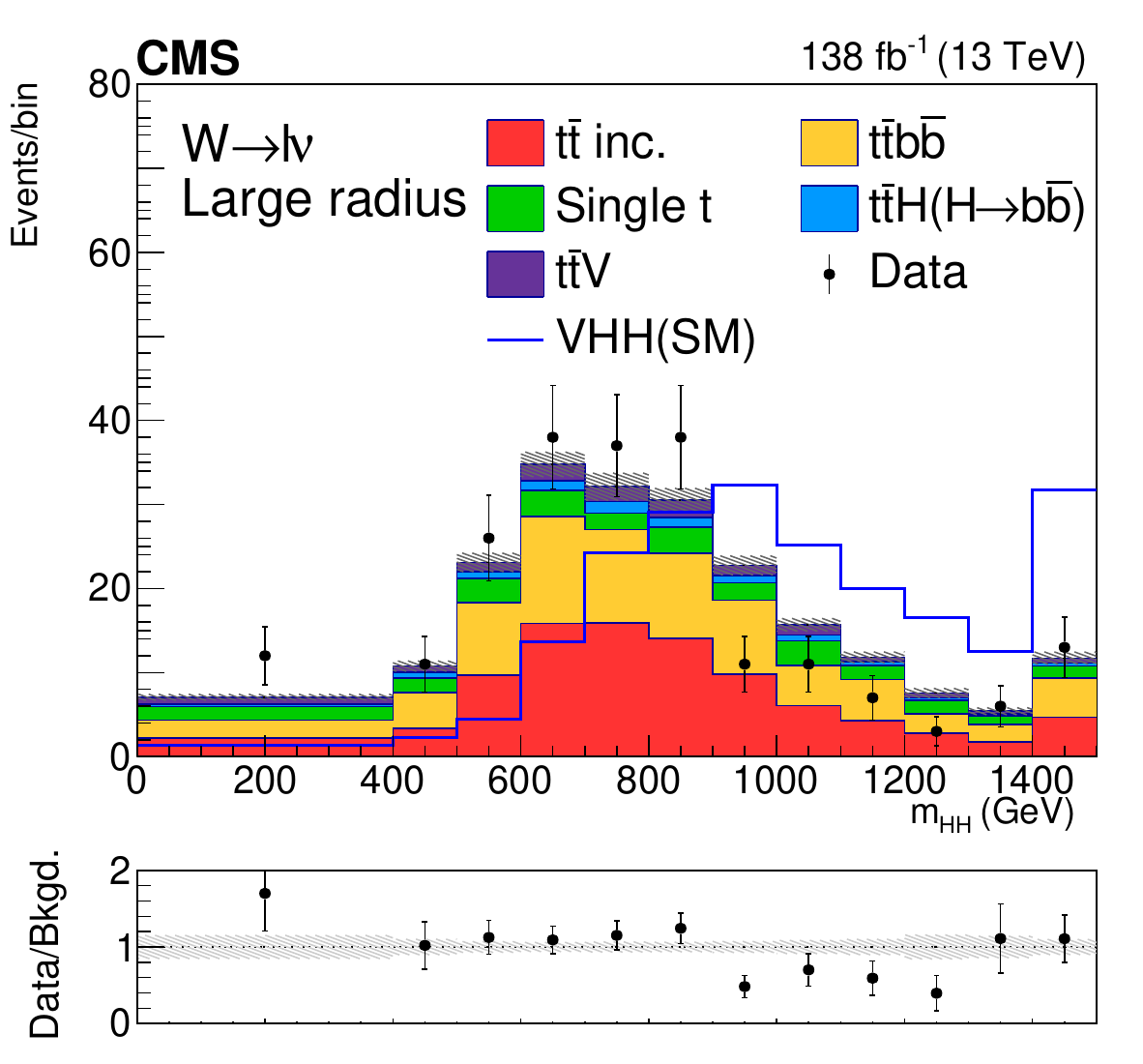}%
\hfill%
\includegraphics[width=0.31\textwidth]{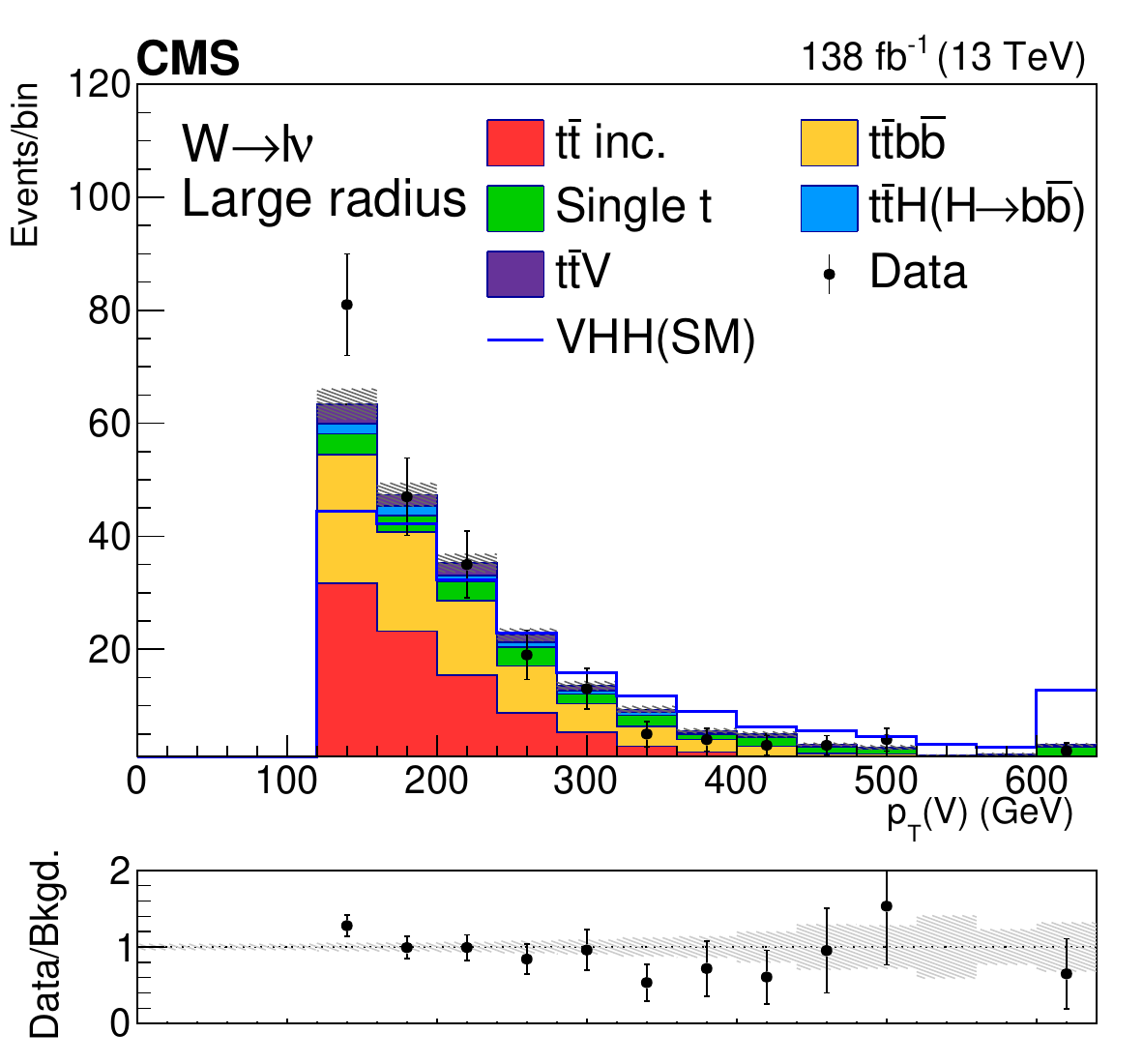}%
\hfill%
\includegraphics[width=0.31\textwidth]{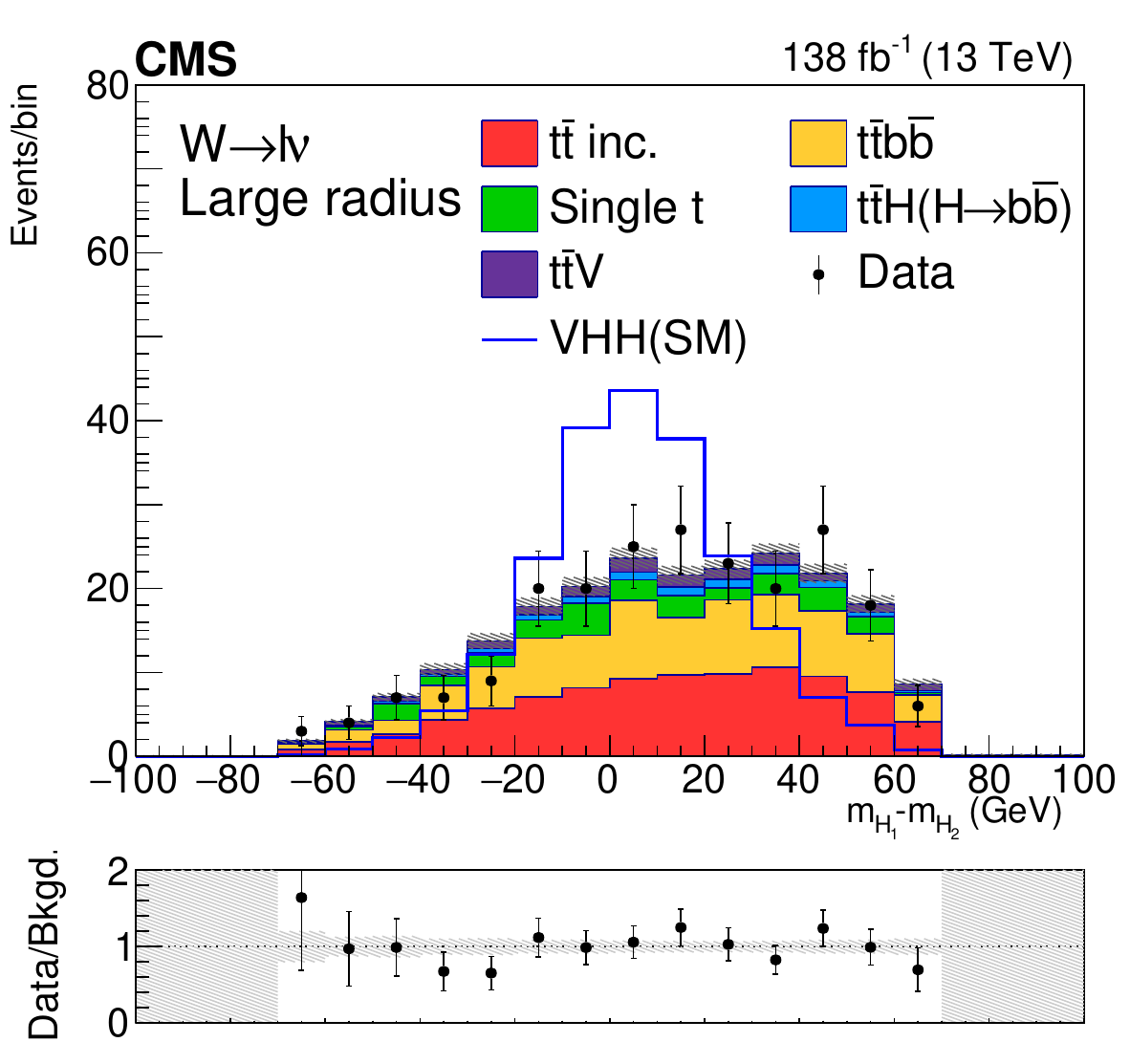}%
\caption{Postfit distributions of kinematic variables in the large-radius jet regions. The upper (lower) row shows the \met (\oneL) channel. The variables in each channel are \mHH, \ptv, and \mHminH. The fit is done with the background-only hypotheses and the final bin in each plot includes overflows. The ratios of data to the total expected background are shown in the lower panel of each plot and the hatched bands are the combined statistical and systematic uncertainties of total background.
The blue lines are SM signal distributions, which are scaled to have the same number of events as the background.}
\label{fig:postfit_bkg_B}
\end{figure}

In the \FH channel, the dominant background processes are QCD multijet followed by \ttbar (including \ttbb) production.
The QCD multijet background is estimated by reweighting $\Nb=3$ data to mimic the $\Nb=4$ QCD
multijet background, while the \ttbar templates are produced using nominally selected simulation directly. The signal purity in $\Nb=4$ is about 18 times higher than in $\Nb=3$, so the signal contamination in the reweighted $\Nb=3$ data is negligible.
A two-step approach is used to reweight $\Nb=3$ to $\Nb=4$ kinematics.  First, a pseudo-tag rate $f$, which is the
probability that an untagged jet would be \PQb-tagged, a normalization factor $t$ and pair-enhancement terms $e$ and $d$~\cite{ZZZH4bRun2}
are used to account for the lower jet multiplicity in the $\Nb=3$ region comparing to the $\Nb=4$ region.
The parameters are derived by fitting the jet multiplicity distribution in the SB region after subtracting the \ttbar contribution,
so that the weight for an event with $n$ non-\PQb-tagged jets is given by:
\begin{equation}
    w_1 = t\sum_{i=1}^{n} \binom{n}{i}  f^{i}(1-f)^{n-i}\times
    \begin{cases}
     1+ \re/n^d & (3+i)\quad\text{even} \\
     1           & (3+i)\quad\text{odd}.
    \end{cases}
\end{equation}
Then, a neural network classifier is used to distinguish the remaining kinematic differences between $\Nb=3$ and $\Nb=4$ regions.
The classifier is trained to predict the probabilities of an event to be $\Nb=4$ data, $\Nb=4$ \ttbar, $\Nb=3$ data or $\Nb=3$ \ttbar
using both data and simulated \ttbar in SB region and the second weight is given by:
\begin{equation}
    w_2 = \frac{P(\Nb=4~\text{data})-P(\Nb=4~\ttbar)}{P(\Nb=3~\text{data})}.
\end{equation}
The final weight is the product of $w_1$ and $w_2$. Since only \ttbar is subtracted explicitly during both fits,
besides the QCD multijet, the contributions of all other processes and potential mismodeling of \ttbar simulation are included in the weight.

Figures~\ref{fig:postfit_bkg_R_0l1l} and \ref{fig:postfit_bkg_R_2lfh} compare the \mHH, \ptv, and \mHminH distributions of data with the expected distributions for events that pass the SR selections (\met, \oneL, and \FH) in the small-radius jet regions, based on their respective background estimates.
The background distributions are obtained from the fit with the background-only hypothesis.
SM signal distributions are also shown and they are scaled to have the same number of events as the background.
Similarly, Fig.~\ref{fig:postfit_bkg_B} shows the distribution of the same variables in the large-radius jet regions.

\begin{figure}[!ht]
    \centering
    \includegraphics[width=0.8\textwidth]{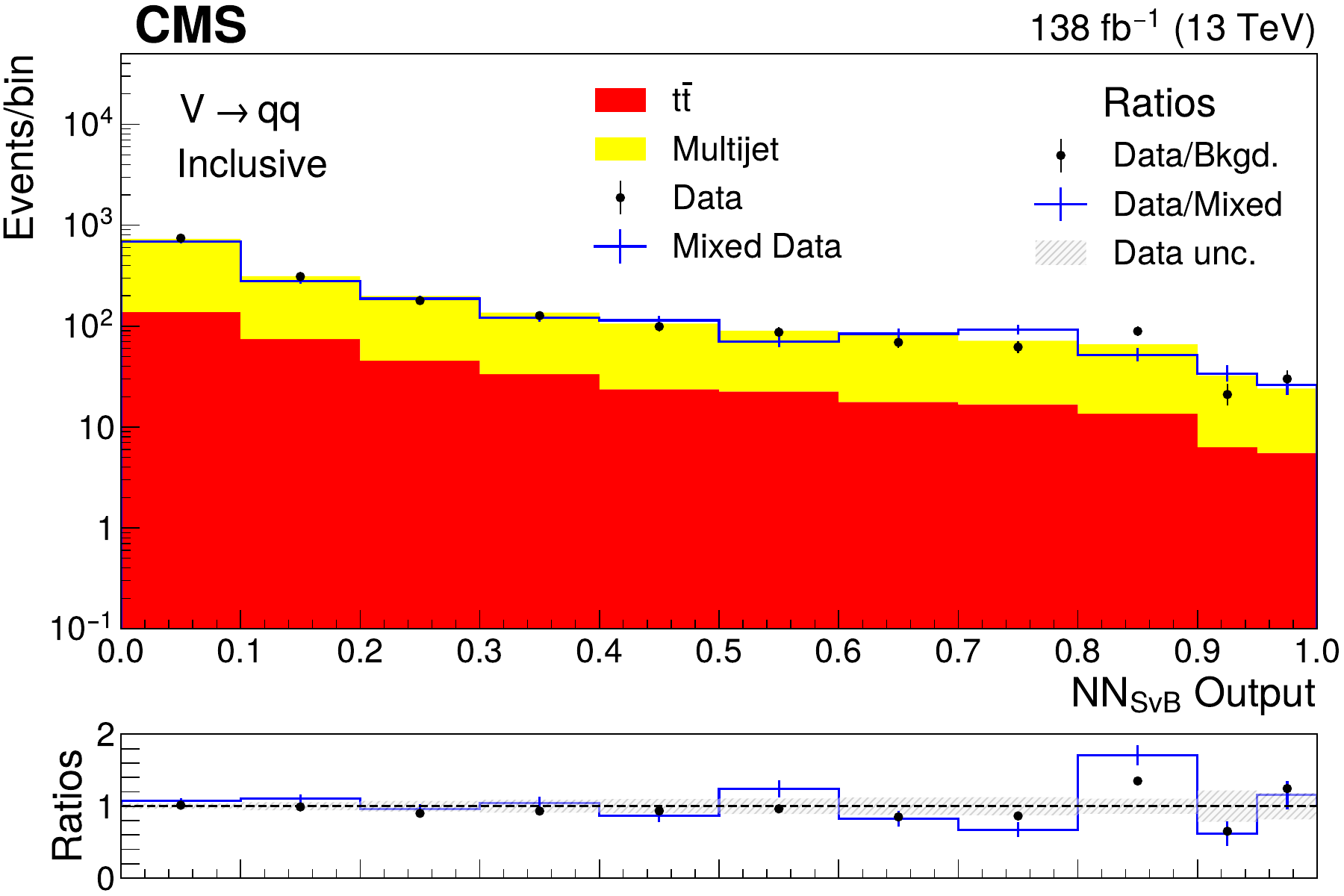}
    \caption{The prefit distribution of \NNSVB output in the \FH channel including both \kl and \kvv enriched categories. The ratios of data to the background model and data to one of the mixed data set is shown in the lower panel where the hatched band is the statistical uncertainty of data. The systematic uncertainty is then estimated based on the remaining discrepancy.}
    \label{fig:fh_mixed_data}
\end{figure}

Moreover, the \FH background model is validated using a synthetic data set, which is generated by splitting
individual events into hemispheres and then combining similar hemispheres from different events following the method described in Refs.~\cite{HH4bRun2,ZZZH4bRun2}.
This mixing procedure removes event-level correlations from any signal events in the data,
while preserving the kinematic distributions of the $\Nb=4$ background. The procedure is to divide events
with $\Nb = 3$ into two hemispheres: one with two \PQb-tagged jets and the other one with two jets where only one is \PQb-tagged.
Similar hemispheres from different data events are combined to produce mixed $\Nb = 4$ events.
There are sufficient combinations of suitable hemispheres for mixing that 15 distinct data sets with
the same statistical precision of the data are created. These samples provide a signal-free,
data-driven data set that can be used to assess the  background model directly in the SR. Figure~\ref{fig:fh_mixed_data} compares one of the 15 synthetic data sets to the background model and data in the SR.
The background procedure is performed treating the mixed data as the four-tag data set. Systematic
uncertainties are derived by comparing the predicted background to the observed yield in the signal region.

\section{Systematic uncertainties}
\label{sec:syst}

Uncertainties are categorized into theoretical and experimental uncertainties,
and are described in Section~\ref{subsec:theorysyst} and Section~\ref{subsec:expsyst}, respectively.
The signal extraction is performed with a maximum likelihood fit~\cite{LHCCLs}. The signal strength
is the additionally fitted normalization parameter relative to the nominal model.  All uncertainties
are implemented as penalty terms, known as ``nuisance parameters'', to a likelihood function.
Systematic uncertainties may affect the normalization of the different background processes,
their shape, or both.
We also include freely-floating nuisance parameters for the \ttbb, inclusive \ttbar, and Drell--Yan background
where a fixed background cross section is not assumed.
Some uncertainties are fully correlated between event categories and data-taking periods, while other
uncertainties only apply to specific categories. Correlations are described in the uncertainty
descriptions.

\subsection{Theoretical uncertainties}
\label{subsec:theorysyst}

\textit{Signal cross section predictions}: The total signal cross section has been calculated at NNLO accuracy, and the total uncertainty is about
3.6\% for both \ZHH and \WHH processes, including the effect of scale and PDF variation~\cite{LHCHiggsCrossSectionWorkingGroup:2016ypw}.

\textit{Branching fraction}: The uncertainty in the $\HH \to \bbbb$ branching fraction is 2.5\%~\cite{LHCHiggsCrossSectionWorkingGroup:2016ypw}.

\textit{Reweighting to include \ggZHH}:  As shown in Fig.~\ref{fig:ggZHH} right, the LO \ZHH simulation is initially scaled to NLO. Subsequently it is kinematically reweighted to NNLO to include the \ggF \ZHH contribution.
These reweighting factors are established prior to any selection criteria being implemented.
Despite this procedure, after the signal region selections, a discernible discrepancy emerges between the reweighted LO \ZHH simulation and the real distributions arising from NNLO simulations. This observed discrepancy is used to produce a shape-based systematic uncertainty.

\textit{NLO Drell--Yan simulations reweighting}: The LO and NLO Drell--Yan simulations are compared with minimal selection, and a shape discrepancy
as a function of generator-level \ptz is observed. The LO simulation is reweighted as a function
of generator-level \ptz to match the NLO generator-level \ptz distribution.  While the reweighted LO
sample matches the NLO with the minimal selection where the reweighting is defined, there is a
discrepancy after full selection between the NLO and LO with reweighting. This discrepancy is
linearly parameterized as a function of \ptz and implemented as a shape uncertainty.

\textit{Factorization and renormalization scales}:  The uncertainty from the choice of the factorization and renormalization scales, \muF and \muR, in the calculation of the matrix
elements of the hard-scattering process is estimated by varying each scale by a factor of 0.5 and 2, excluding
nonphysical anticorrelated combinations due to large logarithmic corrections $\abs{\ln(\muR/\muF)} > 1$.
The effects on different signal and background processes are considered uncorrelated.
The normalization effect on the uncertainty due to the choice of scales is already included in the cross section uncertainty.

\textit{Proton PDF uncertainties}:  In order to estimate an impact on the limit due to the uncertainty in the proton PDFs, event weights corresponding
to the set of NNPDF3.1~\cite{EXT:NNPDF-2017} MC replicas were applied to the simulation as a shape systematic
uncertainty.

\textit{Parton shower uncertainty}:  In order to evaluate the impact of the \alpS choice in the parton shower simulation, the renormalization scales in the
shower simulation are varied by a factor 0.5 and 2 via weights obtained directly from the
generator information. This is done independently for the initial- and final-state radiation showers, and treated as a shape systematic uncertainty.

\subsection{Experimental uncertainties}
\label{subsec:expsyst}

Some of the most significant experimental systematic uncertainties come from the limited size of the simulation and data samples used to build the templates for each process.
To account for this, the Barlow-Beeston-lite approach is used~\cite{BARLOW1993219}. A dedicated Gaussian nuisance parameter is assigned to each histogram bin in every analysis region.
The prior uncertainty of each nuisance parameter is set to the total statistical uncertainty obtained by combining all background processes in the corresponding bin.

\textit{Integrated luminosity}: The integrated luminosity uncertainties for 2016, 2017, and 2018 are 1.2, 2.3, and 2.5\%,
respectively~\cite{CMS-LUM-17-003,CMS-PAS-LUM-17-004,CMS-PAS-LUM-18-002}.  A correlation scheme is used for the
three sets of uncertainties based on correlated features in calibration methods, measurements, and data set.
Effectively, the uncertainty is about 1.6\% for the full data set.

\textit{Lepton and trigger efficiencies}: Uncertainties from the choice of the lepton identification and reconstruction criteria in the baseline selection, as
well as in the trigger efficiency in the leptonic channels, are also modeled as shape uncertainties. The efficiency corrections applied to
simulated samples have uncertainties from bin-by-bin differences using alternative samples, alternative
selections, alternative models, and the limited number of events in bins of $\eta$ and \pt when measuring the
efficiencies in data~\cite{CMS:2020uim,CMS:2018rym}. All these sources are combined, and they are considered
fully correlated across all bins of lepton $\eta$ and \pt. These uncertainties are mostly in the range of 1--2\%.
They are uncorrelated by analysis channel and by year.

\textit{Trigger efficiencies in the FH channel}:
Trigger efficiency uncertainties in the FH channel are evaluated for the signal, based on the systematic uncertainties arising from measured trigger turn-ons.
There is an additional, small, non-closure uncertainty associated with the calculation of per-event trigger efficiencies using the measured per-jet efficiencies.
The non-closure uncertainty was derived by comparing the event-level trigger efficiencies in simulation to those derived using the per-jet efficiencies as measured in simulation.
The total trigger efficiency uncertainties in the FH channel are in the range of 1--2\% and are uncorrelated by year.

\textit{Pileup}: The systematic uncertainty on the signal and background shapes introduced by the pileup reweighting procedure
is quantified by varying the effective total inelastic cross section of 69.2\unit{mb} within its $\pm4.6\%$ uncertainty.  This is a shape
uncertainty.

\textit{Small-radius jet}: To evaluate the effect from the jet energy scale (JES) and resolution (JER) on
the signal and background shapes,
alternative templates (\ie, one standard deviation shape uncertainty templates) of all analysis region observables
are obtained by varying absolute JES and JER within their uncertainties and propagating the
events through the full reconstruction chain~\cite{CMS:2016lmd}.  In total, 11 different sources of systematic
uncertainty affecting the JES and one affecting JER were considered.
A \PQb jet energy regression is applied to improve small-radius \PQb JER~\cite{breg}.
A further 2\% scale uncertainty and 10\% resolution uncertainty are also assigned to all small-radius jets.

\textit{\PQb tagging}: The \PQb tagging algorithm plays a crucial role in distinguishing between jets that stem from \PQb or \PQc quarks (referred to as heavy-flavor jets) and those that originate from light-flavor quarks or gluons (referred to as light-flavor jets).
The related efficiency and misidentification correction factors are applied per event with efficiencies measured in bins
of jet \pt, $\eta$, and \DeepJet \PQb tagging score.  There are uncertainties from the fitting procedure used to derive the
applied correction factors.
The contamination from jets originating from non-\PQb (\PQc or \PQb) partons but identified as heavy- (light-)flavor jets,
and the statistical precision in both data and simulation samples are used to evaluate the uncertainty in these measurements~\cite{BTV-16-002}.
That uncertainty is applied in a fully correlated manner for \PQb and light-flavor jets and \PQc jets have separate uncertainty.
The resulting alternative templates are implemented as one standard deviation shape systematic uncertainties.

\textit{Large-radius jet}: The \Dbb tagger is used to select events in the large-radius channels and the correction factors of efficiency are applied in bins of jet \pt and \Dbb score. The correction factors and associated uncertainties are consistent with those in the previous publications~\cite{HH4bboostedRun2} where \Dbb boundaries are aligned.
Similar to \PQb jet energy regression, a dedicated
graph neural network mass regression is used in the large-radius jet analyses~\cite{CMS-DP-2021-017}.  An additional 1\% mass scale
uncertainty and 5\% resolution uncertainty are included for the regressed mass (\mreg).  These are all shape uncertainties.

\textit{Normalization}: The normalizations of the primary backgrounds in the leptonic channels are free parameters that are extracted
and constrained in the likelihood fit.  The \ttbar process normalization is uncorrelated in each channel.
However, the ratio of 4FS \ttbb to 5FS \ttbar is correlated among all channels.  In the \twoL channel, the normalization of the
Drell--Yan process background is also free and fit to data.

\textit{Inverted sample reweighting}: Where the BDT reweighting procedure from inverted-to-nominal region is used (\ie, \met and
\oneL large-radius channels and \twoL small-radius channel), two associated systematic uncertainties, as described in
Section~\ref{sec:background}, are implemented independently per channel. These uncertainties are independent per region and
correlated among years.

\textit{Top quark \pt}: The top quark \pt distribution in \ttbar MC samples has a higher mean value in simulation than in data.
To correct for this bias, a linear variation to the top quark \pt floats during the signal extraction fit. This correction
is constrained in the SBs and in the \ttbar region in the \twoL channel where the observable is the reconstructed \ptv,
which is correlated with the top quark \pt. This uncertainty is correlated across all regions.

\textit{\FH background estimation cross check}: The uncertainties in \FH background estimation are derived by comparing the
predicted background to the observed yield in the SR of the corresponding mixed data set, as described in Section~\ref{sec:background}.
The uncertainties are parameterized using a Fourier basis on the SR \NNSVB shape. Two Fourier
terms ($\sin\pi x$, $\cos\pi x$) provide sufficient flexibility and no risk of spurious signal as estimated with an
$F$-test~\cite{dda9a3b2-7e66-38cd-8ce8-7d36cd60b878}. Together with a constant term these three terms are orthonormalized to
eliminate correlations among them. These systematic uncertainties are among the most significant for this channel.

For systematic variation templates of processes modeled with inverted reweighting, the alternative shapes are taken from propagating systematic uncertainties in the inverted samples directly. Comparisons with the variations from the passing events were performed. The size of the variations are compatible in nearly all cases, and the differences are negligible when there is any discrepancy.

Table~\ref{tab:sys_unc_breakdown} presents the contributions of various uncertainty sources to the overall uncertainty in the signal strength,
when such signal strength is derived assuming a SM-like signal shape.
Each group of uncertainties is quantified with respect to the total uncertainty as shown in the final line
of the table. The impactful sources of uncertainty include statistical uncertainty, background modeling, \PQb tagging, and JES/JER affecting
the small-radius jets.
The statistical uncertainty accounts for the limited size of both data and MC samples.
The \PQb tagging uncertainty corresponding to the contamination of light-quark-flavored jets in heavy flavor regions
is the most impactful uncertainty, and it is pulled by about $1\sigma$. Other impactful uncertainties that are pulled within $1\sigma$ include
top quark \pt reweighting, \PQb energy regression scale uncertainty, and inverted sample reweighting uncertainty.

\begin{table}[ht!]
\centering
\topcaption{The contribution of each group of uncertainties is quantified relative to the total uncertainty in the
signal strength, which is listed in the final line. To compute the relative contributions, the group of nuisance parameters
is fixed to the best fit value while the likelihood is scanned again profiling all other nuisance parameters.  The reductions
in the upper and lower variations are shown in each line.  The likelihood shape is asymmetric, and so the upper and lower variations are quantified separately.}
\label{tab:sys_unc_breakdown}
\renewcommand{\arraystretch}{1.1}
\cmsTable{\begin{tabular}{lllr@{/}lr@{/}lr@{/}lr@{/}lr@{/}l}
    \hline
    \multicolumn{3}{l}{Uncertainty sources} & \multicolumn{2}{c}{\twoL} & \multicolumn{2}{c}{\oneL} & \multicolumn{2}{c}{\met} & \multicolumn{2}{c}{\FH} & \multicolumn{2}{c}{Combined} \\
    \hline
    \multicolumn{3}{l}{Systematic uncertainty} & $+54\%$ & $-40\%$ & $+47\%$ & $-40\%$ & $+64\%$ & $-45\%$ & $+51\%$ & $-36\%$ & $+68\%$ & $-49\%$ \\
    & \multicolumn{2}{l}{Theory} & $+16\%$ & $-3\%$ & $+3\%$ & $-12\%$ & $+23\%$ & $-10\%$ & $+15\%$ & $-2\%$ & $+17\%$ & $-7\%$ \\
    & \multicolumn{2}{l}{Integrated luminosity} & $+6\%$ & $-0\%$ & $+5\%$ & $-1\%$ & $+8\%$ & $-1\%$ & $+4\%$ & $-0\%$ & $+6\%$ & $-4\%$ \\
    & \multicolumn{2}{l}{Lepton} & $+2\%$ & $-1\%$ & $+4\%$ & $-1\%$ & $+0\%$ & $-1\%$ & $+0\%$ & $-0\%$ & $+3\%$ & $-4\%$ \\
    & \multicolumn{2}{l}{Pileup} & $+3\%$ & $-6\%$ & $+4\%$ & $-2\%$ & $+8\%$ & $-7\%$ & $+3\%$ & $-0\%$ & $+9\%$ & $-14\%$ \\
& \multicolumn{2}{l}{Small-radius jet} & $+17\%$ & $-5\%$ & $+15\%$ & $-5\%$ & $+26\%$ & $-23\%$ & $+21\%$ & $-2\%$ & $+26\%$ & $-16\%$ \\
    & \multicolumn{2}{l}{\PQb tagging} & $+41\%$ & $-4\%$ & $+35\%$ & $-3\%$ & $+56\%$ & $-29\%$ & $+36\%$ & $-1\%$ & $+62\%$ & $-34\%$ \\
    & \multicolumn{2}{l}{Large-radius jet} & $+2\%$ & $-0\%$ & $+12\%$ & $-18\%$ & $+3\%$ & $-3\%$ & $+1\%$ & $-0\%$ & $+5\%$ & $-17\%$ \\
    & \multicolumn{2}{l}{Background modeling} & $+53\%$ & $-38\%$ & $+37\%$ & $-19\%$ & $+54\%$ & $-29\%$ & $+44\%$ & $-19\%$ & $+62\%$ & $-40\%$ \\
    & & Normalization & $+40\%$ & $-12\%$ & $+34\%$ & $-4\%$ & $+52\%$ & $-25\%$ & $+35\%$ & $-0\%$ & $+58\%$ & $-32\%$ \\
    & & Reweighting & $+34\%$ & $-36\%$ & $+13\%$ & $-17\%$ & $+22\%$ & $-13\%$ & $+12\%$ & $-1\%$ & $+25\%$ & $-19\%$ \\
    & & Kinematic & $+11\%$ & $-10\%$ & $+17\%$ & $-3\%$ & $+13\%$ & $-4\%$ & $+24\%$ & $-24\%$ & $+19\%$ & $-14\%$ \\
    \multicolumn{3}{l}{Statistical uncertainty} & $+84\%$ & $-91\%$ & $+88\%$ & $-92\%$ & $+77\%$ & $-89\%$ & $+86\%$ & $-93\%$ & $+73\%$ & $-87\%$ \\
\multicolumn{3}{l}{Signal strength and uncertainty} & \multicolumn{2}{c}{$101^{+136}_{-99}$} & \multicolumn{2}{c}{$12^{+111}_{-83}$} & \multicolumn{2}{c}{$283^{+161}_{-123}$} & \multicolumn{2}{c}{$190^{+163}_{-132}$} & \multicolumn{2}{c}{$145^{+81}_{-63}$} \\
    \hline
\end{tabular}}
\end{table}

\section{Results}
\label{sec:results}

For the signal extraction we performed a binned maximum likelihood fit using the 59 regions for signal extraction and for background control altogether, as shown in the Table~\ref{tab:cats}.
All regions for signal extraction are either \kl- or \kvv-enriched by construction.  The observables are machine
learning scores, \ie, from BDT or neural networks, that distinguish signal from background.
In the background control regions, the reconstructed \ptv and \mreg of the subleading Higgs boson candidate are used as fitting variables in the small-radius and large-radius regions, respectively.
The BDT distributions are divided to ensure a uniform distribution for the signal and also maintain background MC statistical uncertainties $<$30\% in each bin.

\begin{figure}[!htp]
\centering
\includegraphics[width=0.31\textwidth]{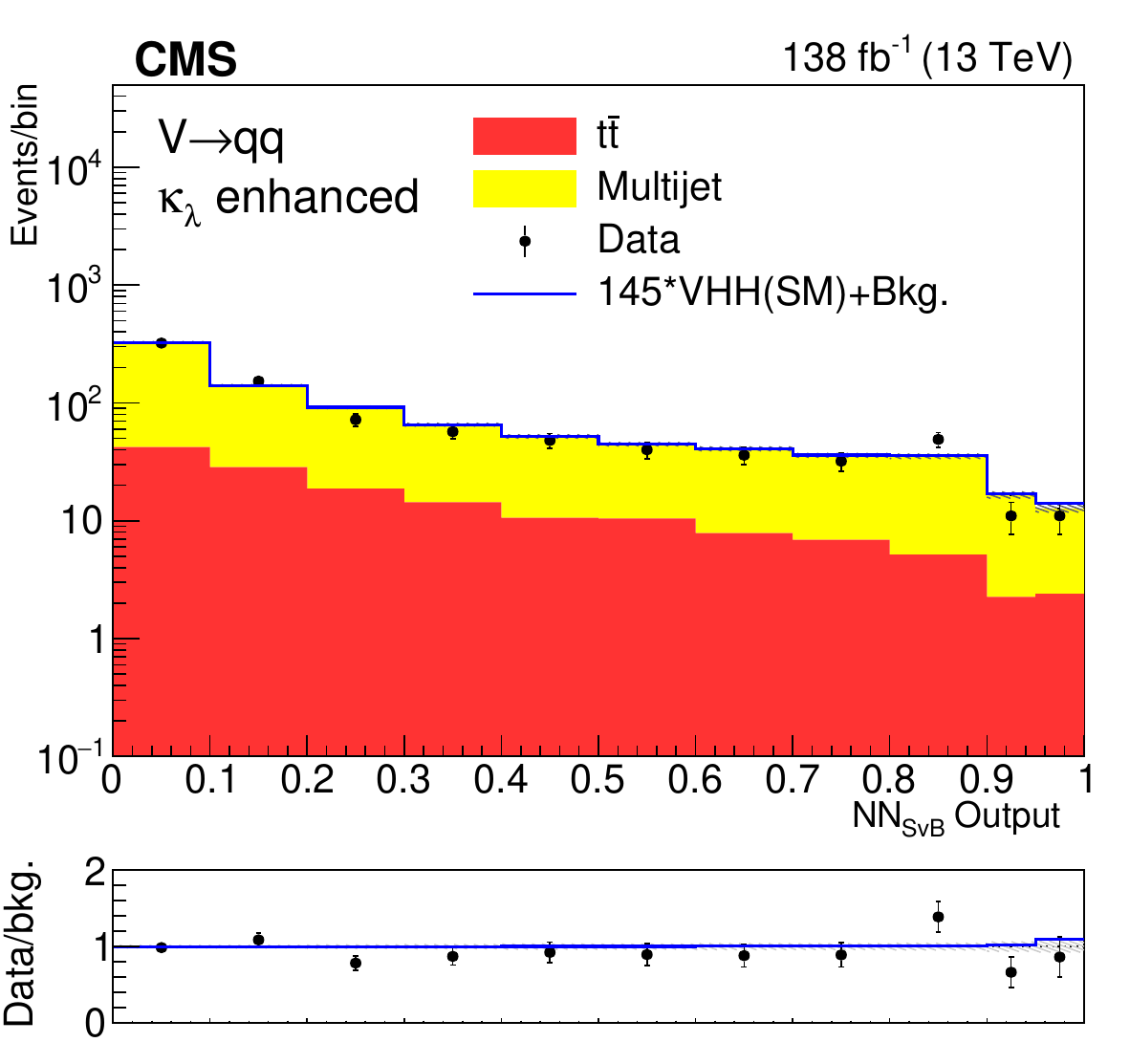}%
\hspace*{0.035\textwidth}%
\includegraphics[width=0.31\textwidth]{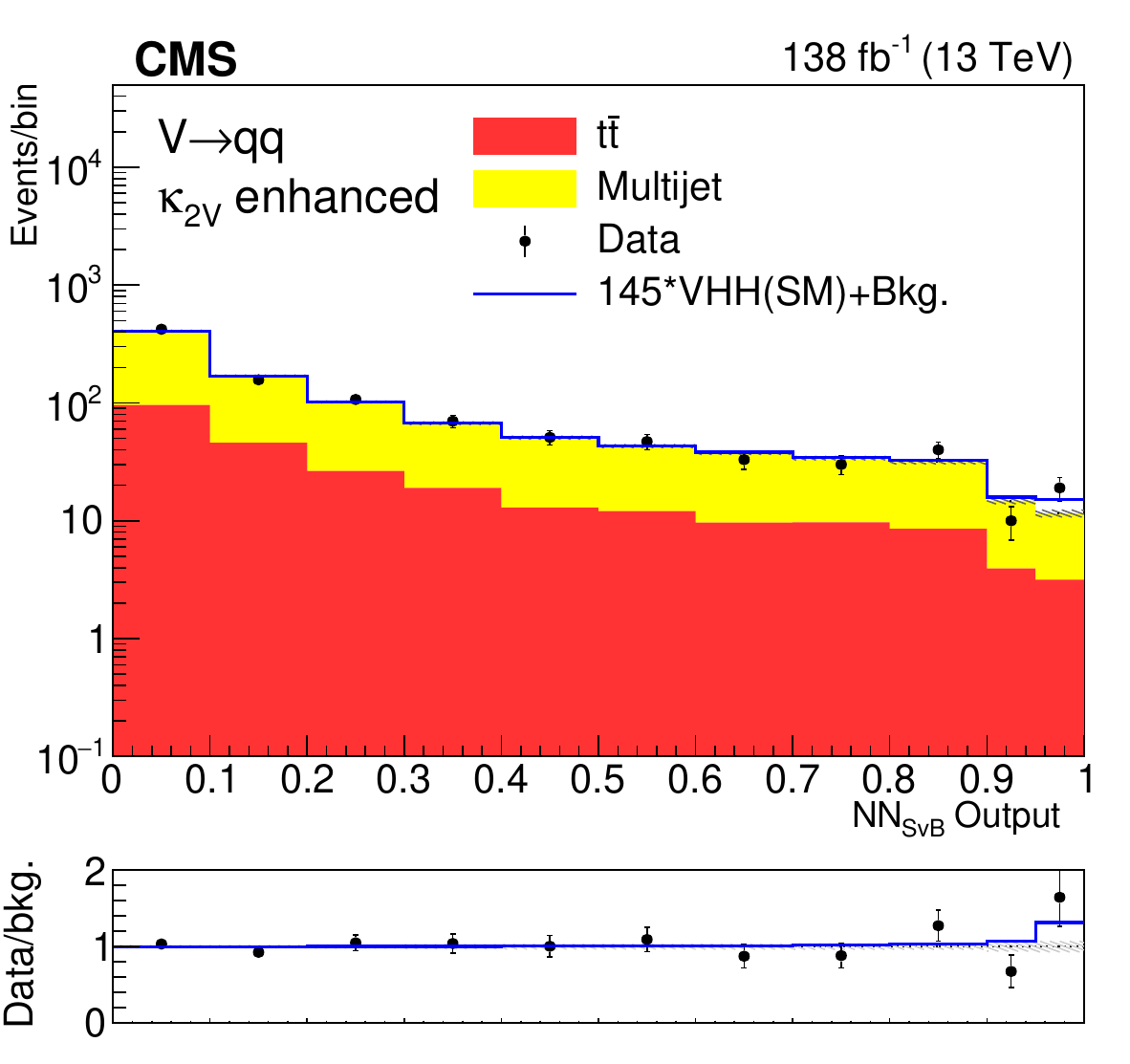} \\
\includegraphics[width=0.31\textwidth]{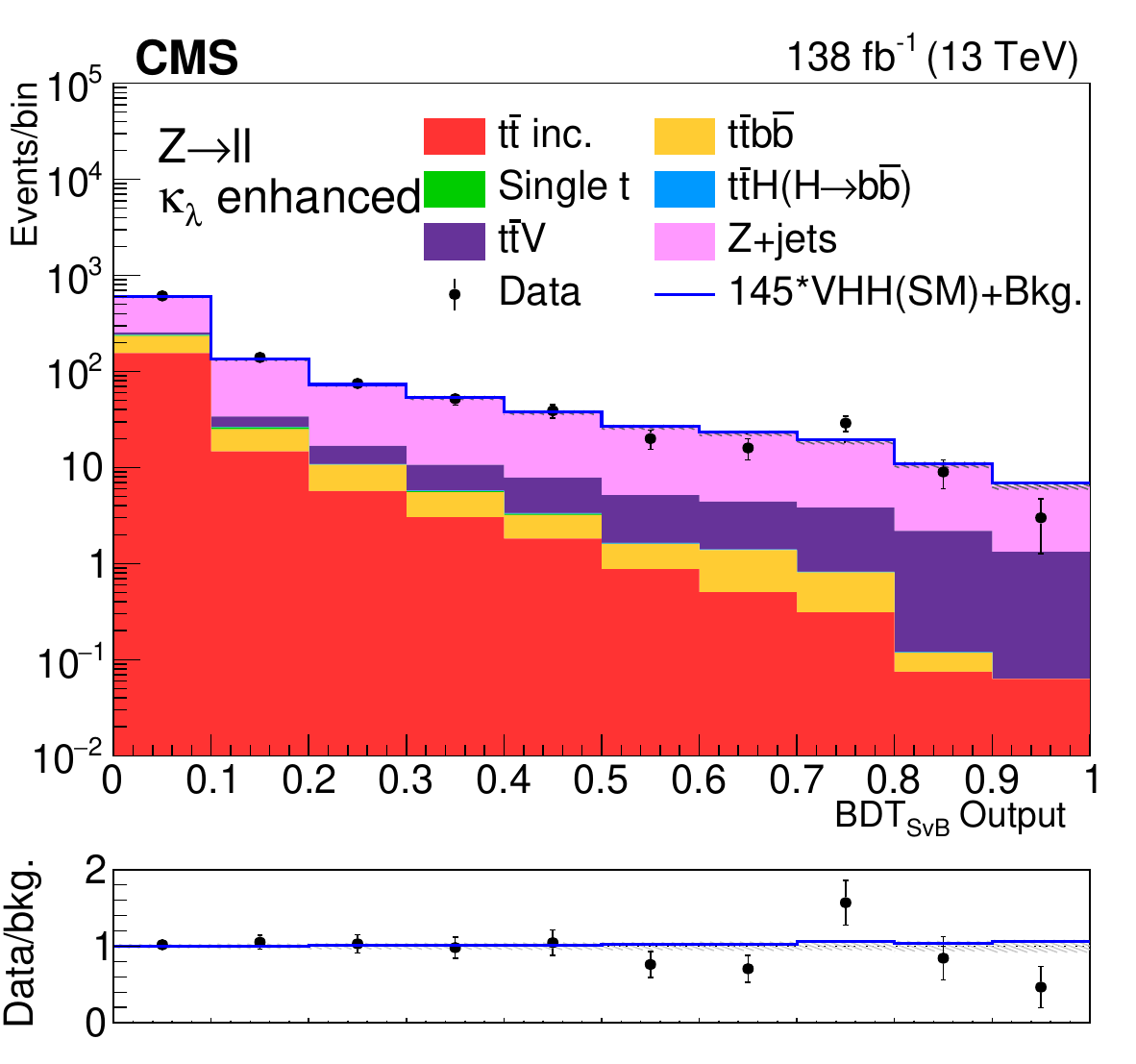}%
\hspace*{0.035\textwidth}%
\includegraphics[width=0.31\textwidth]{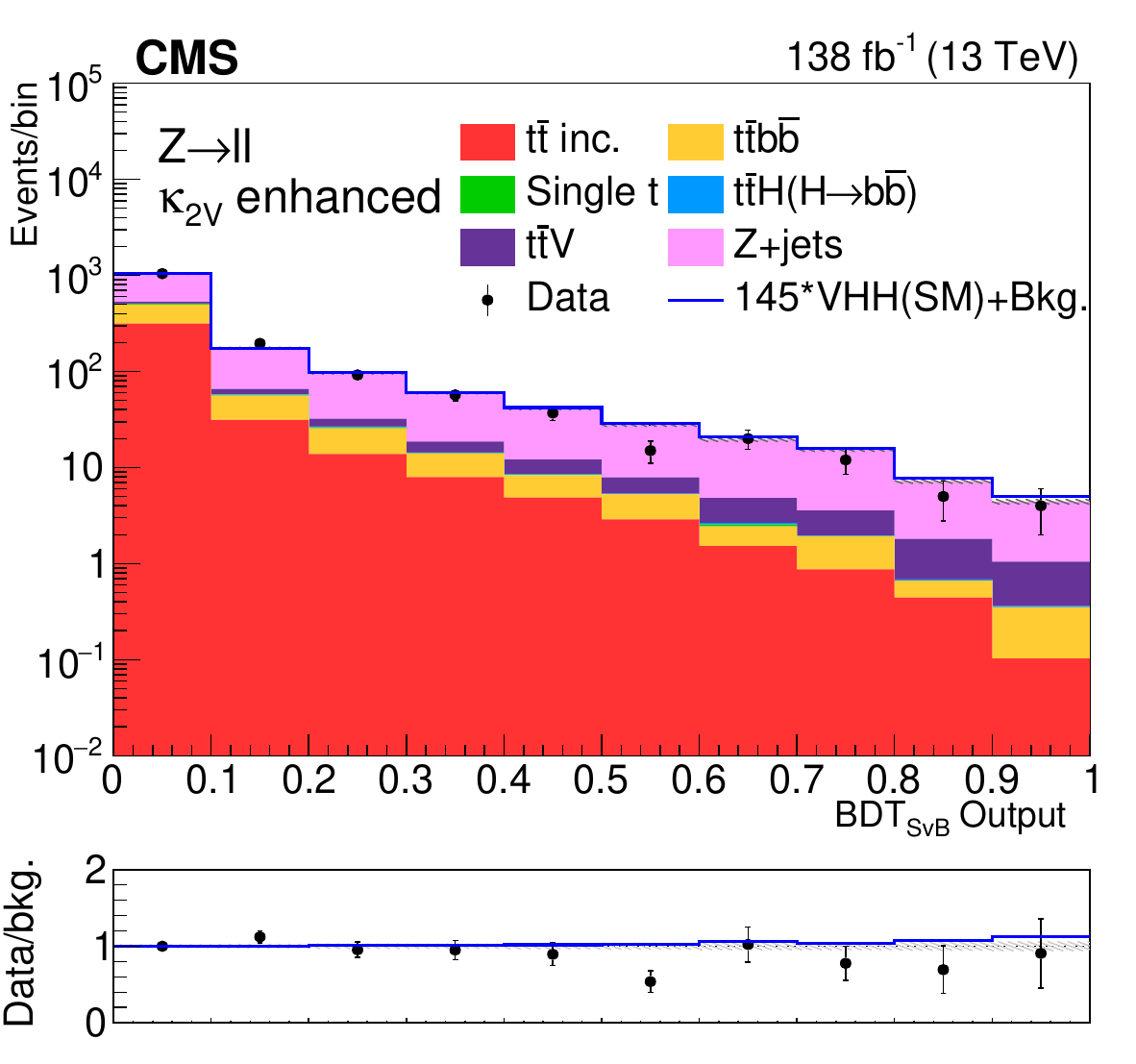}%
\caption{Postfit BDT distributions with the signal-plus-background hypotheses of the FH and 2L channels.}
\label{fig:postfit_bdt_FH_2l}
\end{figure}

\begin{figure}[!htp]
\centering
\includegraphics[width=0.31\textwidth]{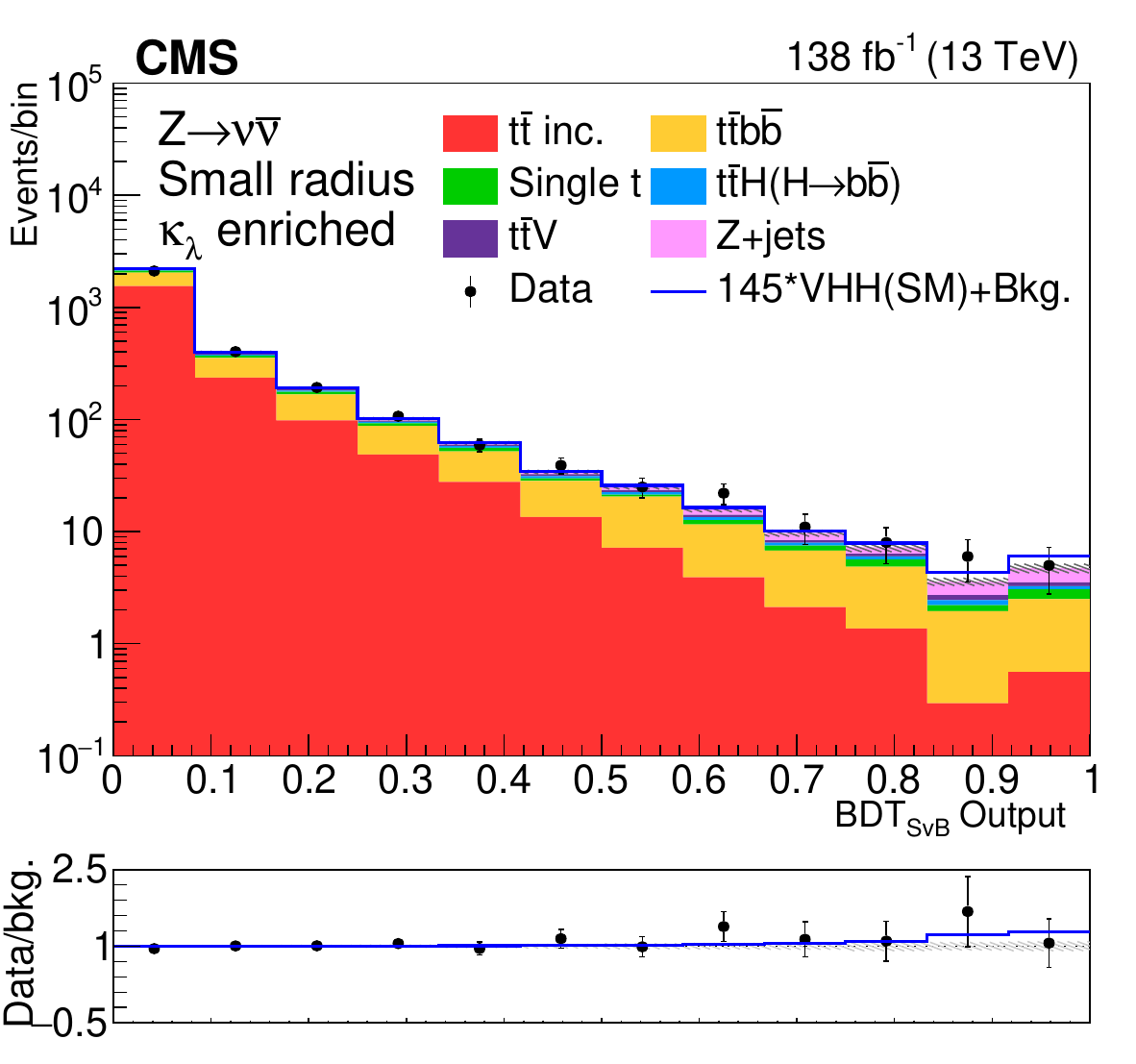}%
\hfill%
\includegraphics[width=0.31\textwidth]{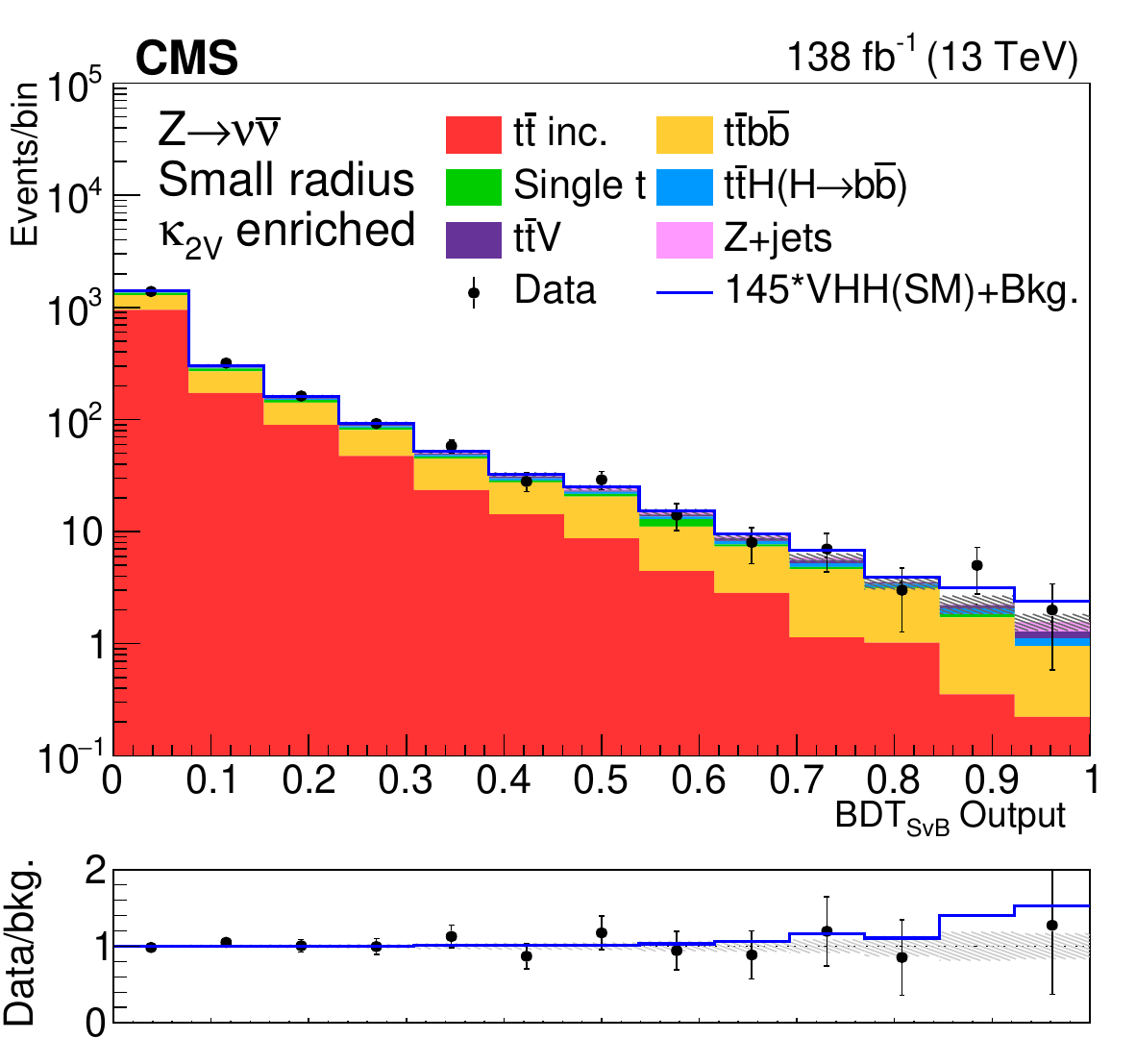}%
\hfill%
\includegraphics[width=0.31\textwidth]{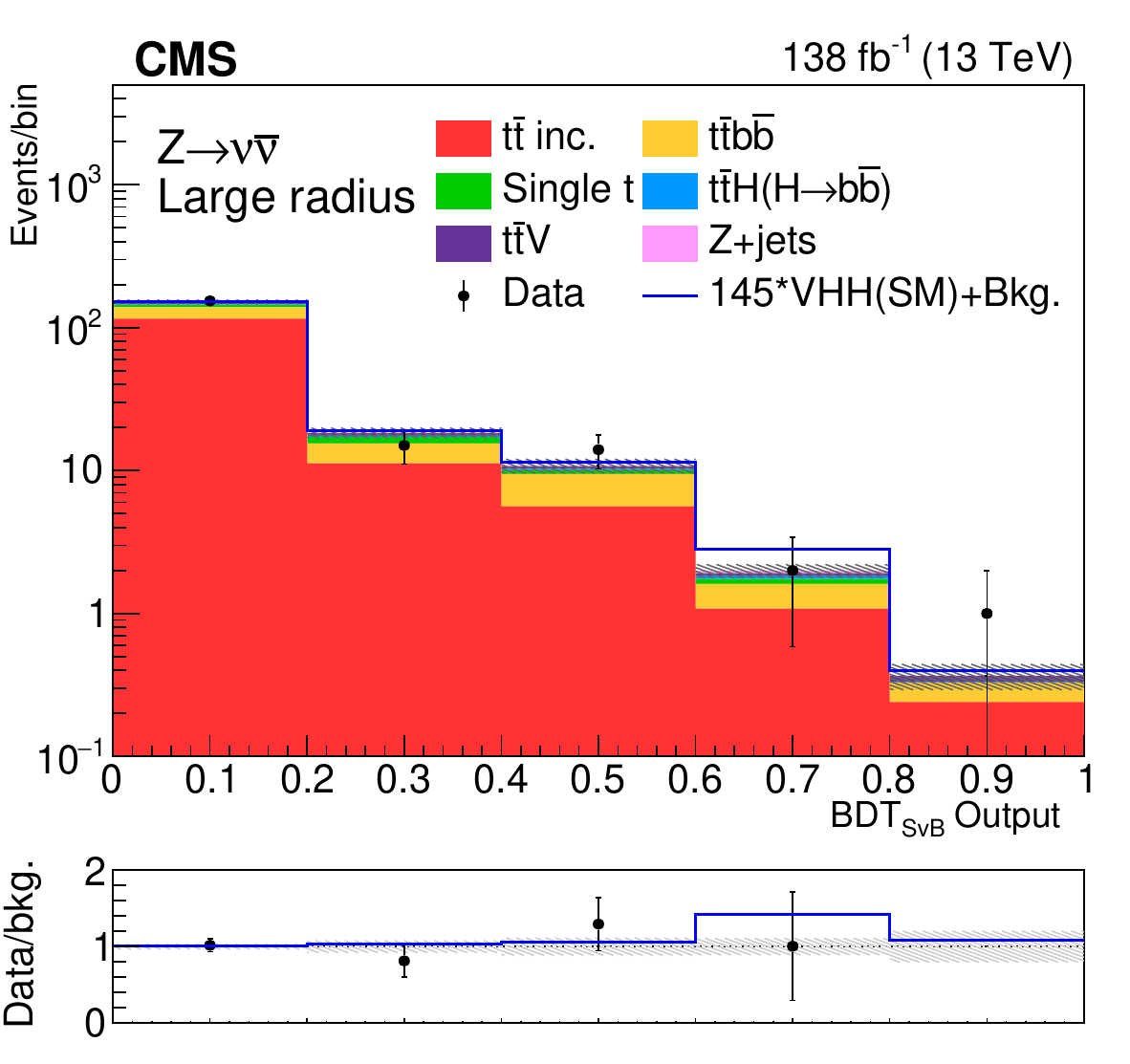} \\
\includegraphics[width=0.31\textwidth]{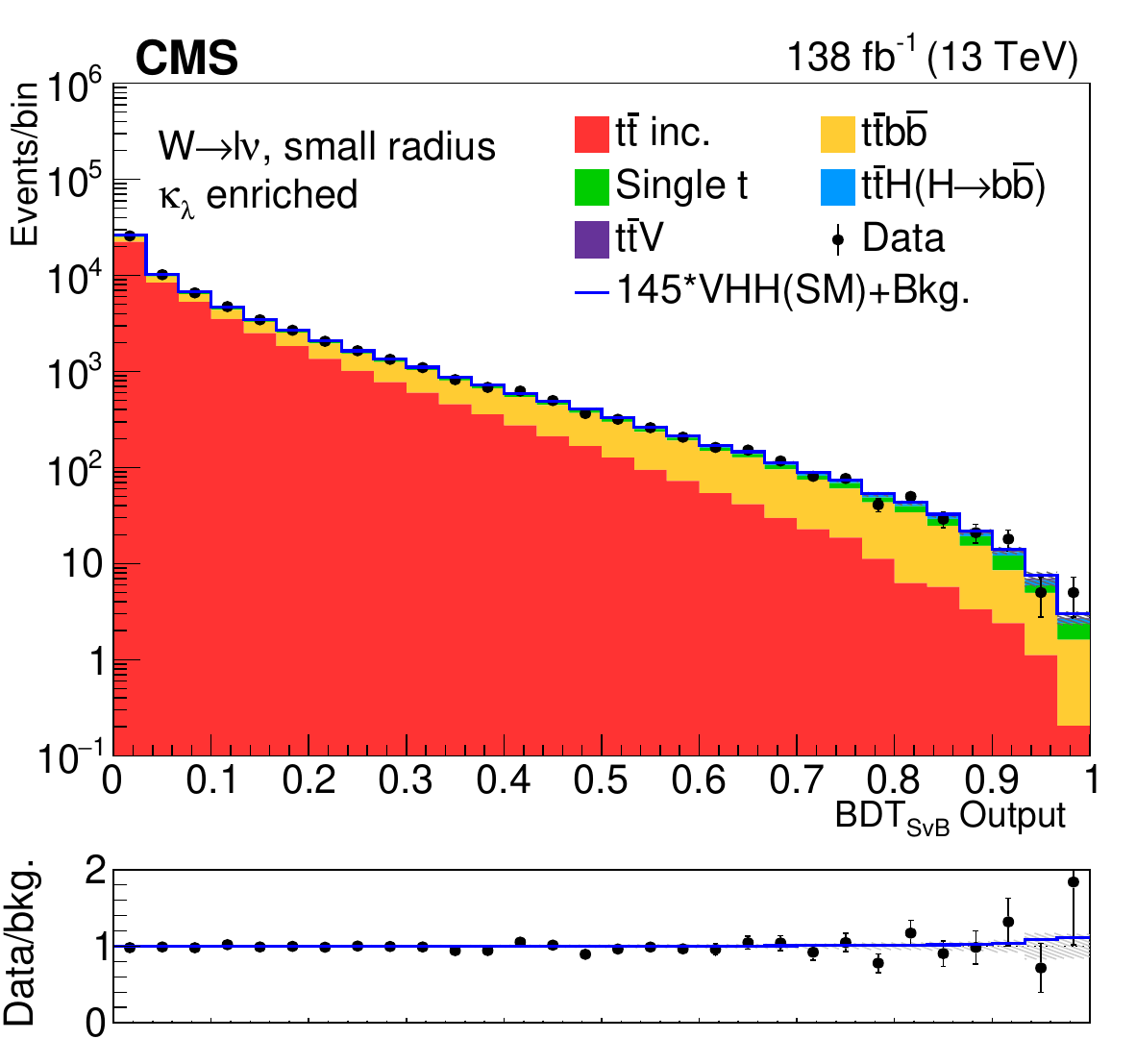}%
\hfill%
\includegraphics[width=0.31\textwidth]{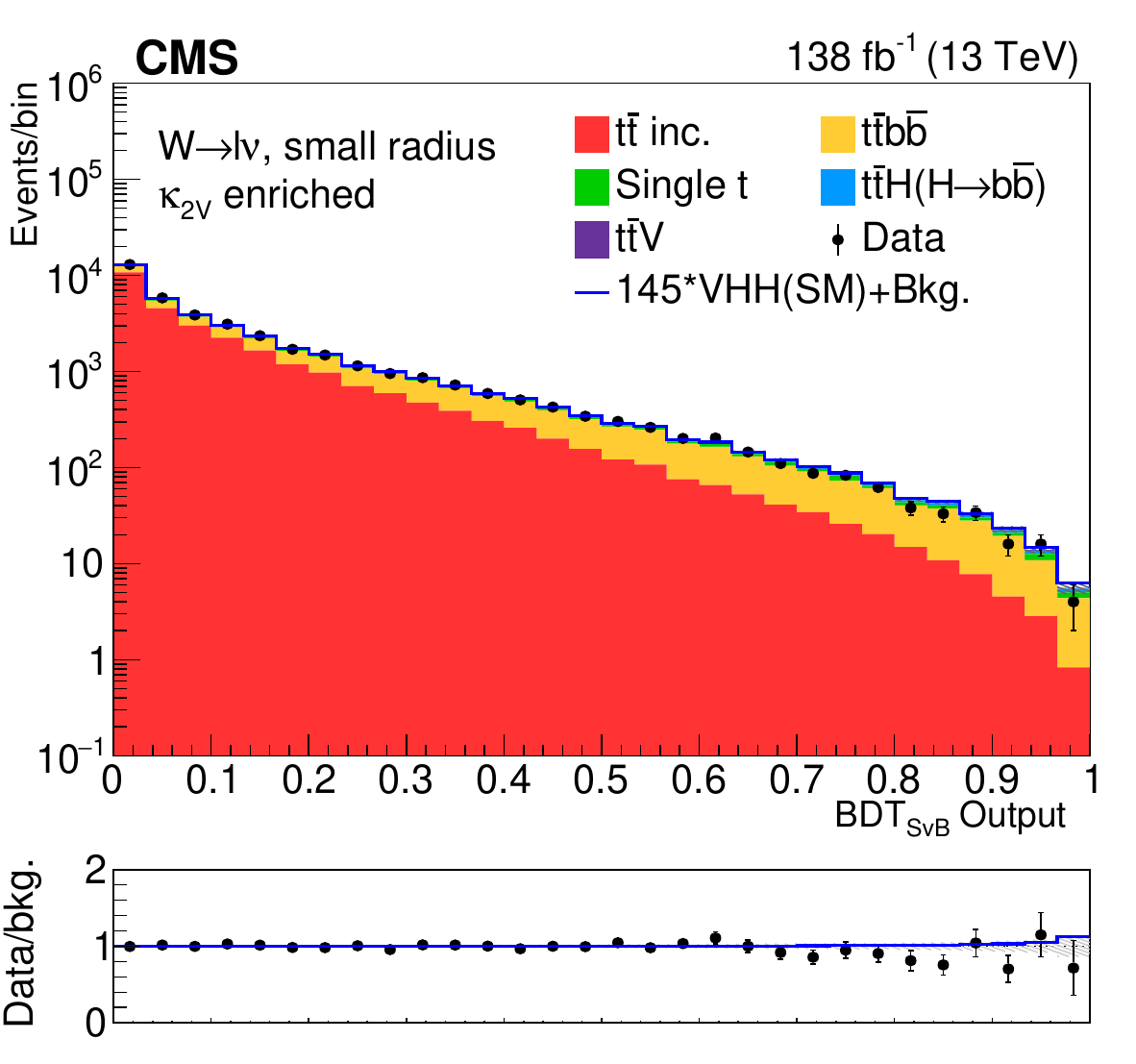}%
\hfill%
\includegraphics[width=0.31\textwidth]{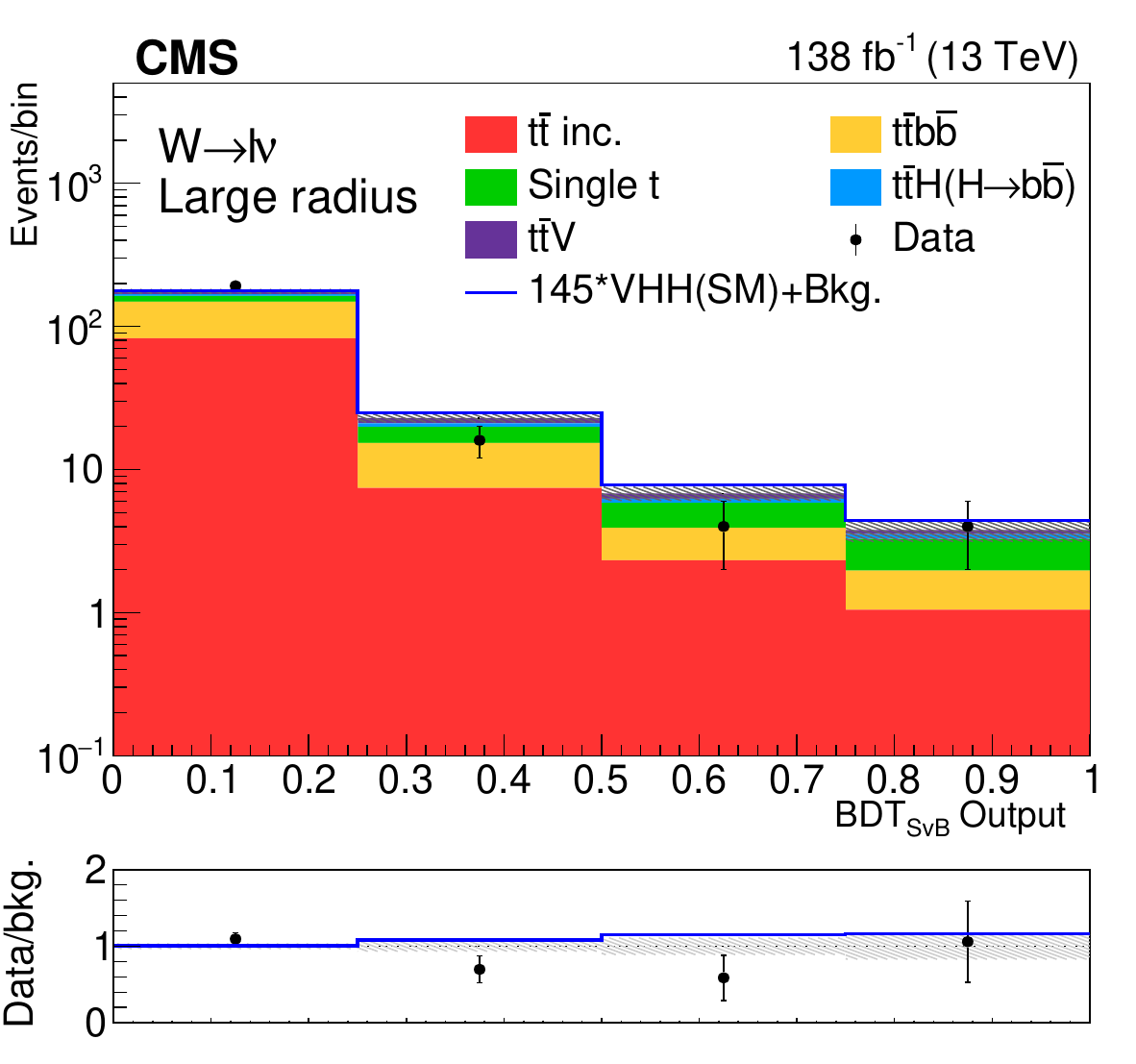}%
\caption{Postfit BDT distributions with the signal-plus-background hypotheses of the MET and 1L channels.}
\label{fig:postfit_bdt_0l_1l}
\end{figure}

\begin{figure}[!ht]
\centering
\includegraphics[width=0.85\textwidth]{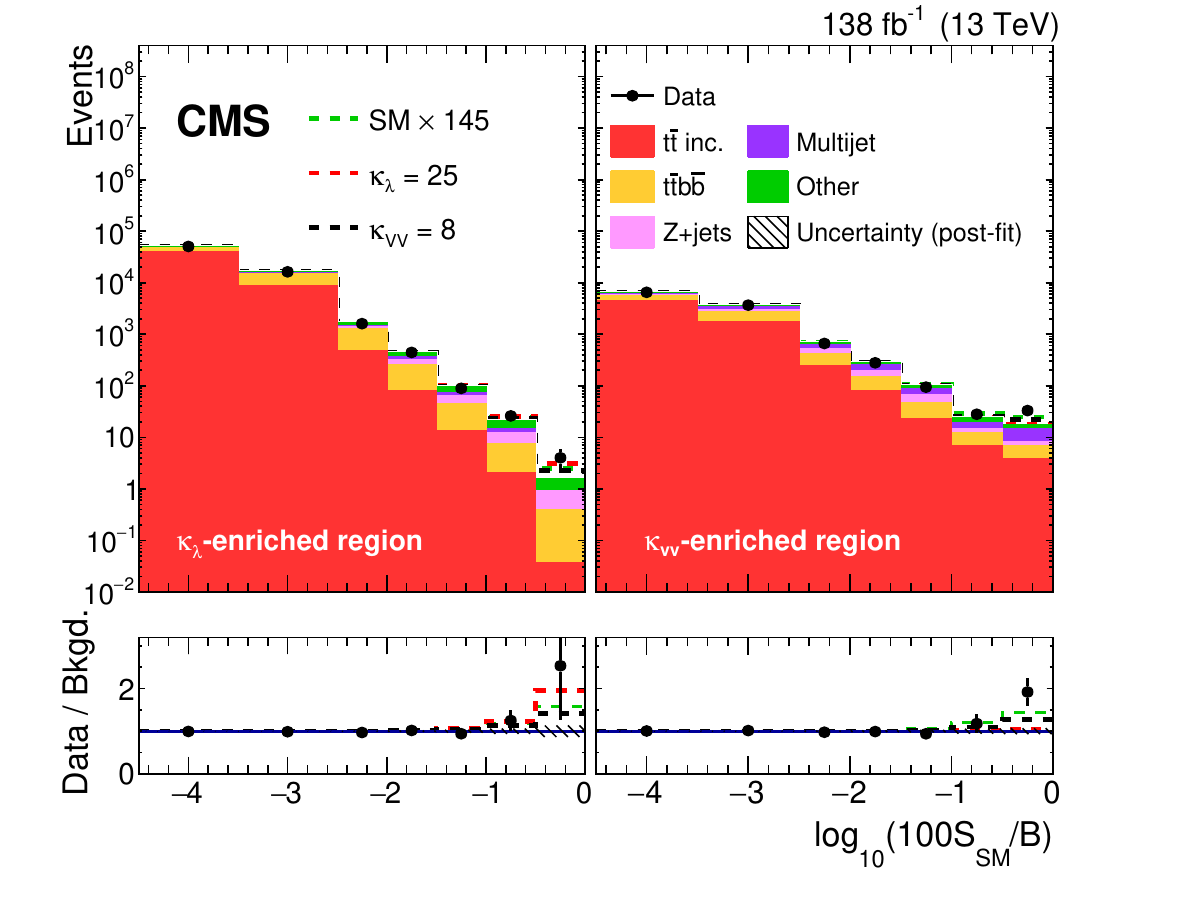}
\caption{Machine learning output distributions are transformed to \SoverB
and summed for \kl- and \kvv-enriched SR samples separately. The filled histograms represent the
postfit simulation. The total postfit uncertainty is represented by the hatched band. The SM
contribution and two signal models near expected exclusion at the 95\% \CL, each assuming the other couplings to be SM-like, are shown with the dashed lines.}
\label{fig:logsb}
\end{figure}

The postfit distributions with the signal-plus-background hypotheses are shown in Figs.~\ref{fig:postfit_bdt_FH_2l}--\ref{fig:postfit_bdt_0l_1l} with SR, CR, \Nb, \Dbb categories, and all data-taking periods summed up.
For an accurate and simultaneous visualization,
the SRs are combined by transforming all the discriminant outputs into bins of increasing signal purity,
\ie, \SoverB, where \Ssm is the signal predicted by
SM and \Bkg is estimated background, which are then summed separately for each enrichment type.
Figure~\ref{fig:logsb} shows these summed distributions, overlaid with data and signal
models within the coupling sensitivity of this analysis.
For better visualization, the single \PQt, \ttH, and \ttV backgrounds are grouped and labeled as ``Other'' in Fig.~\ref{fig:logsb}.

\begin{figure}[!htp]
\centering
\includegraphics[width=0.6\textwidth]{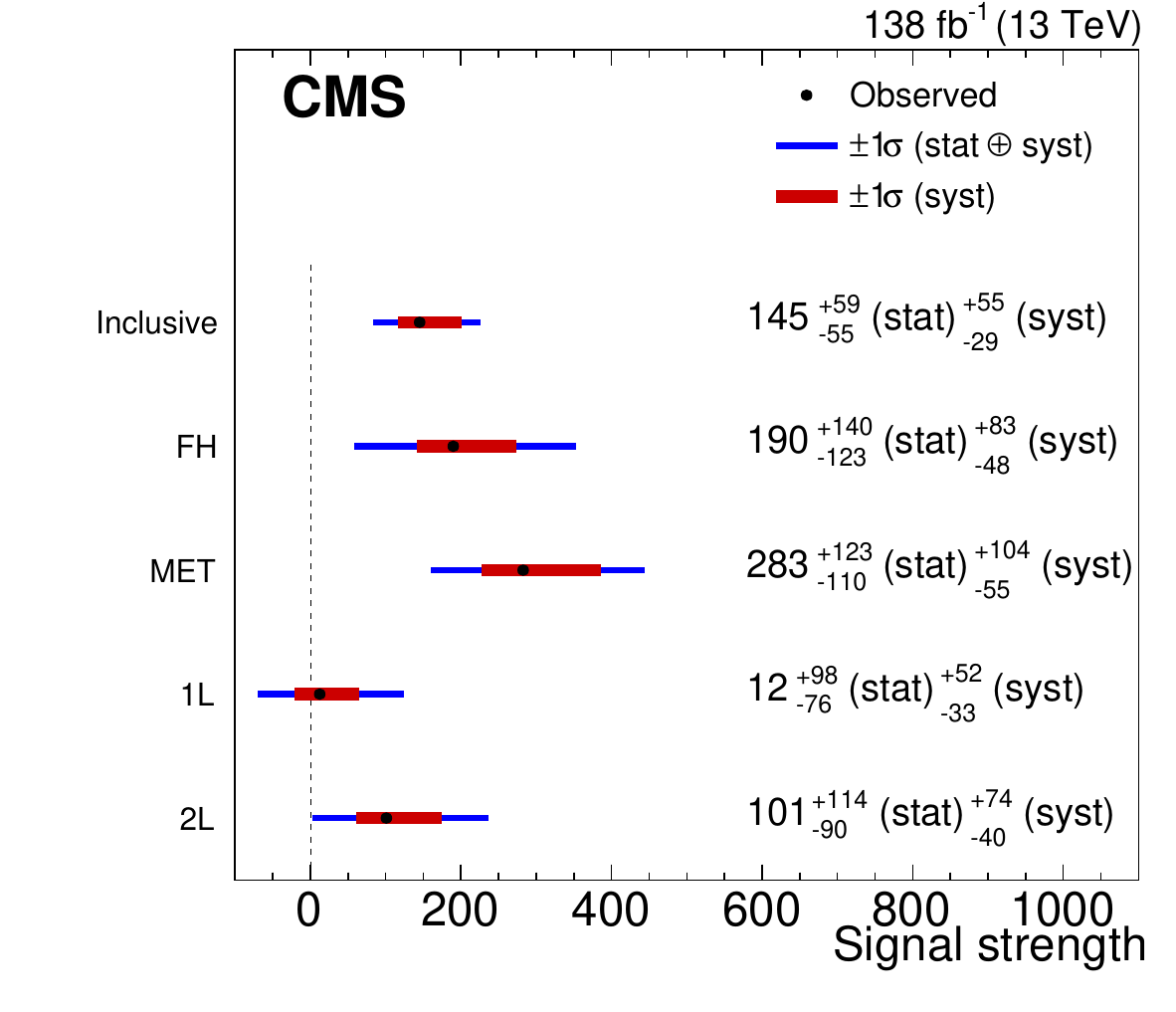}
\caption{Results of two maximum likelihood fits.  The top entry, labeled ``Inclusive'',
is the result of a single signal strength fit of all channels.  The other four entries are from a fit of the
same regions but with independent signal strengths in each channel.  The thinner blue bands are one
standard deviation from the full likelihood scan in that parameter, while the thicker red bands are one standard deviation
bands of the systematic uncertainties only. The mutual compatibilty of the multi-signal strength fit with the inclusive fit
is 38\%.}
\label{fig:sm_mu_fitted}
\end{figure}

A test statistic based on the profile likelihood ratio~\cite{LHCCLs} is used to determine the signal strength, with the combined signal strength as the parameter of interest (POI).
Systematic uncertainties are incorporated as additional nuisance parameters. In the likelihood calculations, each of the nuisance parameter adds an additional multiplicative term to the total likelihood.
Alternatively four signal strengths are assigned as POI to each channel and compared to the combined signal strength.
Figure~\ref{fig:sm_mu_fitted} shows the signal strengths per analysis channel, as well as the combined signal strengths.

A small excess of data over the background-only expectation in the most signal-like bins in \kvv-enrichment region can
be seen in Fig.~\ref{fig:logsb}, which is primarily from the \met channel \kvv-enrichment region SRs shown in Fig.~\ref{fig:postfit_bdt_0l_1l}.
For an SM-like signal, the observed significance is 2.6 standard
deviations with a fitted signal strength of $145^{+81}_{-63}$ times the SM signal when all coupling modifiers are equal to 1.
Figure~\ref{fig:sm_mu_fitted} shows the results of two fits.  The first one is the inclusive fit with a single signal strength
modifier, and the second fit has separate signal strength modifiers per channel.

\begin{figure}[!htp]
\centering
\includegraphics[width=0.49\textwidth]{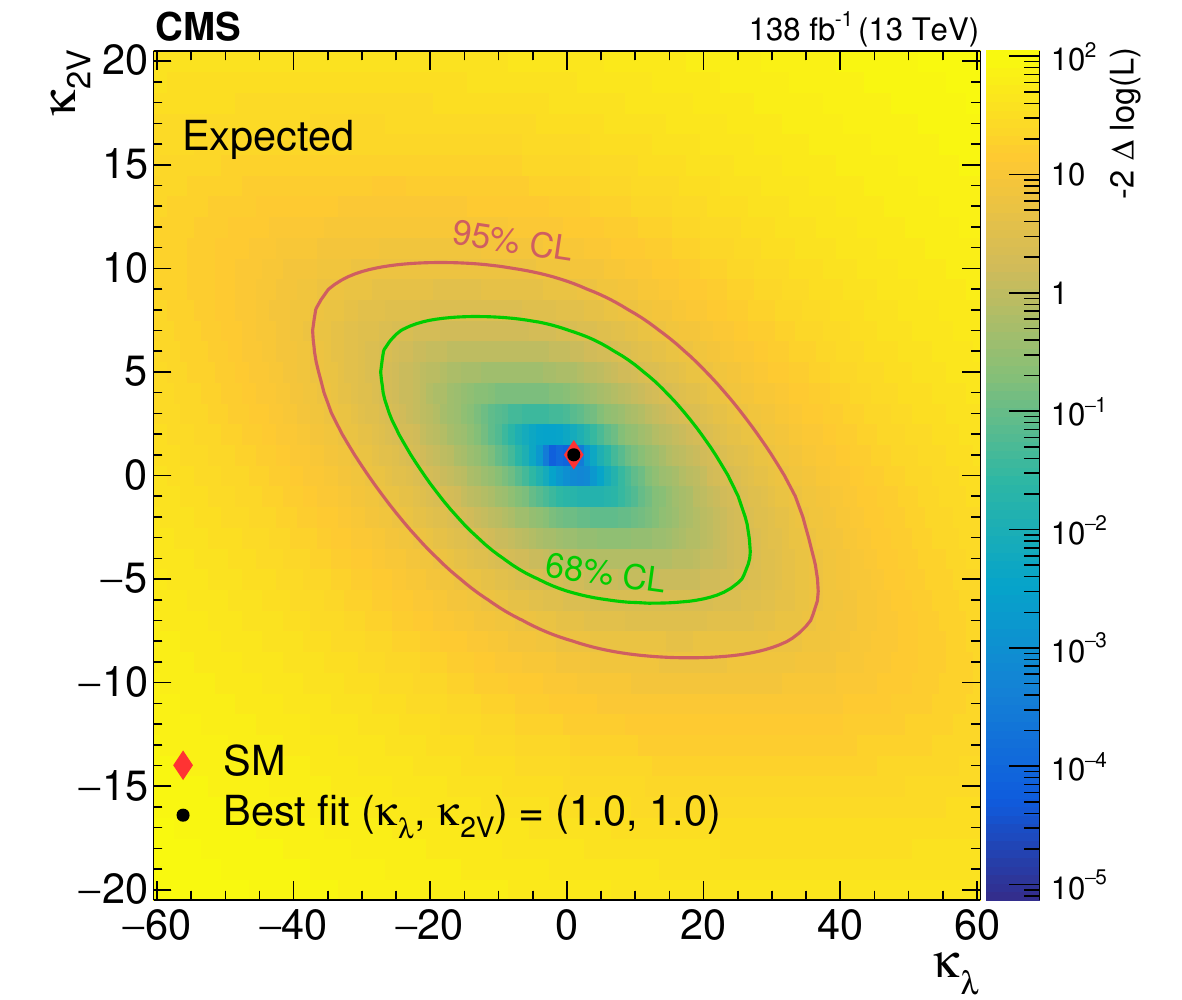}%
\hfill%
\includegraphics[width=0.49\textwidth]{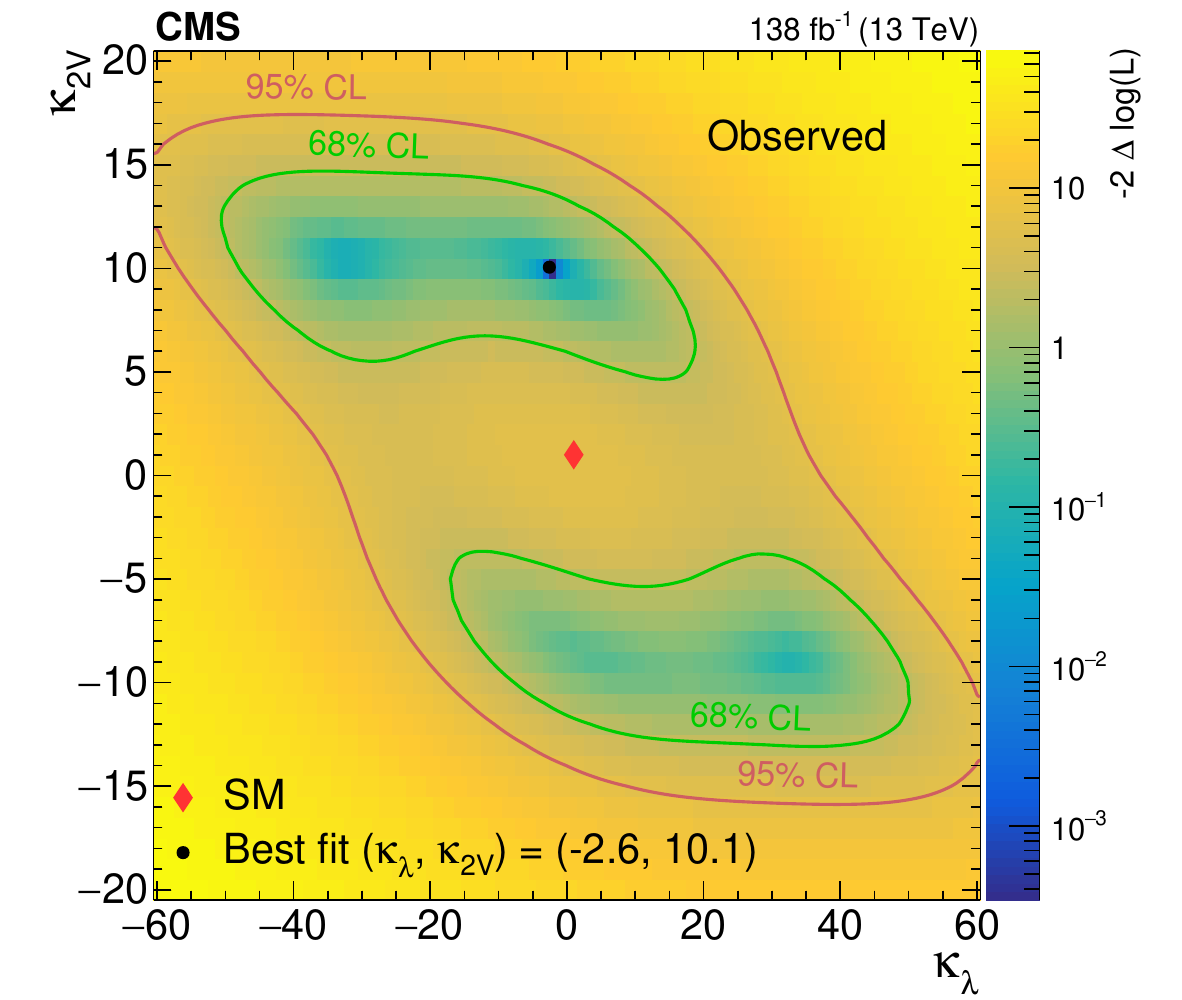}%
\caption{Expected (left) and observed (right) likelihood scans in \kl versus \kvv are shown, with other couplings
fixed to the SM predicted strength. The excess is most prominent in the \kvv-enriched region, and so
the most likely point of the scan at $\kvv=10.1$ and $\kl=-2.6$ is pulled from the SM mostly in the \kvv dimension.}
\label{fig:kappascans_kl_kvv}
\end{figure}

\begin{figure}[!htp]
\centering
\includegraphics[width=0.49\textwidth]{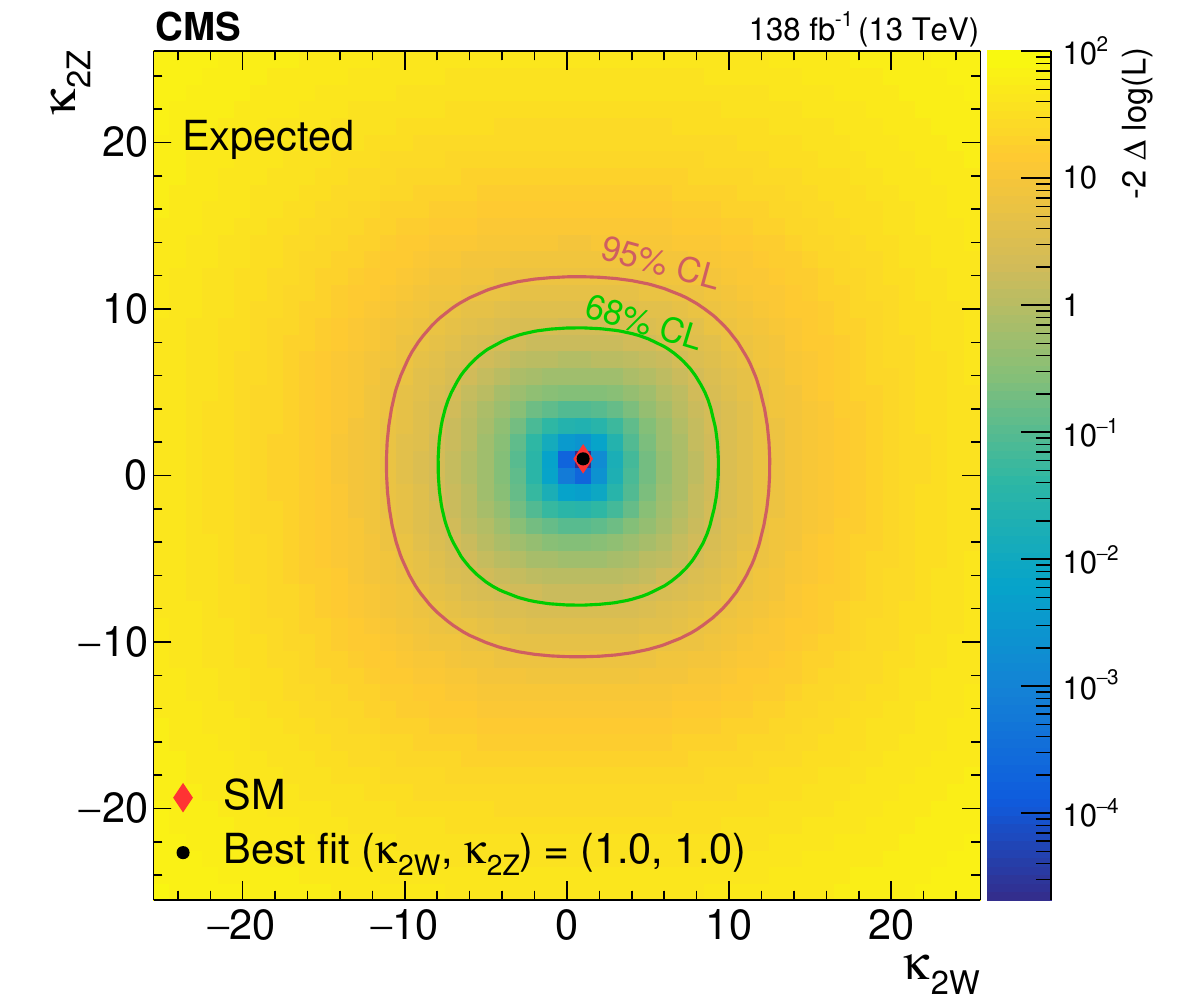}%
\hfill%
\includegraphics[width=0.49\textwidth]{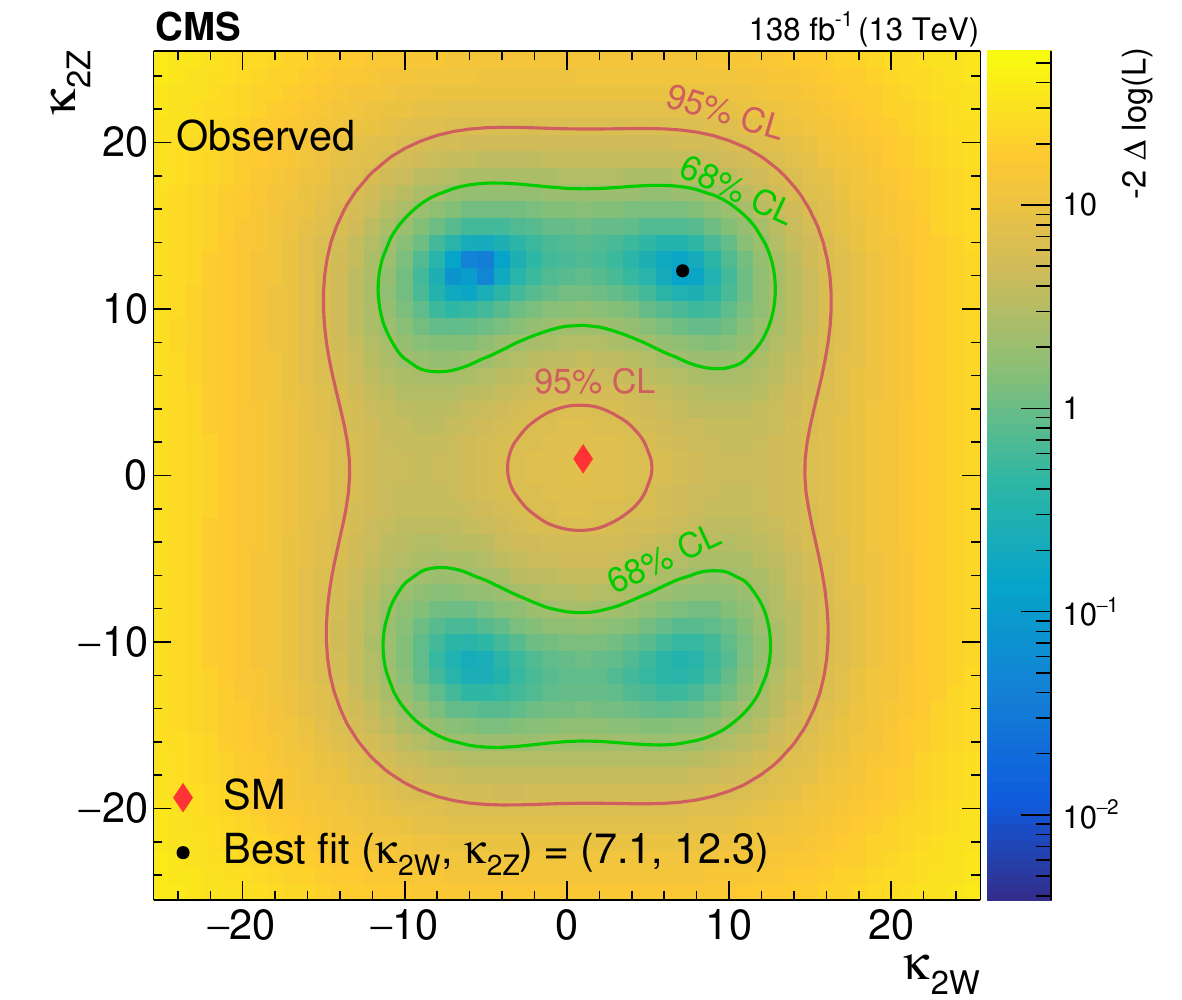}%
\caption{Expected (left) and observed (right) likelihood scans of \kww versus \kzz are shown, with other couplings
fixed to the SM predicted strength.  The excess is most prominent in the \met channel, and so
the most likely point of the scan at $\kww=7.1$ and $\kzz=12.3$ is pulled from the
SM mostly in the \kzz dimension, to which the signal in the \met channel is solely sensitive.}
\label{fig:kappascans_kww_kzz}
\end{figure}

Two-dimensional likelihood scans of \kl versus \kvv and \kzz versus \kww are shown in Fig.~\ref{fig:kappascans_kl_kvv} and \ref{fig:kappascans_kww_kzz}, respectively.  Other couplings are fixed to their SM values.

\begin{figure}[!htp]
\centering
\includegraphics[width=0.48\textwidth]{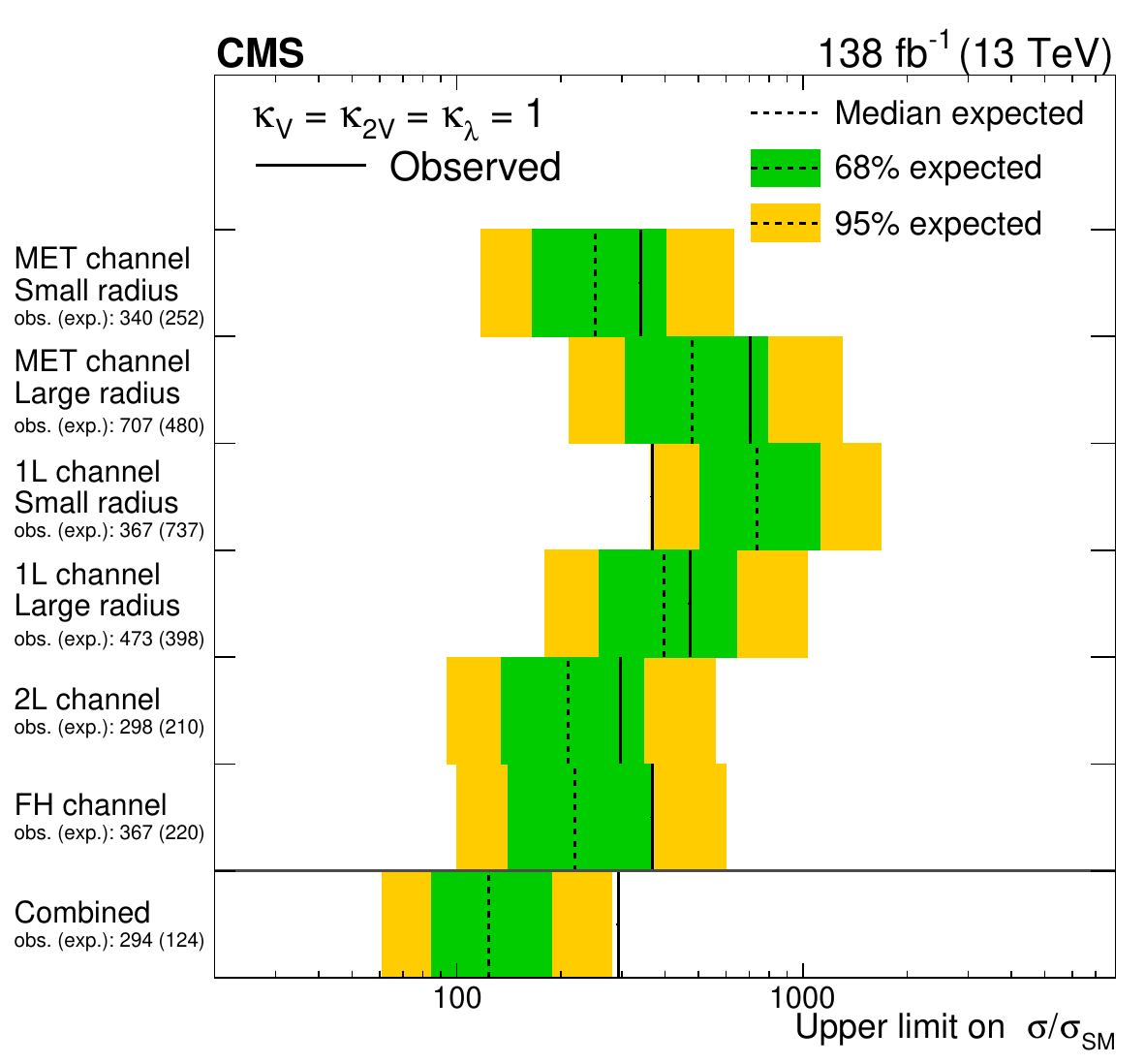}%
\hfill%
\includegraphics[width=0.48\textwidth]{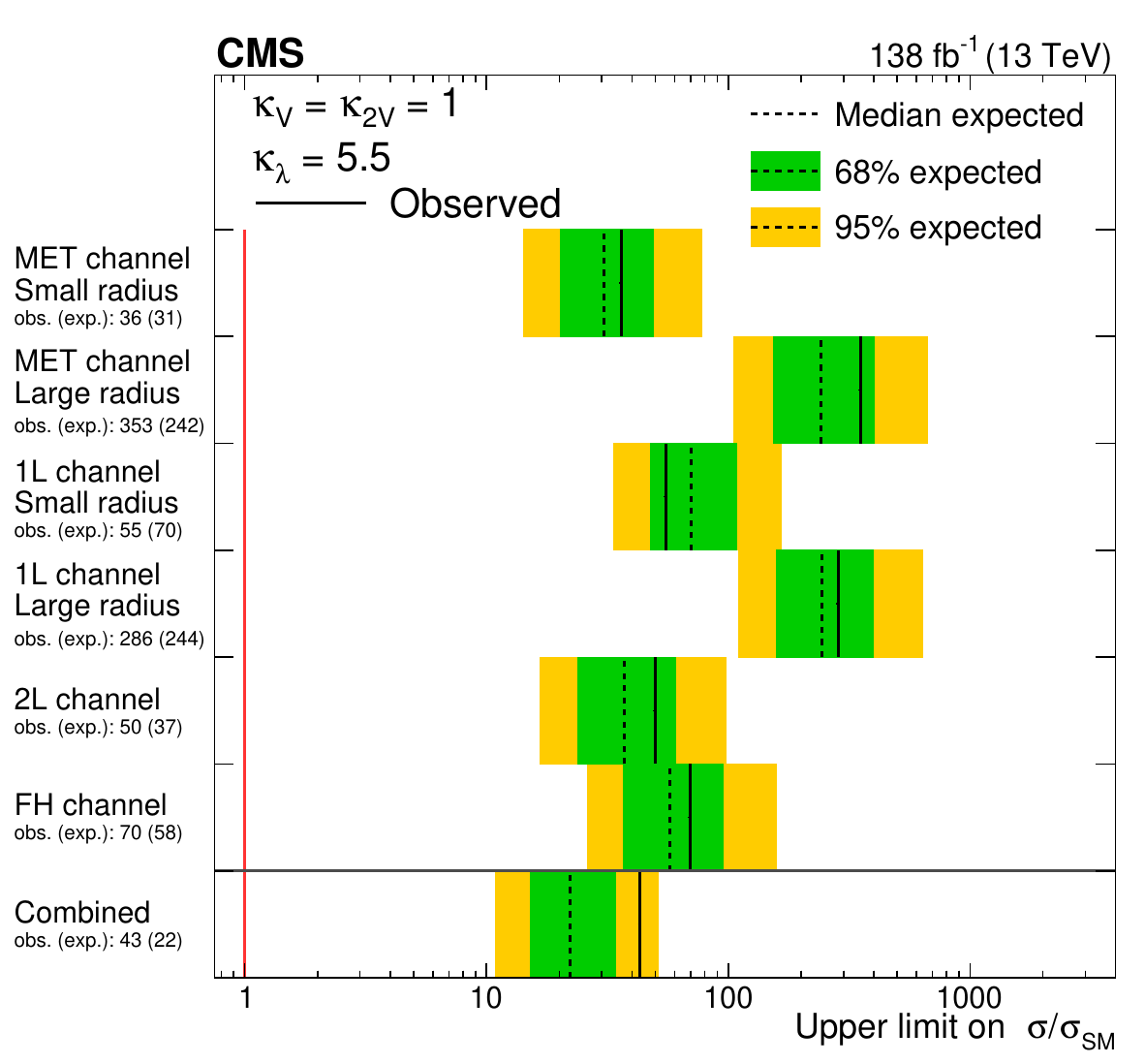}%
\caption{The left plot shows the \VHH cross section upper limits per channel and combined for SM value couplings, while results with $\kl=5.5$ and
$\kvv=\kv=1.0$ are shown on the right.}
\label{fig:limits}
\end{figure}

\begin{figure}[!tph]
\centering
\includegraphics[width=0.48\textwidth]{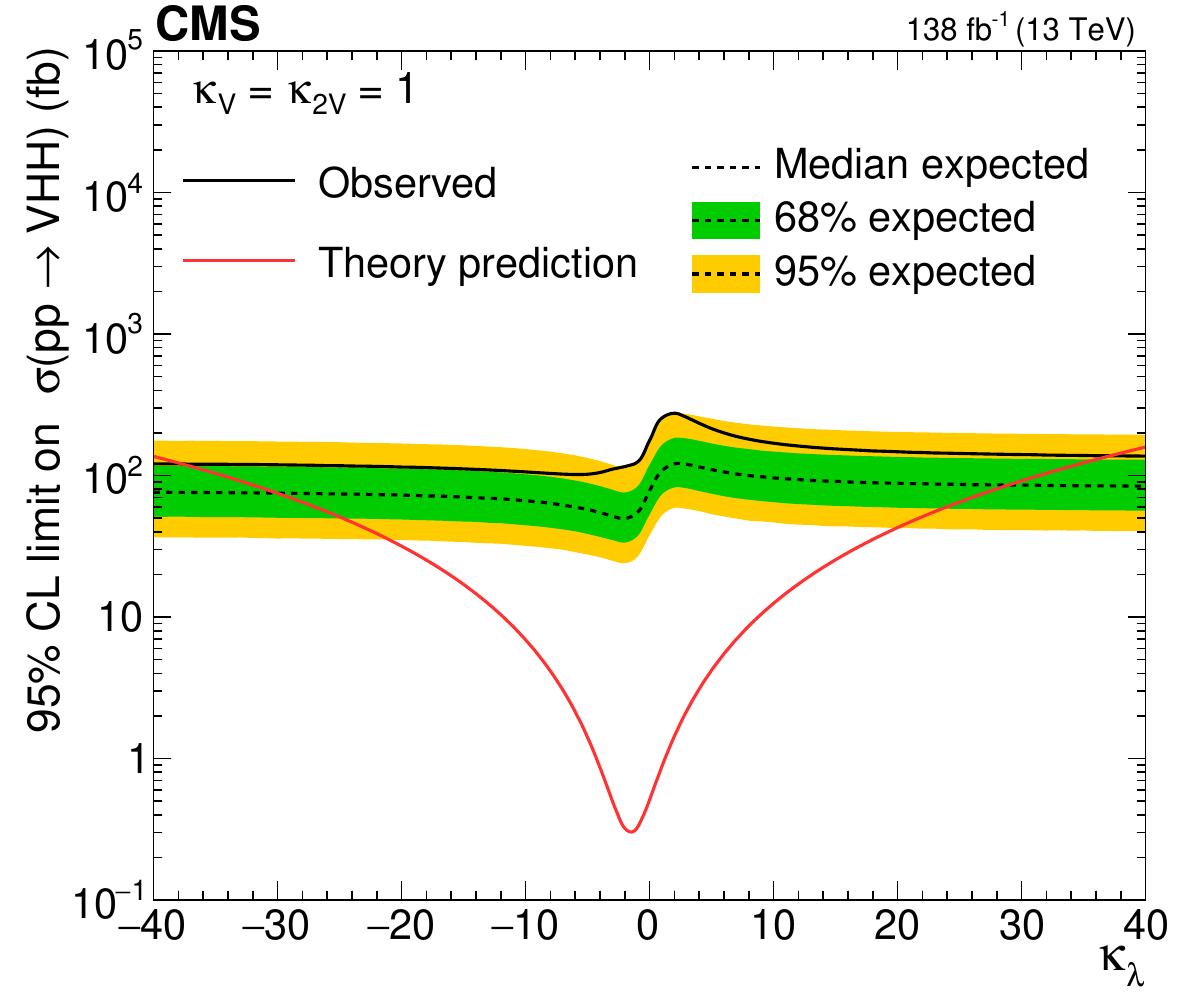}%
\hfill%
\includegraphics[width=0.48\textwidth]{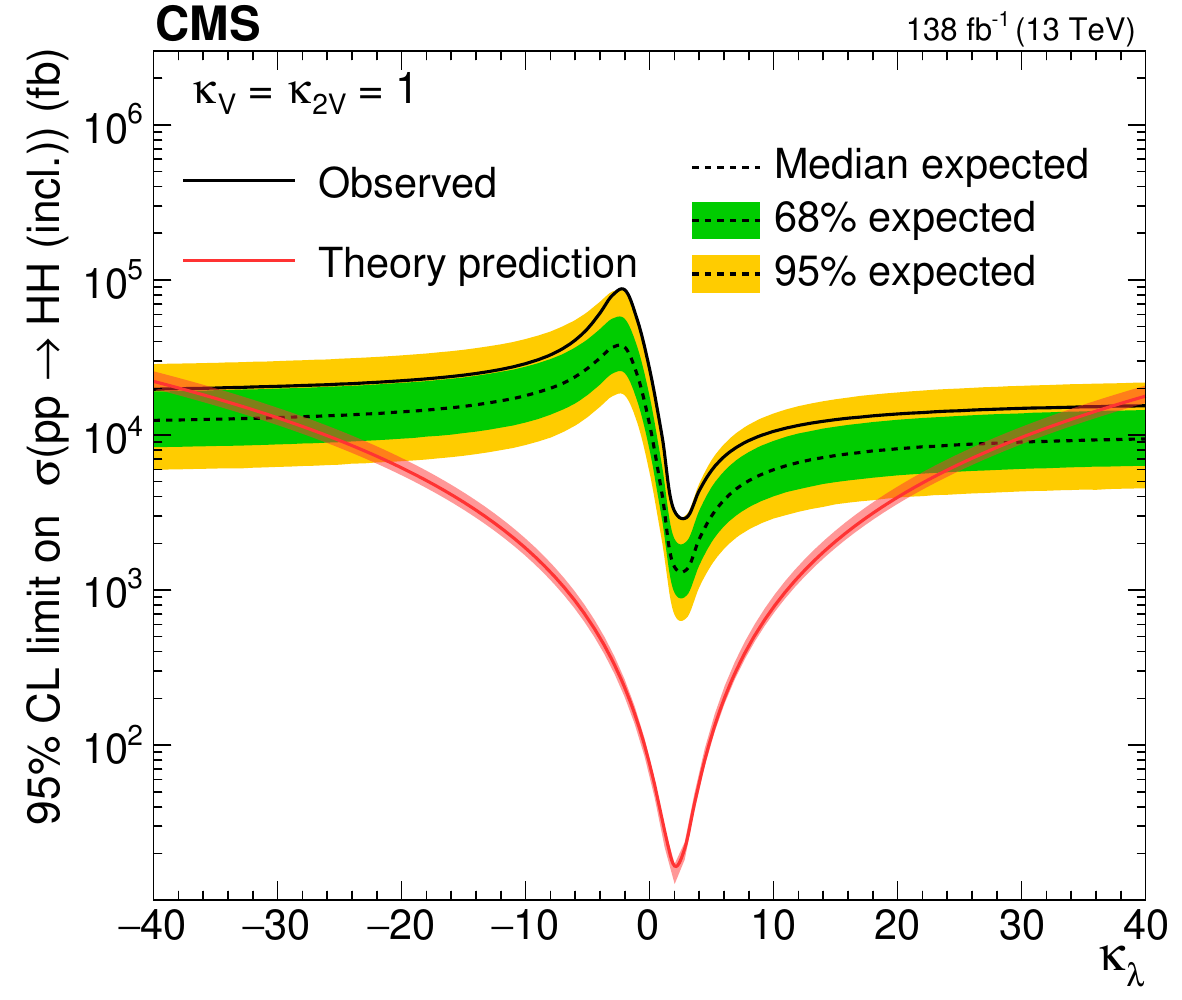}%
\caption{Upper 95\% \CL limits on \VHH (left) and \HH (right) signal cross section scanned over
the \kl parameter while fixing the \kvv and \kv to their SM-predicted values. The independent axis is
the scanned \kl parameter, and the dependent axis is the 95\% \CL upper limit on signal cross section.
The theoretic prediction of \VHH (left) and \HH (right) production cross sections are shown with the red lines.}
\label{fig:klscan}
\end{figure}

\begin{figure}[!tph]
\centering
\includegraphics[width=0.48\textwidth]{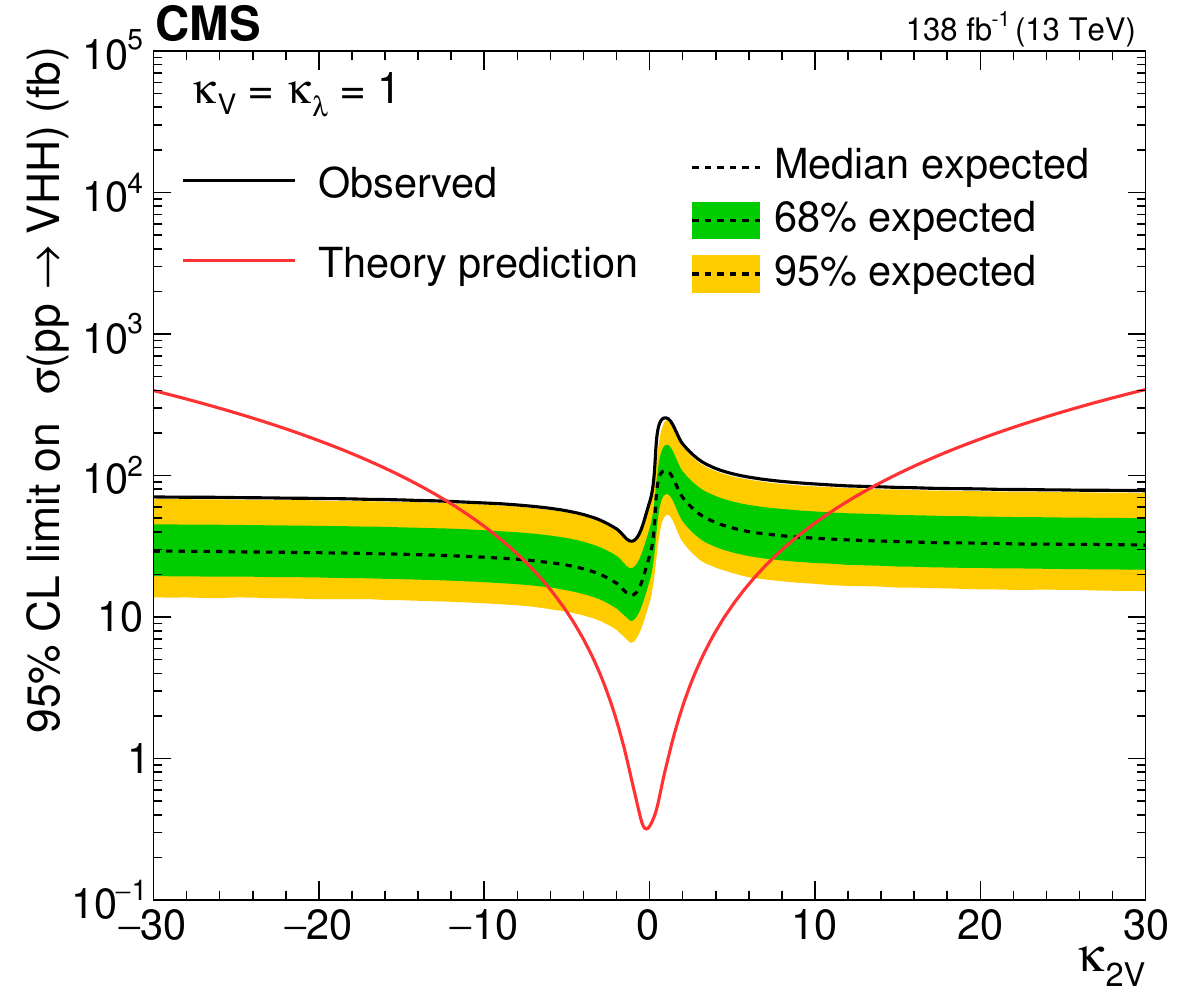}%
\hfill%
\includegraphics[width=0.48\textwidth]{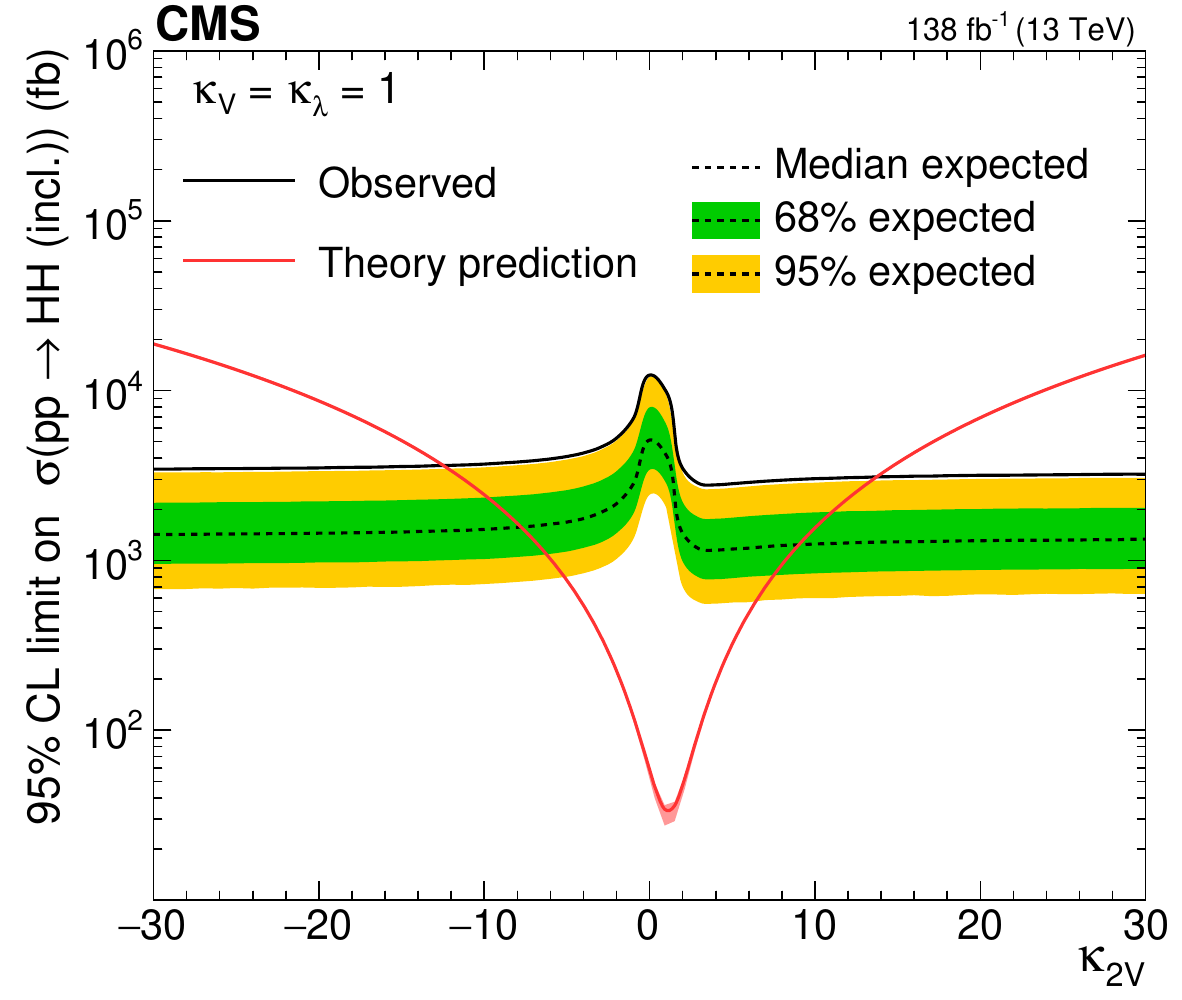}%
\caption{Upper 95\% \CL limits on \VHH (left) and \HH (right) signal cross section scanned over
the \kvv parameter while fixing the \kl and \kv to their SM-predicted values. The independent axis is
the scanned \kvv parameter, and the dependent axis is the 95\% \CL upper limit on signal cross section.
The theoretic prediction of \VHH (left) and \HH (right) production cross sections are shown with the red lines.}
\label{fig:kvvscan}
\end{figure}

\begin{figure}[!tph]
\centering
\includegraphics[width=0.48\textwidth]{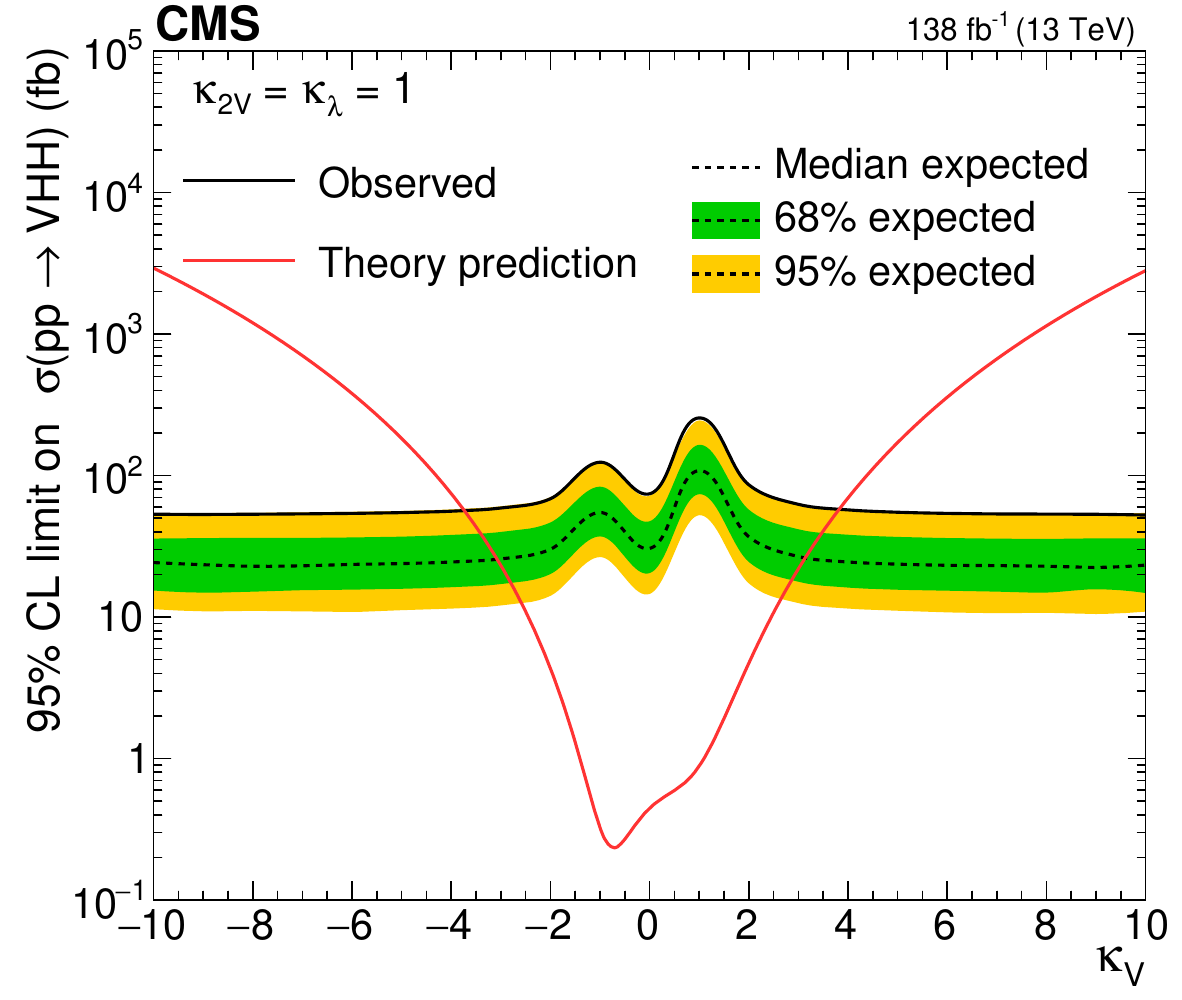}%
\hfill%
\includegraphics[width=0.48\textwidth]{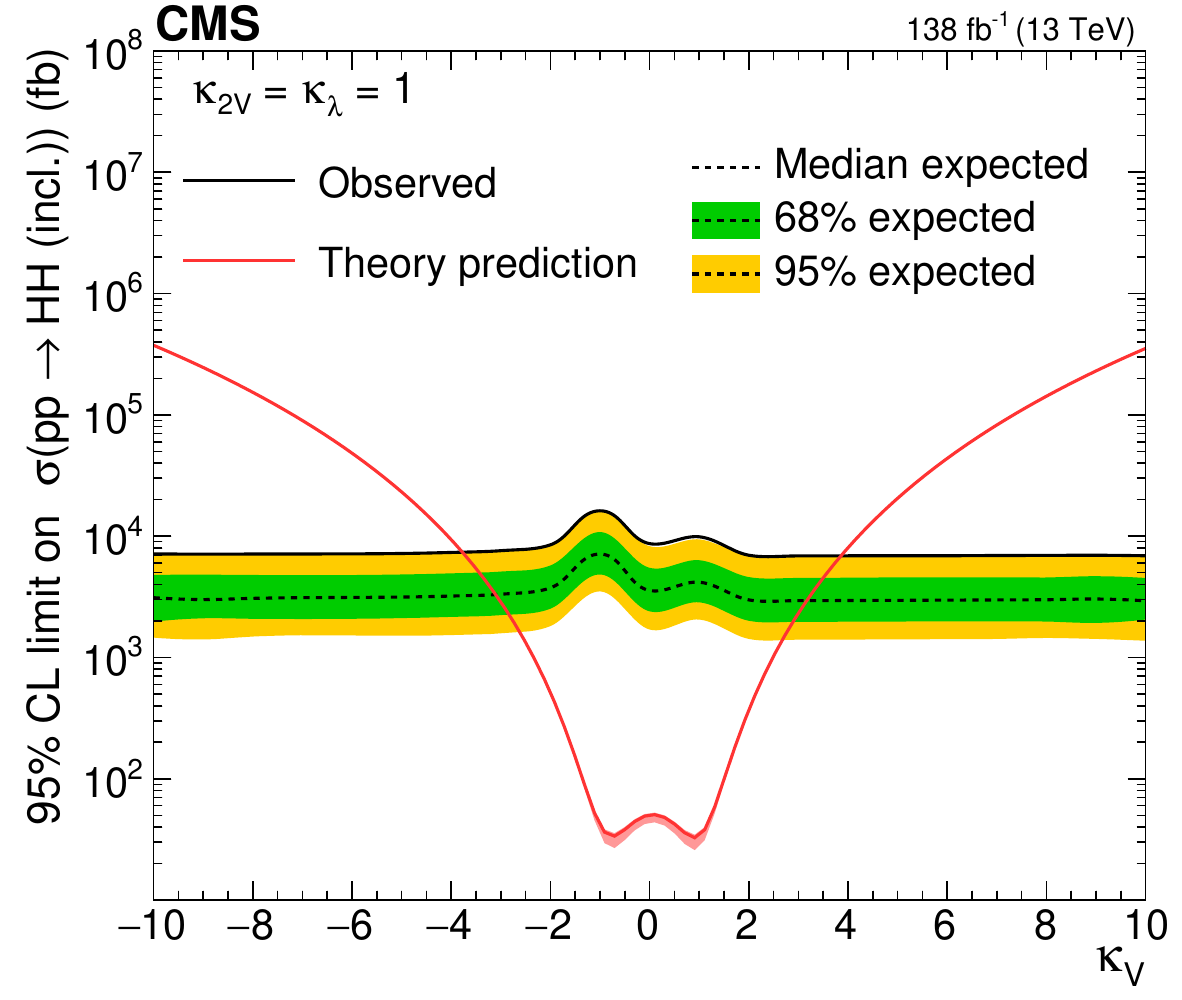}%
\caption{Upper 95\% \CL limits on \VHH (left) and \HH (right) signal cross section scanned over
the \kv parameter while fixing the \kvv and \kl to their SM-predicted values. The independent axis is
the scanned \kv parameter, and the dependent axis is the 95\% \CL upper limit on signal cross section.
The theoretic prediction of \VHH (left) and \HH (right) production cross sections are shown with the red lines.}
\label{fig:kvscan}
\end{figure}

Based on the \CLs criterion~\cite{CLS2,CLS1} and asymptotic formulas~\cite{Cowan:2010js},
the upper limits on the \VHH cross section at 95\% \CL are extracted both with the SM couplings and with scans on the coupling modifiers.
The upper limit at 95\% \CL of the \VHH production cross section is observed (expected) to be at 294 (124) times the SM prediction.
Because of destructive interference with positive \kl in leading \HH production modes (\ggF and VBF), \VHH searches can
make a significant contribution to the overall \HH program near the corresponding minimum in \HH sensitivity.
In particular, in the range of $4<\kl<7$ this search has a sensitivity near other \HH searches.
Scaling cross sections by the ratio of $\pp\to\HH$ to $\pp\to\VHH$ with the same coupling modifiers
is done to interpret the \VHH results in the greater \HH context.
For example, the $\pp\to\HH$ cross section 95\% \CL
expected limit from this \VHH search is about 3--4 times the expected \HH cross section limit from the \bbtt search in this range on the
equivalent data set~\cite{HHbbtautaRun2}. Figure~\ref{fig:limits} shows the SM 95\% \CL cross section upper limits,
as well as those for $\kl=5.5$, which is in the middle of the highlighted region, with other coupling
modifiers set to unity. The upper limits on \VHH and total \HH cross section as functions of \kl, \kvv,
and \kv are shown in Figs~\ref{fig:klscan}, \ref{fig:kvvscan}, and \ref{fig:kvscan}.
The theoretic prediction of \VHH and inclusive \HH production cross sections are shown with the red lines.
The 95\% \CL upper limits are summarized in Table~\ref{tab:95CLlimits}.

\begin{table}[!th]
\centering
\topcaption{Observed and expected 95\% \CL upper limits on the coupling modifiers.}
\label{tab:95CLlimits}
\renewcommand\arraystretch{1.2}
\begin{tabular}{l>{(}r@{, }r<{)}>{(}r@{, }r<{)}>{(}r@{, }r<{)}>{(}r@{, }r<{)}>{(}r@{, }r<{)}}
    \hline
    & \multicolumn{2}{c}{\kl} & \multicolumn{2}{c}{\kvv} & \multicolumn{2}{c}{\kv} & \multicolumn{2}{c}{\kzz} & \multicolumn{2}{c}{\kww}  \\
    \hline
    Observed & $-37.7$ & 37.2 & $-12.2$ & 13.5 & $-3.7$ & 3.8 & $-17.4$ & 18.5 & $-14.0$ & 15.4 \\
    Expected & $-30.1$ & 28.9 & $-7.2$ & 8.9 & $-3.1$ & 3.1 & $-10.5$ & 11.6 & $-10.2$ & 11.6 \\
    \hline
\end{tabular}
\end{table}

\clearpage
\section{Summary}
\label{sec:summary}

A search for Higgs boson pair production in association with a vector boson (\VHH) using a data set
of proton-proton collisions at $\sqrt{s}=13\TeV$, corresponding to an integrated luminosity of 138\fbinv, is presented.
Final states including Higgs boson decay to bottom quarks are analyzed in events where the \PW or \PZ boson decay
to electrons, muons, neutrinos, and hadrons.
An observed (expected) upper limit at 95\% confidence level of \VHH production cross section is set at 294 (124) times
the standard model prediction. Coupling modifiers, defined relative to the standard model
coupling strengths, are scanned and constrained for the Higgs boson trilinear coupling (\kl)
and the coupling between two \PV bosons with two Higgs bosons (\kvv).
The observed (expected) 95\% confidence level limits constrain \kl and \kvv to be
$-37.7<\kl<37.2$ ($-30.1<\kl<28.9$) and $-12.2<\kvv<13.5$ ($-7.2<\kvv<8.9$),
respectively, where each of these constrains assumes the other couplings to be SM-like.

\begin{acknowledgments}
 We congratulate our colleagues in the CERN accelerator departments for the excellent performance of the LHC and thank the technical and administrative staffs at CERN and at other CMS institutes for their contributions to the success of the CMS effort. In addition, we gratefully acknowledge the computing centers and personnel of the Worldwide LHC Computing Grid and other centers for delivering so effectively the computing infrastructure essential to our analyses. Finally, we acknowledge the enduring support for the construction and operation of the LHC, the CMS detector, and the supporting computing infrastructure provided by the following funding agencies: SC (Armenia), BMBWF and FWF (Austria); FNRS and FWO (Belgium); CNPq, CAPES, FAPERJ, FAPERGS, and FAPESP (Brazil); MES and BNSF (Bulgaria); CERN; CAS, MoST, and NSFC (China); MINCIENCIAS (Colombia); MSES and CSF (Croatia); RIF (Cyprus); SENESCYT (Ecuador); ERC PRG, RVTT3 and MoER TK202 (Estonia); Academy of Finland, MEC, and HIP (Finland); CEA and CNRS/IN2P3 (France); SRNSF (Georgia); BMBF, DFG, and HGF (Germany); GSRI (Greece); NKFIH (Hungary); DAE and DST (India); IPM (Iran); SFI (Ireland); INFN (Italy); MSIP and NRF (Republic of Korea); MES (Latvia); LMTLT (Lithuania); MOE and UM (Malaysia); BUAP, CINVESTAV, CONACYT, LNS, SEP, and UASLP-FAI (Mexico); MOS (Montenegro); MBIE (New Zealand); PAEC (Pakistan); MES and NSC (Poland); FCT (Portugal); MESTD (Serbia); MCIN/AEI and PCTI (Spain); MOSTR (Sri Lanka); Swiss Funding Agencies (Switzerland); MST (Taipei); MHESI and NSTDA (Thailand); TUBITAK and TENMAK (Turkey); NASU (Ukraine); STFC (United Kingdom); DOE and NSF (USA).

\hyphenation{Rachada-pisek} Individuals have received support from the Marie-Curie program and the European Research Council and Horizon 2020 Grant, contract Nos.\ 675440, 724704, 752730, 758316, 765710, 824093, 101115353,101002207, and COST Action CA16108 (European Union); the Leventis Foundation; the Alfred P.\ Sloan Foundation; the Alexander von Humboldt Foundation; the Science Committee, project no. 22rl-037 (Armenia); the Belgian Federal Science Policy Office; the Fonds pour la Formation \`a la Recherche dans l'Industrie et dans l'Agriculture (FRIA-Belgium); the Agentschap voor Innovatie door Wetenschap en Technologie (IWT-Belgium); the F.R.S.-FNRS and FWO (Belgium) under the ``Excellence of Science -- EOS" -- be.h project n.\ 30820817; the Beijing Municipal Science \& Technology Commission, No. Z191100007219010 and Fundamental Research Funds for the Central Universities (China); the Ministry of Education, Youth and Sports (MEYS) of the Czech Republic; the Shota Rustaveli National Science Foundation, grant FR-22-985 (Georgia); the Deutsche Forschungsgemeinschaft (DFG), under Germany's Excellence Strategy -- EXC 2121 ``Quantum Universe" -- 390833306, and under project number 400140256 - GRK2497; the Hellenic Foundation for Research and Innovation (HFRI), Project Number 2288 (Greece); the Hungarian Academy of Sciences, the New National Excellence Program - \'UNKP, the NKFIH research grants K 131991, K 133046, K 138136, K 143460, K 143477, K 146913, K 146914, K 147048, 2020-2.2.1-ED-2021-00181, and TKP2021-NKTA-64 (Hungary); the Council of Science and Industrial Research, India; ICSC -- National Research Center for High Performance Computing, Big Data and Quantum Computing, funded by the EU NexGeneration program (Italy); the Latvian Council of Science; the Ministry of Education and Science, project no. 2022/WK/14, and the National Science Center, contracts Opus 2021/41/B/ST2/01369 and 2021/43/B/ST2/01552 (Poland); the Funda\c{c}\~ao para a Ci\^encia e a Tecnologia, grant CEECIND/01334/2018 (Portugal); the National Priorities Research Program by Qatar National Research Fund; MCIN/AEI/10.13039/501100011033, ERDF ``a way of making Europe", and the Programa Estatal de Fomento de la Investigaci{\'o}n Cient{\'i}fica y T{\'e}cnica de Excelencia Mar\'{\i}a de Maeztu, grant MDM-2017-0765 and Programa Severo Ochoa del Principado de Asturias (Spain); the Chulalongkorn Academic into Its 2nd Century Project Advancement Project, and the National Science, Research and Innovation Fund via the Program Management Unit for Human Resources \& Institutional Development, Research and Innovation, grant B05F650021 (Thailand); the Kavli Foundation; the Nvidia Corporation; the SuperMicro Corporation; the Welch Foundation, contract C-1845; and the Weston Havens Foundation (USA).
\end{acknowledgments}

\bibliography{auto_generated}
\cleardoublepage \appendix\section{The CMS Collaboration \label{app:collab}}\begin{sloppypar}\hyphenpenalty=5000\widowpenalty=500\clubpenalty=5000\input{HIG-22-006-public-authorlist.tex}\end{sloppypar}
%%% END EDITABLE REGION %%%
% skeleton_end
\end{document}

%% file: HIG-22-006-public-authorlist.tex
\cmsinstitute{Yerevan Physics Institute, Yerevan, Armenia}
{\tolerance=6000
A.~Hayrapetyan, A.~Tumasyan\cmsAuthorMark{1}\cmsorcid{0009-0000-0684-6742}
\par}
\cmsinstitute{Institut f\"{u}r Hochenergiephysik, Vienna, Austria}
{\tolerance=6000
W.~Adam\cmsorcid{0000-0001-9099-4341}, J.W.~Andrejkovic, T.~Bergauer\cmsorcid{0000-0002-5786-0293}, S.~Chatterjee\cmsorcid{0000-0003-2660-0349}, K.~Damanakis\cmsorcid{0000-0001-5389-2872}, M.~Dragicevic\cmsorcid{0000-0003-1967-6783}, A.~Escalante~Del~Valle\cmsorcid{0000-0002-9702-6359}, P.S.~Hussain\cmsorcid{0000-0002-4825-5278}, M.~Jeitler\cmsAuthorMark{2}\cmsorcid{0000-0002-5141-9560}, N.~Krammer\cmsorcid{0000-0002-0548-0985}, D.~Liko\cmsorcid{0000-0002-3380-473X}, I.~Mikulec\cmsorcid{0000-0003-0385-2746}, J.~Schieck\cmsAuthorMark{2}\cmsorcid{0000-0002-1058-8093}, R.~Sch\"{o}fbeck\cmsorcid{0000-0002-2332-8784}, D.~Schwarz\cmsorcid{0000-0002-3821-7331}, M.~Sonawane\cmsorcid{0000-0003-0510-7010}, S.~Templ\cmsorcid{0000-0003-3137-5692}, W.~Waltenberger\cmsorcid{0000-0002-6215-7228}, C.-E.~Wulz\cmsAuthorMark{2}\cmsorcid{0000-0001-9226-5812}
% ...
\par}
\vskip\cmsinstskip
\dag:~Deceased\\
$^{1}$Also at Yerevan State University, Yerevan, Armenia\\
$^{2}$Also at TU Wien, Vienna, Austria\\
% ...
